\documentclass[aps,prapplied,twocolumn,superscriptaddress,longbibliography,nofootinbib]{revtex4-2}
\usepackage{amsmath,amssymb,bm,mathtools,booktabs,graphicx,microtype,xcolor,array,float}
\usepackage[colorlinks=true,linkcolor=blue,citecolor=blue,urlcolor=blue]{hyperref}
\graphicspath{{../figures/}}
\usepackage{orcidlink}
\newcommand{\ii}{\imath}
\newcommand{\Ep}{E_{\mathrm p}}
\newcommand{\Nfa}[1]{\mathcal N_a^{(#1)}}
\newcommand{\Nfb}[1]{\mathcal N_b^{(#1)}}
\newcommand{\mel}[3]{\langle #1|#2|#3\rangle}

\usepackage[normalem]{ulem}

\usepackage{comment}
\usepackage{cancel}

\begin{document}
\raggedbottom

\title{Occupation-selective photon-pair interactions in a flux-pumped SQUID resonator}
\author{L. N. Ferreira~\orcidlink{0009-0005-8306-1478}}
\affiliation{Departamento de Física, Universidade Estadual de Ponta Grossa 84030-900 Ponta Grossa, Paraná, Brazil}

\author{D. Z. Rossatto~\orcidlink{0000-0001-9432-1603}}
\affiliation{Universidade Estadual Paulista (UNESP), Instituto de Ciências e Engenharia, 18409-010 Itapeva, São Paulo, Brazil}

\author{A. S. M. de Castro~\orcidlink{0000-0002-1521-9342}}
\affiliation{Departamento de Física, Universidade Estadual de Ponta Grossa 84030-900 Ponta Grossa, Paraná, Brazil}

\author{G. D. de Moraes Neto~\orcidlink{0000-0003-4273-8380}}
\email{gdmneto@gmail.com}
\affiliation{Department of Fundamental Sciences, Hainan Bielefeld University of Applied Sciences, Danzhou, Hainan 578101, China}
\begin{abstract}
	Flux-pumped Josephson circuits provide a versatile platform for parametric photon generation and multimode bosonic control. 
	We identify an occupation-conditioned pair interaction that arises when two harmonic normal modes participate in the same driven SQUID coordinate.  
	We consider a degenerate pair process in which two photons are created or annihilated in a signal mode while the occupation of a second harmonic mode remains unchanged. 
	Although it does not exchange excitations, this controller mode modifies both the resonance frequency and the strength of the pair transition. 
	The two effects have different physical origins: diagonal Josephson nonlinearities shift the pair resonance, whereas the nonlinear Josephson matrix element makes the avoided-crossing gap itself depend on the controller occupation.  
	Starting from the full Josephson cosine, we develop complementary time-dependent Schrieffer--Wolff and Magnus descriptions that expose the direct and virtual processes responsible for this interaction, and benchmark the resulting effective theory against a numerically converged Floquet treatment of the unexpanded two-mode circuit Hamiltonian.  
	The occupation dependence produces a two-dimensional pair spectrum that enables selective transitions within the bosonic Fock lattice.  
	When the controller is prepared in a coherent superposition, the same interaction converts this conditional dynamics into signal--controller entanglement and controller-dependent non-Gaussian state generation. Observable-level comparisons further show that the variation of the pair-transition strength cannot be reproduced by occupation-dependent resonance shifts alone.  
	These results establish a boson-controlled parametric interaction in which the state of one harmonic mode controls both the spectral position and the coupling strength of a nonlinear process in another.
\end{abstract}
\maketitle

\section{Introduction}
\label{sec:introduction}

Bosonic modes, particularly those implemented in high-$Q$ superconducting microwave cavities, are attractive quantum degrees of freedom because they combine large, formally infinite-dimensional Hilbert spaces with coherence times that can substantially exceed those of nonlinear ancillary qubits~\cite{Ma2021,Joshi2021,Reagor2016,Milul2023}. 
In circuit QED, superconducting microwave resonators can preserve quantum states on timescales substantially longer than the coherence times of the Josephson qubits used to control them, while remaining rapidly addressable through nonlinear circuit elements~\cite{Blais2021,Reagor2016,Ma2021BosonicControl}. 
This combination has made bosonic modes important resources for quantum memories, encoded quantum information, and hardware-efficient implementations of quantum error correction~\cite{CampagneIbarcq2020,Sivak2023}. 
Exploiting the full oscillator Hilbert space, however, requires more than long coherence: one must be able to introduce nonlinear interactions that distinguish and coherently connect selected states without sacrificing the advantages of the bosonic mode.

Selective interactions provide a route to achieve this control. The basic idea has a long history in theoretical proposals in cavity QED and trapped-ion physics, where Stark shifts and dispersive energy corrections can be used to bring a chosen transition into resonance while neighboring transitions remain off-resonant~\cite{Franca2001,Solano2005}. 
In superconducting cavities, the same general principle appears in photon-number-selective phase gates~\cite{Heeres2015}, ancilla-assisted photon-number-dependent Hamiltonian engineering~\cite{Wang2021PND}, stimulated Josephson nonlinearities~\cite{Vrajitoarea2020} and spectrally selective control in coupled Kerr oscillators~\cite{Damas2026}. 
Parametric two-photon driving provides another powerful control mechanism since it changes the excitation number by two and couples states within a fixed-parity sector of Fock space~\cite{Puri2017,Grimm2020,Eriksson2024}. 
Together, these developments show that the regular spectrum of a harmonic mode need not prevent selective quantum control once an appropriate nonlinear resource is introduced.

The multimode setting is richer because the state of one bosonic mode can influence operations performed on another. 
In ancilla-mediated architectures, this dependence can be a source of coherent crosstalk: photons occupying a nominally inactive mode Stark shift the shared nonlinear ancilla and thereby modify the control Hamiltonian seen by the target mode~\cite{You2024Crosstalk}. 
This problem has motivated the development of control pulses that remain robust against the occupation of neighboring modes, while experiments have also demonstrated fast multioscillator control with low crosstalk~\cite{Diringer2024}. 
At the same time, controlled interactions between bosonic modes are themselves valuable resources, as demonstrated by deterministic entangling operations between superconducting cavities~\cite{Gao2019}. 
This suggests a complementary viewpoint: rather than treating every state dependence generated by a shared nonlinear element as an imperfection to be suppressed, one can ask whether it can be promoted to the interaction responsible for the control.

Flux-pumped Josephson circuits provide a natural setting for this question. 
When a SQUID terminates a resonator, an ac flux modulates its Josephson energy and therefore the electrical boundary condition. 
For a discrete cavity mode, modulation near twice the mode frequency activates degenerate parametric photon-pair generation~\cite{Wilson2010}; in an open transmission line, the related sum-frequency process provides the superconducting-circuit realization of the dynamical Casimir effect
~\cite{Johansson2009,Wilson2011,Schneider2020}. 
Related flux-pumped Josephson devices support a much broader multimode parametric toolbox, including nondegenerate parametric amplification and oscillation~\cite{Wustmann2017,Simoen2015,Bengtsson2018}, multimode squeezing and entanglement~\cite{Chang2018,Schneider2020}, synthetic-frequency hopping and pairing~\cite{Lee2020,Busnaina2024}, and higher-order wave-mixing processes~\cite{Chang2020ThreePhoton,Mundhada2019,Eriksson2024}. 
A common feature of these devices is that several electromagnetic normal modes can participate in the same flux-modulated Josephson coordinate~\cite{Nigg2012,Minev2021}.
\begin{figure*}[t]
\centering
\includegraphics[width=1.0\textwidth]{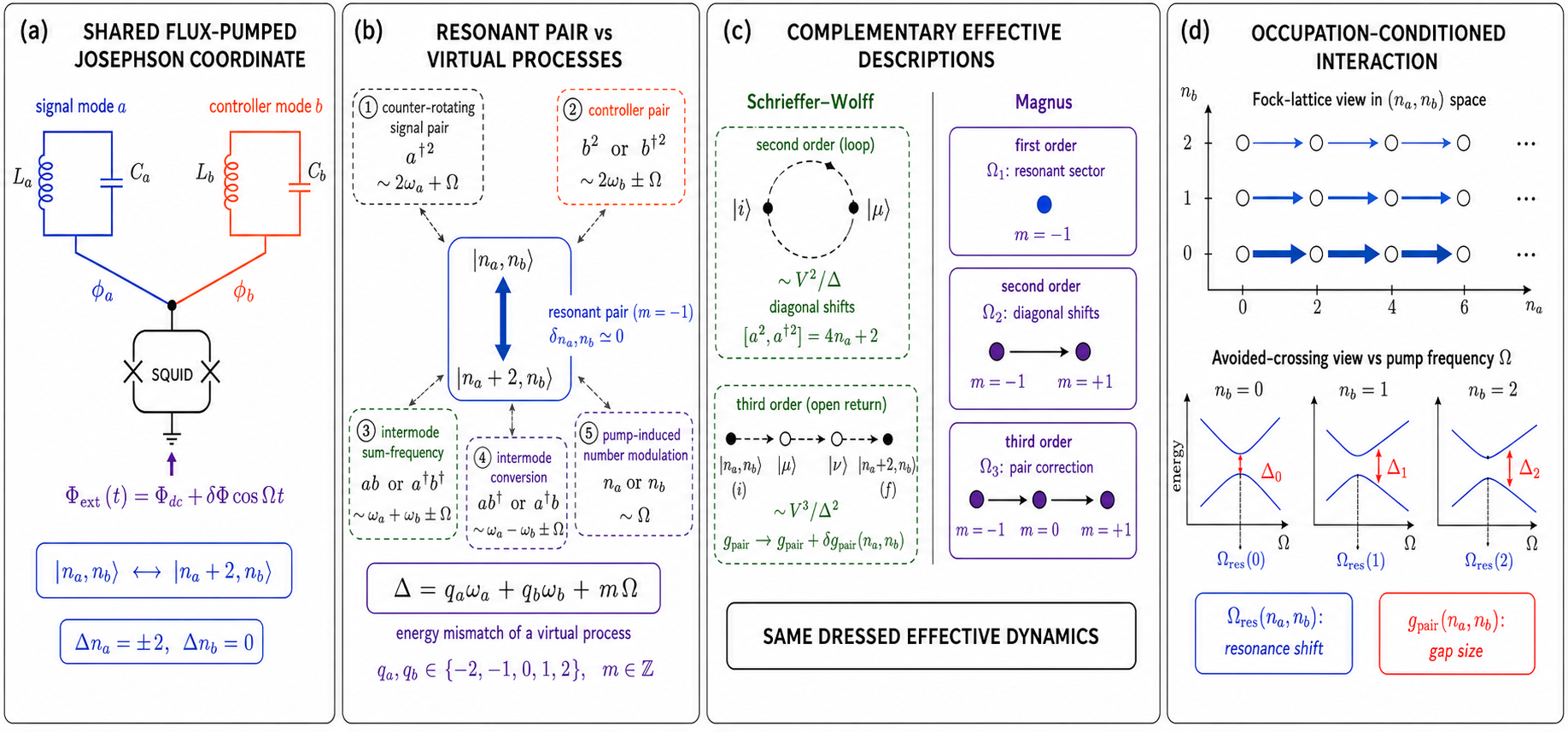}
\caption{\textbf{From a shared flux-pumped Josephson coordinate to an occupation-conditioned pair interaction.}
	(a) The signal mode $a$ and controller mode $b$ have finite zero-point phase across the same SQUID, modulated by $\Phi_{\mathrm{ext}}(t)$. The target process creates or annihilates a signal-mode photon pair while leaving the controller occupation unchanged, $\Delta n_a=\pm2$ and $\Delta n_b=0$. 
	(b) Near the pair resonance, the $m = -1$ transition $|n_a,n_b\rangle\leftrightarrow|n_a+2,n_b\rangle$, with $\delta_{n_{a}, n_{b}} \simeq 0$, is retained explicitly. Counter-rotating signal-pair, controller-pair, intermode sum--frequency, intermode-conversion, and pump-induced number-modulation processes contribute through off-resonant virtual excursions with mismatch $\Delta=q_a\omega_a+q_b\omega_b+m\Omega$, where $q_{a}, q_{b} \in \{-2, -1, 0, 1, 2\}$ and $m \in \mathbb{Z}$. 
	(c) Time-dependent Schrieffer--Wolff and Magnus expansions organize the same microscopic Fourier vertices in complementary ways. In the Schrieffer--Wolff picture, second-order closed virtual loops predominantly renormalize the diagonal spectrum, including $[a^2,a^{\dagger2}]=4n_a+2$. While third-order paths can return to the resonant pair sector and renormalize the pair vertex, $g_{\mathrm{pair}}\rightarrow g_{\mathrm{pair}}+\delta g_{\mathrm{pair}}(n_a,n_b)$. In the Magnus picture, the corresponding first-, second-, and third-order contributions arise from the time ordering of the same resonant and off-resonant Fourier processes. After dressing or micromotion reconstruction, both descriptions yield the same physical effective dynamics. 
	(d) The resulting interaction is resolved over the $(n_a,n_b)$ Fock lattice. The controller occupation changes both the pair-resonance position $\Omega_{\mathrm{res}}(n_a,n_b)$ and the resonant pair coupling $g_{\mathrm{pair}}(n_a,n_b)$, producing controller-dependent crossing positions $\Omega_{\mathrm{res}}(0,1,2)$ and unequal avoided-crossing gaps $\Delta_{0,1,2}$.}
\label{fig:mechanism}
\end{figure*}

Once several modes participate in the same Josephson phase, their occupations can enter the nonlinear matrix elements themselves. Cross-Kerr coupling provides the familiar diagonal manifestation: the occupation of one mode shifts transition frequencies in another~\cite{Nigg2012,Minev2021,Schuster2007}. 
Beyond such diagonal shifts, exact Josephson matrix elements can produce occupation-dependent transition amplitudes~\cite{Armour2015,Baskov2025}. 
High-order exchange processes have been experimentally realized through Raman-assisted eight-wave mixing and, more recently, through a four-to-one conversion used to produce photon-number-selective quartet dissipation~\cite{Mundhada2019,Vanselow2026}. 
Here we focus on a complementary coherent pair process in which the controller mode does not change its occupation during the resonant transition,
\begin{equation}
	|n_a,n_b\rangle	\longleftrightarrow |n_a+2,n_b\rangle ,
	\label{eq:intro-transition}
\end{equation}
where mode $a$ is the signal and mode $b$ is a harmonic controller. 
Related theoretical works have proposed qubit-controlled squeezing operations~\cite{DelGrosso2025,Blumenthal2026,Hope2026}; in contrast, the controller in the present work is itself a harmonic bosonic mode, and its Fock occupation $n_b$ remains unchanged during the pair transition. 
Figure~\ref{fig:mechanism} summarizes the physical mechanism developed throughout this work: both modes participate in the same flux-pumped Josephson coordinate, the desired pair process is retained while off-resonant Josephson processes generate virtual corrections, and complementary Schrieffer--Wolff and Magnus descriptions~\cite{SchriefferWolff1966,Bravyi2011,Magnus1954,Blanes2009} connect these microscopic processes to the occupation-dependent resonance positions and pair-transition strengths obtained in the two-dimensional Fock-space representation.

The occupation $n_b$ enters this pair transition in two physically distinct ways. 
First, diagonal Josephson nonlinearities make the energy difference between the two states depend on the controller occupation, so that different $n_b$ sectors become resonant at different pump frequencies. 
Second, the off-diagonal Josephson matrix element that couples the two states also depends on $n_b$, so that the avoided-crossing gap changes from one controller sector to another even after each transition is individually tuned to resonance. 
The first effect is spectral conditioning and includes the familiar cross-Kerr contribution. 
The second is matrix-element conditioning and represents a change in the strength of the resonant pair process itself. 
Resolving both the resonance position and the avoided-crossing gap across the $(n_a,n_b)$ Fock lattice therefore provides a direct way to distinguish the two mechanisms.

This matrix-element dependence is already present in the unexpanded Josephson cosine. 
Because the phase across the nonlinear element is a sum of commuting normal-mode coordinates, the Fock-state matrix element factorizes mode by mode. 
For the transition in Eq.~\eqref{eq:intro-transition}, the unchanged controller contributes a number-dependent diagonal factor, so the pair matrix element of mode $a$ retains information about $n_b$ even though $\Delta n_b=0$. 
We use this microscopic result as the starting point for two complementary analytical constructions. 
A time-dependent Schrieffer--Wolff transformation organizes the interaction in terms of resonant and virtual processes and makes the relevant intermediate states and energy mismatches explicit. 
An independent Magnus expansion organizes the same Fourier vertices through their time ordering. Together they separate the direct Josephson contribution from virtual Stark shifts and higher-order renormalizations of the pair vertex.

The analytical hierarchy is benchmarked against a numerically converged Floquet calculation that retains the unexpanded two-mode Josephson cosine and the exact Fourier harmonics of the flux modulation. 
The resulting occupation-resolved spectrum supports the selective excitation of an interior pair transition in the two-dimensional Fock lattice. 
When the controller is prepared in a coherent superposition, the occupation-conditioned interaction acquires a genuinely quantum role: different controller branches drive different signal dynamics, generating signal--controller entanglement and controller-dependent non-Gaussian states. 
We also compare reduced observable-level models across multiple controller sectors to determine whether occupation-dependent resonance shifts alone can reproduce the calculated pair gaps.

Finally, we connect the reduced two-mode description to experimentally established multimode SQUID circuits.  
The normal-mode frequencies and zero-point phases used below are treated as representative circuit-level parameters rather than as the output of a newly designed fabrication geometry. 
In a concrete implementation, these quantities can be obtained from the measured or simulated electromagnetic spectrum and the participation of each mode in the common Josephson coordinate.  
Additional electromagnetic and Josephson-plasma modes then enter through their pump-assisted detunings and coupling matrix elements, providing a quantitative criterion for when the occupation-conditioned pair dynamics can be isolated in a realistic multimode device.

The remainder of this article is organized as follows.  
Section~\ref{sec:model} introduces the multimode Josephson circuit, the shared nonlinear coordinate, and the two-mode Hamiltonian used throughout the work.  
Section~\ref{sec:pairmechanism} develops the occupation-conditioned pair interaction, beginning with the exact Josephson matrix element and then using complementary Schrieffer--Wolff and Magnus constructions to identify the direct and virtual processes that shape the effective coupling.  
Section~\ref{sec:spectrum} analyzes the resulting occupation-resolved pair spectrum, while Sec.~\ref{sec:validation} establishes the quantitative regime of validity through comparisons with the unexpanded Floquet model and systematic convergence tests.  
Section~\ref{sec:control} applies the interaction to selective pair transfer inside the Fock lattice.  
Sections~\ref{sec:qcontroller} and \ref{sec:nonGaussian} promote the controller to a quantum degree of freedom and study the resulting entanglement and controller-conditioned non-Gaussian dynamics.  
Section~\ref{sec:comparator} tests the occupation dependence at the observable level using reduced comparator models, and Sec.~\ref{sec:implementation} connects the two-mode theory to multimode circuit implementations and spectator-mode isolation.  
Section~\ref{sec: conclusion} summarizes the main results and their implications. 
The Appendices collect the circuit reduction, analytical derivations, numerical procedures, convergence tests, control definitions, spectator analysis, and reproducibility details used throughout the article.

\section{Multimode Josephson circuit and two-mode reduction}
\label{sec:model}

\subsection{Shared Josephson coordinate}
\label{subsec:shared-coordinate}

The starting point is a multimode superconducting circuit in which several electromagnetic normal modes have finite field amplitude across the same flux-tunable Josephson element.  
The circuit interpretation is illustrated in Fig.~\ref{fig:mechanism}(a).  
At the dc operating point, the SQUID contributes a linear inductance to the electromagnetic boundary condition and must therefore be included when the normal modes of the circuit are determined.  
Black-box and energy-participation quantization provide general routes from a distributed electromagnetic structure to this normal-mode description~\cite{Nigg2012,Minev2021}, while quarter-wavelength ($\lambda/4$) transmission-line resonators terminated by a SQUID provide a particularly straightforward realization~\cite{Bourassa2012,Simoen2015,Chang2018}.

Once the linearized circuit has been solved, the dimensionless phase across the nonlinear element can be expanded in the resulting normal-mode coordinates as
\begin{equation}
    \hat\varphi_J = \sum_\mu \varphi_\mu \left( a_\mu+a_\mu^\dagger \right),
    \label{eq:phase-multimode}
\end{equation}
where $a_\mu$ annihilates a normal mode of angular frequency $\omega_\mu$ and $\varphi_\mu$ is its dimensionless zero-point phase across the Josephson coordinate.  
Physically, $\varphi_\mu$ measures how strongly the vacuum fluctuations of the mode $\mu$ sample the nonlinear element.  
It is therefore the natural parameter controlling the strength with which that mode enters the Josephson nonlinearity.

We retain two modes with appreciable participation in the driven nonlinear coordinate and denote them by $a$ and $b$,
\begin{equation}
    \hat\varphi_J = \varphi_a(a+a^\dagger) + \varphi_b( b + b^\dagger), 
    \,\,\,
    n_a=a^\dagger a,
    \,\,\,
    n_b=b^\dagger b.
    \label{eq:phase}
\end{equation}
Mode $a$ will be the signal mode whose degenerate pair transitions are driven, while mode $b$ acts as the harmonic controller.  
Before the nonlinear Josephson potential is restored, both are ordinary harmonic normal modes; their self-Kerr, cross-Kerr, and occupation-dependent parametric interactions all originate from their common participation in $\hat\varphi_J$.

A symmetric SQUID, with identical junction energies $E_{J0}$ and negligible loop inductance, behaves as a single effective Josephson element with a flux-tunable Josephson energy~\cite{Sandberg2008,Wustmann2013},
\begin{equation}
    E_J(t) = 2E_{J0} \cos\!\left[ \pi\Phi_{\rm ext}(t)/\Phi_0 \right],
    \label{eq:EJexact}
\end{equation}
where $\Phi_{0} = h / (2e)$ is the quantum magnetic flux. A time-dependent magnetic flux can then be used to parametrically modulate the Josephson energy and, consequently, the effective inductance of the circuit~\cite{Sandberg2008,Wilson2010}. 
We take the applied flux to be
\begin{equation}
    \Phi_{\rm ext}(t) = \Phi_{\rm dc} + \delta \Phi \cos\Omega t ,
    \label{eq:flux}
\end{equation}
where $\Omega$ is the external pump frequency. 
In the weak-pump regime, retaining only terms linear in the modulation amplitude, the Josephson energy may be approximated as~\cite{Wustmann2013}
\begin{equation}
    E_{J}(t) \simeq E_{0} + E_{p} \cos\Omega t.
    \label{eq:EJweak}
\end{equation}

For the analytical description in the weak-pump regime, we write the zero-pump dc Josephson energy as
\begin{equation}
    E_0 \equiv 2E_{J0} \cos\!\left( \pi\Phi_{\rm dc}/\Phi_0 \right)
    \label{eq:E0}
\end{equation}
and denote by $\Ep$ the leading cosine harmonic at the pump frequency.  
To first order in the modulation amplitude,
\begin{equation}
    \Ep \simeq -2E_{J0} \frac{\pi\delta\Phi}{\Phi_0} \sin\!\left( \pi\Phi_{\rm dc}/\Phi_0 \right).
    \label{eq:Epweak}
\end{equation}
The numerical Floquet calculations do not employ this weak-modulation approximation. 
Instead, they retain the complete Bessel-function expansion of Eq.~\eqref{eq:EJexact}, including the pump-induced correction to the zero-frequency component and all higher harmonics. 
Such higher-order corrections can become relevant outside the perturbative pump regime~\cite{Kochetov2015,Boutin2017,Baskov2025}. 
The corresponding Fourier coefficients and the circuit-to-mode reduction are given in Appendix~\ref{app:circuit}.

\subsection{Residual nonlinear Hamiltonian}
\label{subsec:residual-hamiltonian}

The separation between the linear circuit and the residual Josephson interaction is important. Expanding the static Josephson potential around $\hat\varphi_J=0$ gives
\begin{equation}
    -E_0\cos\hat\varphi_J = - E_0 + \frac{E_0}{2}\hat\varphi_J^2 
                            - \frac{E_0}{24}\hat\varphi_J^4 +\cdots .
    \label{eq:static-cosine-expansion}
\end{equation}
The quadratic term defines the linearized SQUID inductance and has already been included in the normal-mode frequencies $\omega_\mu$.  
Reintroducing it into the interaction Hamiltonian would therefore count the same linear physics twice. The static Josephson contribution that remains after solving the normal-mode problem begins in quartic order.

The ac modulation plays a different role.  
Because it modulates the complete Josephson energy, it acts on the full cosine and consequently generates both a quadratic parametric drive and higher-order nonlinear driven terms.  
Keeping the first pump harmonic explicitly, the analytical two-mode Hamiltonian becomes
\begin{eqnarray}
   H(t) & = & \hbar\omega_a n_a + \hbar\omega_b n_b \nonumber\\
        & + &  E_0 \left( 1-\cos\hat\varphi_J -\frac{\hat\varphi_J^2}{2} \right) \nonumber\\
        & + & \Ep\cos\Omega t \left( 1-\cos\hat\varphi_J \right),
    \label{eq:fullH} 
\end{eqnarray}
where $\hat{\varphi}_{J} = \pi \hat{\Phi} / \Phi_{0}$. 
Equation~\eqref{eq:fullH} is the analytical starting point used throughout the perturbative treatment.  
The residual static interaction begins with the familiar quartic Josephson nonlinearity, while the pumped part begins quadratically and already contains the degenerate pair process of mode $a$.  
The same driven cosine also contains quartic, sextic, and higher-order terms, which make the parametric matrix elements depend on the occupations of the participating modes.

Two distinct representations of this Hamiltonian can serve different purposes. 
Expanding $\cos\hat\varphi_J$ in powers of the zero-point phase exposes the microscopic origin and perturbative order of the effective interactions.  
In particular, it distinguishes direct Josephson terms from virtual processes generated by off-resonant vertices.  
By contrast, retaining the full cosine preserves the exact Fock-state matrix elements when the occupied state explores a range of phases over which a low-order expansion is no longer quantitatively sufficient.  
We therefore use the phase-expanded Hamiltonian to interpret the interaction and the unexpanded two-mode cosine as the quantitative Floquet reference.

Both descriptions start from the same normal-mode frequencies, zero-point phases, dc flux bias, and pump waveform. 
We compare them quantitatively after accounting for the corresponding perturbative dressing or Floquet micromotion, as discussed in Sec.~\ref{sec:validation} and Appendix~\ref{app:sw}.

\subsection{Representative mode-level benchmark}
\label{subsec:benchmark}

To make the different contributions visible on experimentally relevant energy scales, we use the representative operating point listed in Table~\ref{tab:benchmark}.  
The chosen zero-point phases, $\varphi_a\simeq 0.23$ and $\varphi_b\simeq 0.31$, are sufficiently large for the higher-order Josephson matrix elements to produce a measurable dependence on the controller occupation, while still allowing the hierarchy of phase order and virtual-process order to be examined systematically.  
The modulation remains weak compared with the static Josephson energy, so the nontrivial occupation dependence arises primarily from the nonlinear structure of the shared Josephson coordinate rather than from an exceptionally strong pump.

The zero-point phases correspond to junction-coordinate modal impedances of approximately $112~\Omega$ and $195~\Omega$ through
\begin{equation}
    \varphi_\mu^2 = \frac{\pi Z_\mu}{R_q},
    \qquad
    R_q=\frac{h}{(2e)^2},
    \label{eq:modal-impedance}
\end{equation}
where $Z_\mu$ refers to the impedance associated with the Josephson coordinate of the normal mode rather than to the characteristic impedance of an uncoupled transmission line.  
Zero-point phase fluctuations on this scale are compatible with modern multimode Josephson circuits~\cite{Makihara2024,Vanselow2026}.
\begin{table}[tbp]
\caption{Representative mode-level benchmark used throughout the calculations. Frequencies $f_j = \omega_j/2\pi$ are cyclic frequencies.  $E_{J0}$ is the Josephson energy of each junction in the symmetric-SQUID convention of Eq.~\eqref{eq:EJexact}.}
\label{tab:benchmark}
\vspace{3pt}
\begin{ruledtabular}
\begin{tabular}{lc}
quantity & value\\
\hline
$f_a$ & $6.4856317$ GHz\\
$f_b$ & $11.8617261$ GHz\\
$\varphi_a$ & $0.233502$\\
$\varphi_b$ & $0.307952$\\
$E_{J0}/h$ & $20$ GHz\\
$\Phi_{\rm dc}/\Phi_0$ & $0.459359$\\
$\delta\Phi/\Phi_0$ & $6.42\times10^{-4}$\\
$\Ep/h$ & $\simeq-80$ MHz
\end{tabular}
\end{ruledtabular}
\end{table}

The same modal parameters can be reproduced by a minimal two-node circuit in which two LC branches share a common SQUID branch~\cite{Felicetti2014}.  
This construction provides a transparent circuit-level realization of the benchmark and fixes the connection between component values, normal-mode frequencies, and zero-point phases. 
The corresponding capacitances, inductances, linearized SQUID inductance, and static junction-asymmetry test are given in Appendix~\ref{app:circuit}.

A distributed resonator generally contains additional electromagnetic and Josephson-plasma modes.  
The two-mode Hamiltonian above should therefore be viewed as a spectrally selected sector of the underlying multimode Josephson circuit: modes $a$ and $b$ are retained because they participate in the near-resonant dynamics of interest, while additional modes remain inactive when their pump-assisted transitions are sufficiently detuned on the scale of their coupling matrix elements.  
The quantitative Floquet-isolation condition for those additional modes is developed in Sec.~\ref{sec:implementation} and Appendix~\ref{app:spectator}.

\section{Occupation-conditioned pair interaction}
\label{sec:pairmechanism}

The pair process of Eq.~\eqref{eq:intro-transition} can be understood at two complementary levels. The unexpanded Josephson cosine directly gives the physical Fock-state matrix element.  
A phase expansion then resolves that matrix element into circuit nonlinearities and virtual paths. 
We begin with the exact result because it makes the controller dependence transparent before any perturbative transformation is introduced.

\subsection{Exact matrix element of the shared cosine}
\label{sec:Exmatel}

To determine how the occupation of one mode can modify the amplitude of a parametric process involving another mode, it is convenient to calculate the matrix element of the Josephson term between Fock states directly. 
The purpose of this analysis is to quantitatively identify how the transition involving the creation or annihilation of a photon pair in mode $a$ depends on the quantum state of a second mode $b$, even when the photon number of this second mode remains unchanged during the transition.

Let us consider the Josephson phase expressed in terms of two bosonic modes, as given in Eq.~\eqref{eq:phase}. 
Since operators associated with distinct modes commute, $[\hat{a},\hat{b}]=[\hat{a},\hat{b}^{\dagger}]=0$, the Josephson exponential operator can be factorized as
\begin{equation}
e^{\ii\hat{\varphi}_{J}} = e^{\ii\varphi_{a}(\hat{a} 
                         + \hat{a}^{\dagger})} e^{\ii\varphi_{b}(\hat{b} + \hat{b}^{\dagger})}.
\label{Jphas}
\end{equation}

This factorization allows the contributions from the two modes to be evaluated separately. 
For a single harmonic oscillator, the matrix element of the displacement operator is known exactly. For $d\geq 0$,
\begin{equation}
\left\langle n + d \left| e^{\ii\varphi(\hat{a} + \hat{a}^{\dagger})} \right|n \right \rangle 
= e^{-\frac{\varphi^{2}}{2}} (\ii\varphi)^{d} \sqrt{\frac{n!}{(n + d)!}} L_{n}^{(d)}(\varphi^{2}),
\label{eq:laguerre-general}
\end{equation}
where $L_{n}^{(d)}$ is an associated Laguerre polynomial. 
This identity makes it possible to treat the Josephson cosine without resorting, at this stage, to the expansion in Eq.~\eqref{eq:static-cosine-expansion}. 
The transition of interest is the creation of a photon pair in mode $a$, $\lvert n_a,n_b\rangle \longrightarrow \lvert n_a+2,n_b\rangle$, such that $\Delta n_a=2$ and $\Delta n_b=0$. 
Mode $b$ therefore does not exchange photons during this transition. 
Nevertheless, because its phase coordinate appears in the same Josephson operator, its occupation can modify the process amplitude. 
It is precisely this dependence that we seek to determine. 
For mode $a$, we have $d=2$. Applying the general expression above,
\begin{align}  
\left\langle n_{a} + 2\left| e^{\ii \varphi_{a}(\hat{a} + \hat{a}^{\dagger})} \right|n_{a} \right\rangle & = e^{\frac{-\varphi_{a}^{2}}{2}} (\ii\varphi_{a})^{2} \notag\\ 
&\times \sqrt{\frac{n_{a}!}{(n_{a} + 2)!}} L_{n_{a}}^{(2)}(\varphi_{a}^{2}).  
\end{align}  

The factor $\varphi_a^2$ is consistent with the creation of two quanta in mode $a$, whereas \begin{equation}
    \sqrt{\frac{n_a!}{(n_a+2)!}},  
\end{equation}
represents the bosonic ladder structure of the transition. 
The function $L_{n_a}^{(2)}$ provides the correction associated with the finite magnitude of the phase fluctuations. 
For mode $b$, the situation is different. Since its occupation remains unchanged, we have $d=0$
\begin{equation}
\left\langle n_{b} \left| e^{\ii \varphi_{b}(\hat{b} + \hat{b}^{\dagger})} \right|n_{b} \right\rangle = e^{\frac{-\varphi_{b}^{2}}{2}} L_{n_{b}}(\varphi_{b}^{2}).  
\end{equation}

Thus, although mode $b$ remains in the same Fock state, its matrix element is not a constant independent of $n_b$. We therefore define
\begin{equation}
C_{b}(n_{b}) = e^{\frac{-\varphi_{b}^{2}}{2}} L_{n_{b}}(\varphi_{b}^{2}),
\label{eq:controllerfactor}
\end{equation}
and $C_b(n_b)$ can be interpreted as the modulation factor of the pair-creation amplitude induced by the Fock state of mode $b$. 
Considering now the Fourier component of the pump that is resonant with the creation of two photons in mode $a$, the corresponding matrix element can be written as
\begin{eqnarray} 
\frac{M_{n_{a}, n_{b}}^{(1)}}{\hbar} & = &  \frac{E_{p}}{2\hbar} e^{-\frac{\varphi_{a}^{2} + \varphi_{b}^{2}}{2}} \varphi_{a}^{2} \sqrt{\frac{n_{a}!}{(n_{a} + 2)!}} \nonumber \\  
&\times&  L_{n_{a}}^{(2)}(\varphi_{a}^{2}) L_{n_{b}}(\varphi_{b}^{2}),
\label{eq:exactpairmatrix}
\end{eqnarray}
where $E_p$ is the amplitude of the modulation of the Josephson element. 
The overall sign arising from $\ii^2=-1$ depends on the convention adopted for the pump term and does not affect the transition rate, which depends on $\lvert M_{n_a,n_b}^{(1)}\rvert^2$. 
The expression above can be written in a particularly transparent form
\begin{equation}
M_{n_{a}, n_{b}}^{(1)} = M_{n_{a}}^{(1)} C_{b}(n_{b}),  
\end{equation}
where
\begin{equation}
    M_{n_{a}}^{(1)} \propto e^{\frac{-\varphi_{a}^{2}}{2}} \varphi_{a}^{2} \sqrt{\frac{n_{a}!}{(n_{a} + 2)!}} L_{n_{a}}^{(2)} (\varphi_{a}^{2}),
\end{equation}
contains the dependence associated with the transition in mode $a$, whereas $C_b(n_b)$, given by Eq.~\eqref{eq:controllerfactor}, contains the entire dependence associated with the occupation of mode $b$. 
This result has an important physical interpretation. 
The transition $\lvert n_a,n_b\rangle \rightarrow \lvert n_a+2,n_b\rangle$ does not involve an actual exchange of photons with mode $b$, since $\Delta n_b=0$. 
Nevertheless, the amplitude of the process depends on $n_b$.
Mode $b$ therefore acts as a quantum controller of the pair-creation amplitude in mode $a$. 
To understand the origin of this dependence within a perturbative description, we can expand the factor $C_b(n_b)$ in powers of $\varphi_b$. 
For $\varphi_b\ll 1$, we obtain
\begin{equation}
C_{b}(n_{b}) = 1 - \left(n_{b} + \frac{1}{2} \right) \varphi_{b}^{2} + \mathcal{O}(\varphi_{b}^{4}),
\label{eq:controllerfactor-expansion}
\end{equation}
and the leading term that explicitly depends on the occupation is $-n_b\varphi_b^2$. 
This expression clearly shows how the population of mode $b$ modifies the amplitude of the pair-creation process in mode $a$. 
Even in the absence of an actual exchange of quanta with mode $b$, the quantum structure of this mode modifies the matrix element of the Josephson operator.

This approach is particularly useful because the Josephson term contains the cosine of the phase rather than merely a low-order polynomial interaction.
By initially retaining the Josephson operator in its unexpanded form, we can obtain an exact expression for the matrix element and thereby directly reveal its dependence on the zero-point phase amplitudes $\varphi_a$ and $\varphi_b$, as well as on the occupation numbers $n_a$ and $n_b$. 
The expansion in powers of the phase can then be performed subsequently, allowing us to identify which circuit nonlinearities are responsible for the observed dependence. 
At the reference point, the exact values of $C_b(n_b)$ for $n_b=0,1,2,3$ are approximately $0.954$, $0.863$, $0.777$, and $0.695$, respectively. 
Thus, the controller changes the pair-transition matrix element by an experimentally detectable amount within the lowest Fock sectors. 
Appendix~\ref{app:exact} derives Eq.~\eqref{eq:exactpairmatrix} and provides the corresponding direct coefficients up to the sixth order in the Josephson phase.

\subsection{Direct nonlinearities and the occupation-resolved normal form}
\label{sec:dirnoloccup}

In the previous section, we showed that the matrix element of the Josephson cosine for the pair transition $\lvert n_a,n_b\rangle \longrightarrow \lvert n_a+2,n_b\rangle$ depends on the occupation $n_b$, even though no quanta are transferred to mode $b$. 
In this section, we recast this result in terms of an effective Hamiltonian to systematically identify how mode occupations modify the pairing amplitude and how these contributions extend to higher orders. 

The diagonal part, obtained from the static expansion of the Josephson cosine, is
\begin{eqnarray}
\frac{H_{\text{diag}}}{\hbar} & = & \bar{\omega}_{a} \hat{n}_{a} + \bar{\omega}_{b} \hat{n}_{b} 
- K_{a} \hat{n}_{a} (\hat{n}_{a} - 1) - K_{b} \hat{n}_{b} (\hat{n}_{b} - 1) \notag \\ 
& - & \chi_{ab} \hat{n}_{a} \hat{n}_{b} + \text{higher occupation terms}, 
\label{eq:diagconvention} 
\end{eqnarray}
with
\begin{equation}
    \hbar K_{a} = \frac{E_{0} \varphi_{a}^{4}}{4}, 
    \quad 
    \hbar K_{b} = \frac{E_{0} \varphi_{b}^{4}}{4}, 
    \quad 
    \hbar \chi_{ab} = E_{0} \varphi_{a}^{2} \varphi_{b}^{2}.
\label{eq:kerrleading}
\end{equation}

The parameters $K_a$ and $K_b$ are the self-Kerr coefficients, whereas $\chi_{ab}$ describes the cross-Kerr interaction. 
The frequencies $\bar{\omega}_j$ already include occupation-independent renormalizations. 
For the resonant dynamics of pair creation and annihilation in mode $a$, it is convenient to separate the diagonal part from the off-diagonal part,
\begin{equation}
    \frac{H_{\text{eff}}}{\hbar} = F (\hat{n}_{a}, \hat{n}_{b}) + \{\hat{a}^{\dagger 2} G(\hat{n}_{a}, \hat{n}_{b}) + \text{H.c.}\},
\label{eq:compact}
\end{equation}
here, $F$ contains the occupation-dependent energies, whereas $G$ is the effective coefficient of the pairing process. 
Since the Josephson nonlinearity can cause the amplitude of this process to depend on the Fock state, we write
\begin{eqnarray}
G (\hat{n}_{a}, \hat{n}_{b}) & = &\Gamma_{00} + \Gamma_{10} \hat{n}_{a} 
+ \Gamma_{01} \hat{n}_{b} + \Gamma_{20} \mathcal{N}_{a}^{(2)} \notag \\
& + & \Gamma_{11} \hat{n}_{a} \hat{n}_{b} + \Gamma_{02} \mathcal{N}_{b}^{(2)} + \ldots, 
\label{eq:G}
\end{eqnarray}
where $\mathcal{N}_{j}^{(2)}=\hat{n}_{j}(\hat{n}_{j}-1)$. 
This expansion distinguishes the dependence of the pairing process on the occupation of each mode and on products of their occupations. 
For a specific transition, the physical matrix element is
\begin{align}
g_{n_{a}, n_{b}} & = \frac{1}{\hbar} \langle \hat{n}_{a} + 2, \hat{n}_{b}| \hat{H}_{\text{eff}} |\hat{n}_{a}, \hat{n}_{b} \rangle,\\
                 & = \sqrt{ (n_{a} + 1) (n_{a} + 2) } G (n_{a}, n_{b}).
\label{eq:physicalg}
\end{align}

The factor $\sqrt{(n_a+1)(n_a+2)}$ is simply the bosonic factor associated with the creation of two quanta. 
Therefore, $G(n_a,n_b)$ contains the additional dependence induced by the circuit nonlinearity. 
At direct quartic order, the leading terms of this dependence can be written as
\begin{equation}
    G(n_{a}, n_{b}) \simeq g_{0} + g_{a} n_{a} + g_{b} n_{b}, 
    \quad 
    \hbar g_{b} = - \frac{E_{p} \varphi_{a}^{2} \varphi_{b}^{2}}{4}.
\label{eq:leadinggb}
\end{equation}

The coefficient $g_0$ represents the bare pairing amplitude, $g_a$ describes its leading correction due to the occupation of mode $a$ itself, and $g_b$ quantifies the modification induced by the occupation of the controller mode $b$. 
Thus, $g_b n_b$ is the perturbative counterpart of the $n_b$-dependence previously obtained from the exact factor given in Eq.~\eqref{eq:controllerfactor}.

Higher orders in the Josephson expansion and virtual processes correct these coefficients and generate additional terms, such as $n_a n_b$ and $n_b(n_b-1)$. 
The Schrieffer–Wolff and Magnus transformations considered below make it possible to determine these corrections without eliminating the resonant pairing process connecting 
$\lvert n_a,n_b\rangle \longrightarrow \lvert n_a+2,n_b\rangle$.

This form is particularly useful because it establishes a direct correspondence between the exact result of the previous section and the perturbative description: the former provides the full dependence on $n_b$, whereas the latter organizes it into a systematic expansion in occupation numbers and orders of the nonlinearity.

\subsection{Which processes are retained and which are virtual}
\label{subsec:vertices}

The phase-expanded Hamiltonian, written as $H(t) = H_0 + V(t)$, where $H_{0} = \hbar \omega_{a} n_{a} + \hbar \omega_{b} n_{b}$, contains many Fourier-resolved bosonic monomials. A term that changes the occupations by $(q_a,q_b)$ and carries the pump harmonic $m$ oscillates in the interaction picture of the linear Hamiltonian with mismatch
\begin{equation}
\Delta_{m;q_aq_b} = q_a\omega_a+q_b\omega_b+m\Omega.
\label{eq:mismatch-main}
\end{equation}
Near the degenerate pair resonance, the slow sector is chosen as
\begin{equation}
\mathcal R = \{(0;0,0),\,(-1;+2,0),\,(+1;-2,0)\}.
\label{eq:slowset-main}
\end{equation}
The first element contains static number-conserving operators, while the other two contain pair creation and annihilation in mode $a$.  
In particular, the small quantity $2\omega_a - \Omega$ never appears in a Schrieffer–Wolf denominator: that process belongs to the effective Hamiltonian and is not integrated out.

The most transparent virtual processes already occur in the quadratic pump term.  
They include a counter-rotating signal pair with mismatch $2\omega_a + \Omega$, controller-pair processes with mismatches $2\omega_b \pm \Omega$, and intermode sum- and difference-frequency processes with mismatches $\omega_a \pm \omega_b \pm \Omega$.  
Table~\ref{tab:virtualvertices-main} lists representative vertices.  
The quartic and sextic Josephson terms generate the same classes dressed by number operators and by additional powers of the zero-point phases.

\begin{table}[tbp]
\caption{Representative off-resonant vertices generated by the quadratic pump.  
The Hermitian-conjugate partner of each term is implicit.  
The resonant $a^{\dagger2}e^{-\ii\Omega t}$ vertex is omitted from the table because it is retained in $\mathcal R$.}
\label{tab:virtualvertices-main}
\vspace{3pt}
\begin{ruledtabular}
\begin{tabular}{ll}
vertex & mismatch \\
\hline 
$a^{\dagger2}e^{+\ii\Omega t}$ & $2\omega_a+\Omega$\\
$(2n_a+1)e^{+\ii\Omega t}$ & $\Omega$\\
$b^{\dagger2}e^{\pm\ii\Omega t}$ & $2\omega_b\pm\Omega$\\
$a^\dagger b^\dagger e^{\pm\ii\Omega t}$ & $\omega_a+\omega_b\pm\Omega$\\
$a^\dagger b e^{\pm\ii\Omega t}$ & $\omega_a-\omega_b\pm\Omega$
\end{tabular}
\end{ruledtabular}
\end{table}

\subsection{Schrieffer--Wolff view: virtual excursions and their return paths}
\label{subsec:sw-main}

The time-dependent Schrieffer––Wolff (SW) transformation provides a systematic way to incorporate the effects of off-resonant processes without retaining them explicitly in the effective Hamiltonian, while preserving the slow manifold in Eq.~\eqref{eq:slowset-main}~\cite{SchriefferWolff1966,Bravyi2011,Malekakhlagh2022}. 
We write the interaction as
\begin{equation}
    V(t) = V_{\text{res}}(t) + V_{\text{off}}(t),
\end{equation}
where $V_{\mathrm{res}}$ contains the processes near the resonance of interest, particularly the pair-creation term, whereas $V_{\mathrm{off}}$ contains the rapidly oscillating processes. 
We introduce a time-dependent unitary transformation in the form
\begin{equation}
    |\psi_{\rm NF}\rangle=e^{S(t)}|\psi\rangle, \quad S^{\dagger} = -S,
\end{equation}
with $S(t) = S_1(t) + S_2(t) + \cdots$. 
The transformed Hamiltonian is
\begin{equation}
    H_{\text{NF}} = e^{S} H  e^{-S} - \ii \hbar e^{S} \partial_{t} e^{-S}.
\label{eq:tdsw-main}
\end{equation}

At first order, $S_1$ is chosen to eliminate $V_{\mathrm{off}}$:
\begin{equation}
    \mathcal{L} S_{1} = V_{\text{off}}, \quad \mathcal{L} X = [H_{0}, X] - \ii \hbar \dot{X}.
\label{eq:homological-main}
\end{equation}
For a Fourier vertex $V_{\alpha}O_{\alpha}e^{im\Omega t}$, this yields
\begin{equation}
    S_{1, \alpha} = \frac{V_{\alpha}}{\hbar \Delta_{\alpha}} O_{\alpha} e^{\ii m \Omega t},
\label{eq:s1-rule-main}
\end{equation}
where $\Delta_{\alpha}$ is the detuning of the process. 
Thus, the amplitude of a virtual excursion is suppressed by the inverse of its detuning. 
Near-resonant processes are not eliminated by the SW transformation and remain in the effective Hamiltonian.

At second order, the eliminated processes reappear through the commutators
\begin{equation}
    A_{2} = [S_{1},V_{\mathrm{res}}] + \frac{1}{2} [S_{1},V_{\mathrm{off}}],
\end{equation}
with the part belonging to the effective sector being retained in $H_{\mathrm{SW}}^{(2)}$. 
For example,
\begin{equation}
    [\hat{a}^{2},\hat{a}^{\dagger 2}] = 4 \hat{n}_{a} + 2, 
    \quad [\hat{a} \hat{b}^{\dagger},\hat{a}^{\dagger} \hat{b}] = \hat{n}_{b} - \hat{n}_{a}. 
\label{eq:basiccomm1}
\end{equation}

Consequently, a virtual excursion involving the creation and annihilation of a pair, or a virtual conversion between the modes, returns to the original sector and produces diagonal corrections proportional to $n_a$ and $n_b$. 
These terms correspond to virtual Stark shifts, with the general structure
\begin{equation}
    \delta \omega_{j}^{(2)} \sim \sum_{\alpha} \frac{|V_{\alpha}|^{2}}{\hbar^{2} \Delta_{\alpha}},
\end{equation}
where each term in the sum corresponds to a distinct virtual path. 
The different detunings $D_{\sigma\tau} = \Omega + \sigma\omega_a + \tau\omega_b$ precisely represent these intermode paths. 
For example, the pair loop involving terms rotating in opposite directions yields
\begin{equation}
    \frac{\delta H_{a,\text{pair}}^{(2)}}{\hbar} 
    = - \frac{E_{p}^{2} \varphi_{a}^{4}}{4 \hbar (\Omega + 2 \omega_{a})} \hat{n}_{a} + \text{const.},
\label{eq:localstark-main}
\end{equation}
whereas an intermode difference-frequency path, $\hat{a}^{\dagger}\hat{b}\leftrightarrow\hat{a}\hat{b}^{\dagger}$, contributes a term proportional to $(\hat{n}_a - \hat{n}_b)/(\omega_a - \omega_b + \Omega)$. 
The complete second-order expression is a coherent sum over local-pair, sum-frequency, and difference-frequency virtual loops; Appendix~\ref{app:sw} provides the closed-form expressions.

Virtual paths affect the pair vertex in two distinct ways at sixth order in the zero-point phase amplitudes. 
A second-order path can combine a static quartic vertex with a quadratic pump vertex, returning to the one-pump pair sector. 
In addition, a third-order loop composed of three quadratic pump vertices can have a net pump harmonic of $\pm 1$ and therefore renormalize the bare pair coefficient. 
Up to sixth order in the zero-point amplitudes, the effective coefficients can be organized as
\begin{eqnarray}
\zeta_{ru}^{(\le6)} & = & \zeta_{ru}^{\rm dir(4)}+\zeta_{ru}^{\rm SW2(p2,p2)}\nonumber\\
& + & \zeta_{ru}^{\rm dir(6)}+\zeta_{ru}^{\rm SW2(p2,p4)},
\label{eq:zeta-provenance-main}\\
\gamma_{ru}^{(\le6)}& = & \gamma_{ru}^{\rm dir(2)}+\gamma_{ru}^{\rm dir(4)}+\gamma_{ru}^{\rm dir(6)}\nonumber\\
& + & \gamma_{ru}^{\rm SW2(s4,p2)}+\gamma_{ru}^{\rm SW3(p2,p2,p2)}.
\label{eq:gamma-provenance-main}
\end{eqnarray}

Here, $p 2$ and $p 4$ denote the quadratic and quartic pump vertices, respectively, and $s 4$ denotes the static quartic vertex. 
Equation~\eqref{eq:gamma-provenance-main} is particularly useful from a physical perspective. 
The direct cosine terms generate the occupation dependence already found in Eq.~\eqref{eq:exactpairmatrix}; the mixed $s 4\text{--}p 2$ virtual paths dress the number-dependent pair operators; and the purely quadratic $p 2\text{--}p 2\text{--}p 2$ loop remains Gaussian and therefore renormalizes only the bare pair coefficient.

The pump harmonics impose a simple selection rule. 
A process containing $j$ first-harmonic pump vertices has net harmonic parity
\begin{equation}
    \Delta m \equiv j \quad \pmod 2.
\label{eq:pumpparity-main}
\end{equation}
Consequently, a diagonal Stark term generated entirely by pump vertices requires an even number of such vertices, whereas three vertices can return to the $\pm 1$ harmonic and therefore contribute to the pair coupling. 
This immediately explains why the purely quadratic SW2 contribution is diagonal and why the purely quadratic SW3 contribution can return to the pair sector. 
Thus, the SW transformation does more than shift the frequencies: it provides the virtual corrections to the occupation-dependent coefficients and to the resonant pair-creation process itself.
Appendix~\ref{app:sw} presents the explicit third-order recursion and a closed-form single-mode example.

\subsection{Magnus view: the same virtual physics organized in time}
\label{subsec:magnus-main}

The Magnus expansion~\cite{Magnus1954,Blanes2009} starts from exactly the same interaction-picture vertices,
\begin{equation}
V_I(t) = \sum_\alpha v_\alpha O_\alpha e^{\ii\Delta_\alpha t},
\label{eq:VI-main}
\end{equation}
so the frequencies $\Delta_\alpha$ in Eq.~\eqref{eq:mismatch-main} are common to both constructions.  
The difference is organizational only.  
SW labels a process by the virtual operator being eliminated and its denominator; Magnus labels the same process by the ordering of its Fourier vertices in time.  
For example, the second Magnus term contains
\begin{equation}
\Omega_2(t) = -\frac{1}{2\hbar^2} \int_0^t dt_1\int_0^{t_1}dt_2 [V_I(t_1),V_I(t_2)],
\label{eq:omega2-main}
\end{equation}
and the counter-rotating sequence $a^{\dagger2}e^{+\ii\Omega t_1}$ followed by $a^2e^{-\ii\Omega t_2}$ generates the same $(4n_a+2)$ Stark structure as Eq.~\eqref{eq:localstark-main}.  
At third order, time-ordered sequences such as $(+1) + (+1) + (-1) = +1$ reproduce the pump-parity rule and feed the one-pump pair sector.

The near-resonant nature of the problem is important here.  
A laboratory-frame high-frequency expansion containing denominators only in integer multiples of $\Omega$ would erase the physical distinction between the deliberately resonant pair channel and the nearby off-resonant circuit processes.  
The interaction-picture form retains the actual mismatches $q_a\omega_a + q_b\omega_b + m\Omega$ and therefore provides a controlled Floquet organization, even though the drive is not \emph{"high frequency"} relative to every transition in the circuit.

SW and Magnus thus provide two views of one perturbative hierarchy.  
Their intermediate coefficients depend on the chosen normal-form or micromotion gauge, whereas reconstructed quasienergies, states, and observables can be compared directly.  Appendix~\ref{app:sw} gives the ordered Magnus integrals and the common-frame reconstruction used in the numerical validation.  
The compact Hamiltonian used below retains direct terms through sixth order in $\hat\varphi_J$ and the corresponding SW hierarchy through third order; we denote it by $H_{\rm eff}^{(6,3)}$.

\section{Two-dimensional occupation-resolved pair spectrum}
\label{sec:spectrum}

\begin{figure*}[t]
\centering
\includegraphics[width=0.96\textwidth]{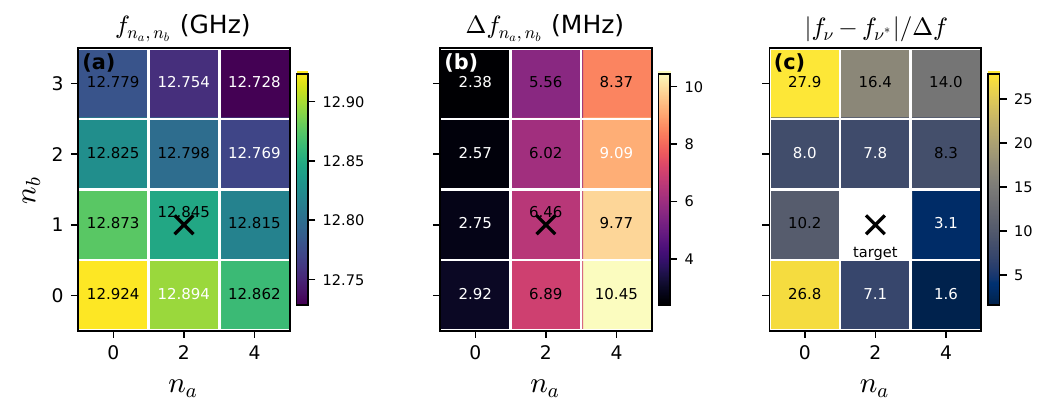}
\caption{\textbf{Two-dimensional occupation-resolved pair spectrum.}
Results are obtained from the numerically converged unexpanded two-mode cosine Floquet calculation for $(n_a,n_b) = (0,2,4)\times(0,1,2,3)$. (a) Pair-resonance frequency $f_{n_a,n_b}$, determined by the diagonal nonlinear spectrum.  (b) Avoided-crossing gap $\Delta f_{n_a,n_b}$, determined by the physical pair-transition matrix element.  The systematic decrease of the gap with increasing $n_b$ is evaluated after each cell has been individually tuned to resonance and therefore reflects matrix-element conditioning rather than a residual cross-Kerr detuning.  (c) Local spectral-separation ratio $R_\nu = |f_\nu-f_{\nu_\star}|/\Delta f_\nu$ relative to the target cell $\nu_\star = (2,1)$ (cross).  Along the signal ladder, bosonic enhancement increases the pair gap while the nonlinear spectrum can simultaneously bring neighboring pair resonances closer in frequency.}
\label{fig:spectrum}
\end{figure*}

Figure~\ref{fig:mechanism}(d) anticipates the central spectroscopic consequence of the shared Josephson coordinate: every pair transition on the $(n_a,n_b)$ Fock lattice is characterized by two distinct observables, namely the pump frequency at which the two states become resonant and the size of the corresponding avoided-crossing gap.  
The first probes the diagonal nonlinear spectrum, whereas the second probes the physical pair-transition matrix element.

The diagonal part of Eq.~\eqref{eq:compact} assigns the diabatic energy
\begin{equation}
    E(n_a,n_b) \equiv \hbar F(n_a,n_b)
    \label{eq:Ediagdef}
\end{equation}
to the Fock state $|n_a,n_b\rangle$ before the resonant pair coupling is diagonalized.  
For the one-pump transition
\begin{equation}
    |n_a,n_b\rangle \longleftrightarrow |n_a+2,n_b\rangle ,
\end{equation}
the resonance condition is
\begin{equation}
    \delta_{n_a,n_b} = E(n_a+2,n_b) - E(n_a,n_b) -\hbar\Omega = 0.
    \label{eq:rescondition}
\end{equation}
At leading nonlinear order, Eq.~\eqref{eq:diagconvention} gives
\begin{equation}
    \Omega_{\rm res}(n_a,n_b) \simeq 2\bar\omega_a - 2 K_a(2n_a+1) - 2 \chi_{ab}n_b + \cdots ,
    \label{eq:leadingres}
\end{equation}
where the omitted terms contain higher occupation polynomials and virtual Stark corrections.

The corresponding off-diagonal coupling is set by Eq.~\eqref{eq:physicalg},
\begin{equation}
    g_{n_a,n_b} = \sqrt{(n_a+1)(n_a+2)}\, G(n_a,n_b).
    \label{eq:spectrum-g}
\end{equation}
The gap also becomes dependent on $n_a$ and $n_b$. 
Thus, measuring the gaps of different transitions in the Fock-state lattice makes it possible, in principle, to directly map the occupation dependence of the pair coupling. Equations~\eqref{eq:leadingres} and \eqref{eq:spectrum-g} make the two-dimensional structure explicit.  
Increasing $n_a$ changes the local pair resonance through the signal-mode $a$ nonlinear spectrum, but it also changes the transition matrix element through the usual bosonic ladder factor and the finite-phase corrections contained in $G$.  
Increasing $n_b$ shifts the pair resonance through cross nonlinearities and, independently, modifies the pair amplitude through the controller, mode $b$, dependence derived in Sec.~\ref{sec:pairmechanism}. 
Each Fock-lattice cell therefore carries a resonance coordinate and a coupling coordinate, both inherited from the same nonlinear boundary but associated with different matrix elements of the Hamiltonian.

Figure~\ref{fig:spectrum} shows these two quantities over the lattice
\begin{equation}
    (n_a,n_b)  = (0,2,4)\times(0,1,2,3)
\end{equation}
using the numerically converged unexpanded two-mode cosine Floquet calculation.  
The resonance surface, shown in Fig.~\ref{fig:spectrum}(a), shifts downward along both occupation directions.  
Along the signal ladder, this primarily reflects the self-nonlinearity of mode $a$, while the shift between controller sectors is dominated at leading order by the cross-Kerr term.  Figure~\ref{fig:spectrum}(b) shows a qualitatively different trend.  
At fixed $n_a$, the avoided-crossing gap decreases systematically as $n_b$ is increased, in agreement with the controller-dependent matrix element derived from the full Josephson cosine.

This distinction is particularly important because every point in Fig.~\ref{fig:spectrum}(b) is evaluated at its own calibrated resonance.  
The variation of the gap with $n_b$ therefore remains after removing the controller-dependent detuning. 
A cross-Kerr shift alone can move an avoided crossing along the frequency axis, but it cannot account for a systematic change in the minimum splitting once the crossing is retuned.  
The two maps thus separate spectral conditioning from matrix-element conditioning at the observable level.

The avoided crossings remain well associated with the intended pair transitions throughout the displayed lattice.  
For each cell, the two Floquet eigenstates forming the selected crossing retain a mean weight larger than $0.953$ in the corresponding bare two-state pair subspace.  
This assignment overlap allows the same branch to be followed continuously as the pump frequency is scanned.  
The complete crossing-tracking algorithm, assignment data, and finite-Sambe convergence are given in Appendix~\ref{app:floquet}.

The signal occupation introduces an additional tradeoff that is already visible in Fig.~\ref{fig:spectrum}.  
Moving upward along the even-parity pair ladder increases the gap through bosonic enhancement: the factor $\sqrt{(n_a+1)(n_a+2)}$ grows with $n_a$.  
At the same time, the nonlinear spectrum brings some transitions closer to neighboring pair resonances.  
Higher occupation can therefore speed up the desired operation while making it harder to isolate spectrally.  
Both effects must be considered together when choosing a working point.

For the selective-control calculations below, we choose the interior cell
\begin{equation}
    \nu_\star=(n_a,n_b)=(2,1).
\end{equation}
Its calibrated pump frequency and avoided-crossing gap in cyclic-frequency units are
\begin{equation}
    f_{\nu_\star} = 12.845263~{\rm GHz},
    \qquad
    \Delta f_{\nu_\star} = 6.46279~{\rm MHz}.
    \label{eq:target-spectrum-values}
\end{equation}
Here and below, $\omega$, $\Omega$, and Hamiltonian coupling coefficients are angular frequencies unless explicitly divided by $2\pi$, whereas numerical resonances $f$ and avoided-crossing gaps $\Delta f$ are quoted as cyclic frequencies.

To quantify how well the target cell is separated from another pair transition $\nu$, we define
\begin{equation}
    \delta f_\nu \equiv f_\nu-f_{\nu_\star},
    \qquad
    R_\nu = \frac{|\delta f_\nu|}{\Delta f_\nu}.
    \label{eq:Rnu}
\end{equation}
The physical meaning of this ratio is clearest for an isolated competing crossing.  If that channel is described in its local rotating frame by
\begin{equation}
\frac{H_\nu}{h} = \frac{\delta f_\nu}{2}\sigma_z + \frac{\Delta f_\nu}{2}\sigma_x ,
\label{eq:local-two-level}
\end{equation}
then an initial diabatic state can reach at most
\begin{equation}
    P_\nu^{\max} = \frac{1}{1+R_\nu^2}
    \label{eq:Rnu-prob}
\end{equation}
under that isolated off-resonant drive.  
Figure~\ref{fig:spectrum}(c) therefore provides a compact local measure of the competition between the detuning of a neighboring pair resonance and its own coupling strength.  
The actual circuit contains several pair channels simultaneously, so this two-state result is used only as a spectral guide; the complete unexpanded-cosine dynamics are evaluated directly in Sec.~\ref{sec:control}.

The gap map also provides the observable signature needed to distinguish the present interaction from a purely diagonal occupation dependence.  
If the controller affected only the resonance frequency, individually retuning each $n_b$ sector would remove its influence on the resonant transition strength.  
The persistent change of the avoided-crossing gap shows that the controller modifies the pair matrix element itself.  
Section~\ref{sec:comparator} tests this distinction quantitatively by fitting reduced models with flexible occupation-dependent diagonal spectra and asking whether the observed gaps can be reproduced without an explicit controller-dependent pair amplitude.

\section{Microscopic validation and regime of validity}
\label{sec:validation}

\begin{figure*}[t]
\centering
\includegraphics[width=0.96\textwidth]{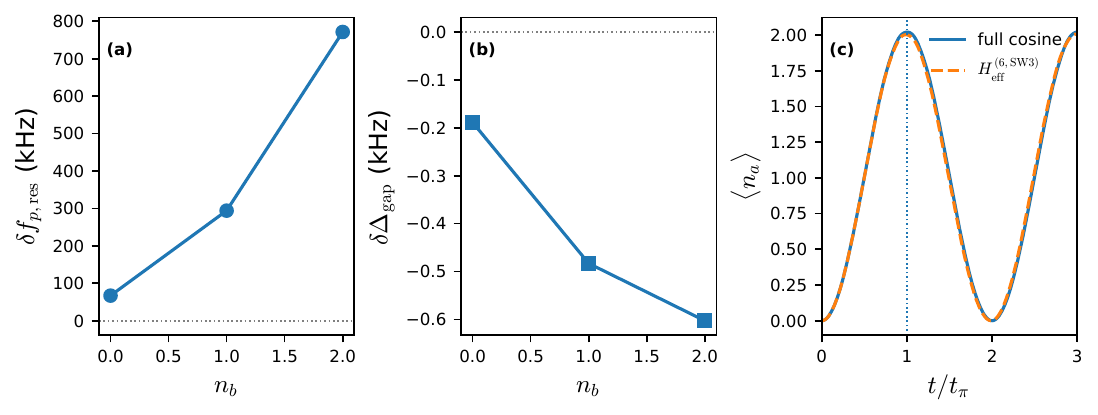}
\caption{\textbf{Validation of $H_{\rm eff}^{(6,3)}$ against the unexpanded two-mode cosine}.  (a) Resonance-frequency error and (b) avoided-crossing-gap error for the vacuum-pair sectors.  (c) Signal occupation at the calibrated $n_b=1$ resonance.  The close dynamical agreement over the first transfer cycle reflects the much smaller error in the pair matrix element than in the accumulated diagonal shift.}
\label{fig:validation}
\end{figure*}

The compact normal form exposes the microscopic origin of the dynamics; the unexpanded two-mode cosine supplies the quantitative reference.  
Their comparison is made at three levels: avoided-crossing spectroscopy, real-time transfer, and convergence with Josephson-phase and SW order.

\subsection{Spectral and dynamical comparison}

For the vacuum-pair transitions with $n_b=0,1,2$, the unexpanded reference gives resonance frequencies $12.923516$, $12.873319$, and $12.824786$ GHz.  
The sixth-phase-order, third-SW-order Hamiltonian $H_{\rm eff}^{(6,3)}$ misses these resonances by approximately $0.067$, $0.294$, and $0.770$ MHz, respectively, while its avoided-crossing-gap errors remain below $1$ kHz, as shown in Figs. \ref{fig:validation}(a) and \ref{fig:validation}(b).  
The different convergence rates are physically natural: the low-lying off-diagonal pair matrix element is already well captured by the finite-order vertex, whereas the diagonal spectrum accumulates several occupation-dependent phase and virtual corrections.

The avoided-crossing gap also defines a natural transfer scale,
\begin{equation}
t_\pi = \frac{1}{2\Delta f_{\rm gap}},\qquad \tau = t/t_\pi.
\label{eq:tpi}
\end{equation}
At $\tau=1$ for the $n_b=1$ vacuum-pair resonance, as shown in Fig.~\ref{fig:validation}(c), the signal occupations obtained from the unexpanded cosine and $H_{\rm eff}^{(6,3)}$ differ by about $1.9\times10^{-2}$ photons.  
This comparison uses the same physical initial state and observable after inverse SW dressing, as described in Appendix~\ref{app:sw}.

\subsection{Separating phase order from SW order}

\begin{figure}[t]
\centering
\includegraphics[width=0.96\columnwidth]{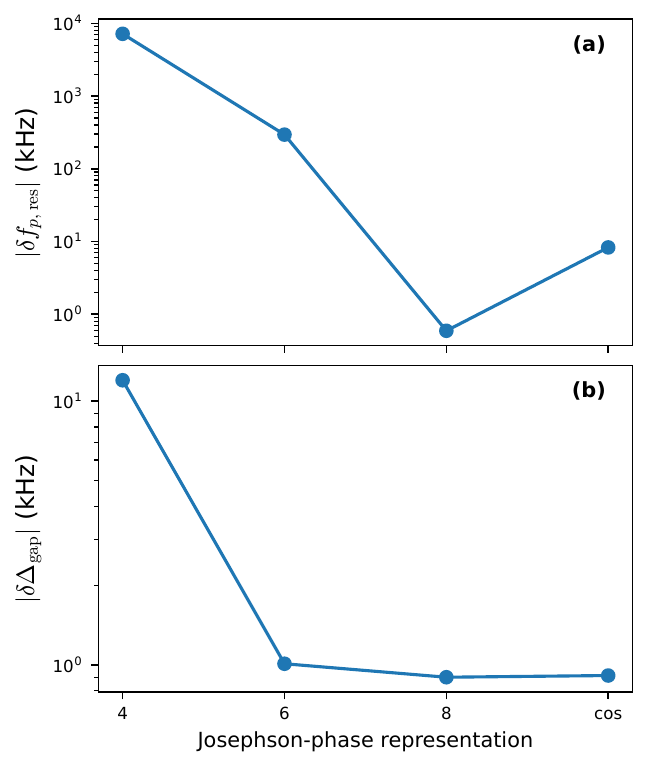}
\caption{Josephson-phase-order convergence at fixed SW3 for the $n_b=1$ transition $|0,1\rangle\leftrightarrow|2,1\rangle$: (a) absolute pump-resonance error and
(b) absolute avoided-crossing-gap error. Here $4$, $6$, and $8$ denote the highest retained powers of $\hat{\varphi}_J$, while ``cos'' denotes the unexpanded matrix cosine.}
\label{fig:error-phase-order}
\end{figure}

A finite Josephson-phase expansion and a finite SW expansion are distinct approximations. 
To separate their contributions, we first hold the virtual-elimination order fixed at SW3 and successively replace the Josephson potential by expansions through $\hat{\varphi}_J^4$, $\hat{\varphi}_J^6$, and $\hat{\varphi}_J^8$, followed by the unexpanded matrix cosine. Figure~\ref{fig:error-phase-order} shows this comparison for the $n_b=1$ benchmark transition $|0,1\rangle\leftrightarrow|2,1\rangle$.

The fourth-order phase expansion gives resonance and gap errors of approximately $7.17$ MHz and $12.0$ kHz, respectively. 
Extending the expansion to sixth order reduces these errors to approximately $295$ kHz and $1.01$ kHz, while the eighth-order expansion reaches approximately $0.59$ kHz and $0.90$ kHz. 
The phase-resummed SW3 calculation gives corresponding errors of approximately $8.29$ kHz and $0.91$ kHz. 
The nonmonotonic sub-kHz resonance error should not be interpreted as a strict superiority of the eighth-order polynomial: it reflects a cancellation between phase-truncation and finite-SW-order errors. 
By contrast, the nearly unchanged gap error at the highest phase orders indicates that the residual discrepancy is no longer controlled primarily by the phase expansion.

To isolate this remaining contribution, we construct a phase-resummed full-ladder SW hierarchy in which the matrix cosine is retained while the virtual-elimination order is increased. 
At the same $n_b = 1$ benchmark resonance, extending this hierarchy to fourth SW order reduces the resonance error to approximately $0.59$ kHz and the gap error to approximately $0.021$ kHz. 
This complementary SW-order test therefore confirms that the residual gap error in Fig.~\ref{fig:error-phase-order} originates predominantly from the finite virtual-elimination order rather than from the Josephson-phase representation.

The accuracy of a finite Josephson-phase expansion is also state-dependent because higher Fock occupations explore larger Josephson-phase fluctuations. 
We monitor this through
\begin{equation}
\eta_\varphi = \langle\hat{\varphi}_J^2\rangle,
\label{eq:etaphi}
\end{equation}
together with the distribution of the off-resonant matrix-element-to-energy-mismatch ratios entering the SW generator. 
For states satisfying $\langle\hat{\varphi}_J\rangle=0$, $\eta_\varphi$ is the phase variance; more generally, it is the second moment of the Josephson phase. 
These diagnostics explain why a compact sixth-order polynomial can be highly accurate for a small set of resolved Fock cells while becoming inadequate for a broad coherent or squeezed state.

For example, a coherent signal state with $\bar n=2$ reaches a minimum reduced-state fidelity of approximately $0.70$ when propagated with the compact $H_{\mathrm{eff}}^{(6,3)}$ during the first operation. 
Retaining the full Josephson phase and enlarging the retained ladder restores the minimum fidelity to approximately $0.998$. 
The application sections therefore use the unexpanded two-mode cosine as the numerical Hamiltonian, whereas the analytical normal form is used to identify the underlying mechanisms and scaling. 
Appendix~\ref{app:floquet} contains the cutoff, phase-order, SW-order, and extended-state convergence tests.

\section{Occupation-selective pair control}
\label{sec:control}

\begin{figure}[b]
\centering
\includegraphics[width=\columnwidth]{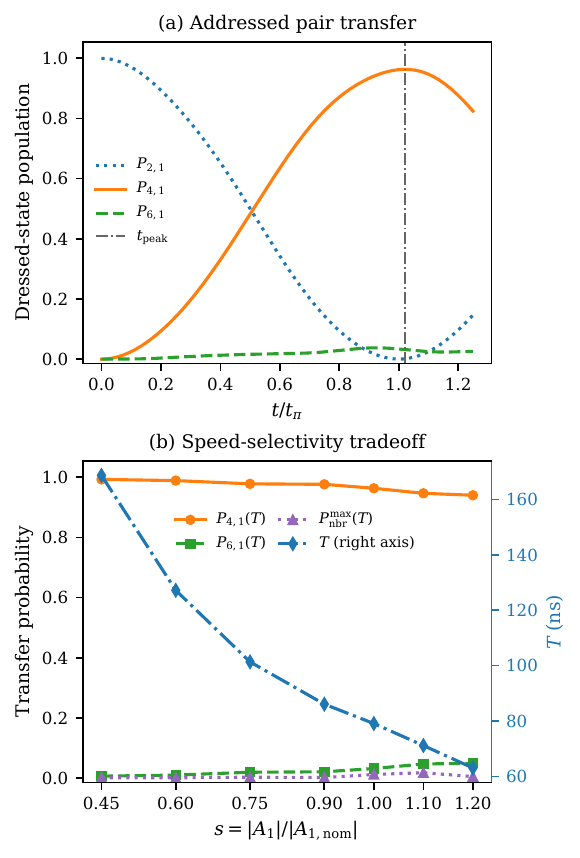}
\caption{\textbf{Occupation-selective pair transfer in the unexpanded two-mode Josephson-cosine model}. Fock labels denote the corresponding static dressed states. (a) Populations of the initial state $|2,1\rangle$ (dotted), target state $|4,1\rangle$ (solid), and higher rung $|6,1\rangle$ (dashed) under the nominal square pump. Time is normalized by $t_\pi = (2\Delta f_{\rm gap})^{-1}$; the vertical line marks the sampled target-population maximum. (b) First-harmonic amplitude sweep, $s=|A_1|/|A_{1,\mathrm{nom}}|$, with the target resonance and pulse duration recalibrated at each point. Circles and squares show $P_{4,1}(T)$ and $P_{6,1}(T)$, respectively. Triangles give the largest final transfer among the separately initialized transitions $|0,1\rangle \to |2,1\rangle$, $|2,0\rangle \to |4,0\rangle$, $|0,0\rangle\to|2,0\rangle$, and $|2,2\rangle  \to |4,2\rangle$, driven by the same pulse. Diamonds show the selected duration $T$ on the right axis.}
\label{fig:gate}
\end{figure}

So far, we have established that each transition $\lvert n_a,n_b\rangle \leftrightarrow \lvert n_a + 2,n_b\rangle$ has its own resonance frequency and avoided-crossing gap. 
Owing to the nonlinearity, different cells of the Fock-state ladder are nondegenerate. 
The question now becomes: can we exploit this spectral resolution to selectively control a single transition? 
The occupation-resolved spectrum permits direct addressing of a transition inside the pair ladder. We use the cell
\begin{equation}
|2,1\rangle\longleftrightarrow|4,1\rangle,
\label{eq:target}
\end{equation}
whose local resonance and gap were identified in Sec.~\ref{sec:spectrum}, mode $b$ remains in the controller sector $n_{b} = 1$, while two quanta are added to mode $a$. 
Close to this avoided crossing, the full complex dynamics can be locally approximated by an effective two-level system
\begin{equation}
\frac{H_{\rm cell}}{h} = \frac{\delta f}{2}\sigma_z + \frac{\Delta f_{\rm gap}}{2}\sigma_x,
\label{eq:cell-twolevel}
\end{equation}
here $\delta f=f_p-f_{\rm res}$, where \(f_{\mathrm{res}}\) denotes the spectral position of the transition and $\Delta f_{\rm gap}$ is the numerically extracted quasienergy splitting, expressed in cyclic-frequency units at resonance. 
On resonance, $\delta f = 0$, the population oscillates between the two rungs, and the first maximum occurs at $t_\pi=1/(2\Delta f_{\rm gap})$, consistent with Eq.~\eqref{eq:tpi}. 
The neighboring pair cells have their own gaps but are detuned by the nonlinear spectrum of Fig.~\ref{fig:spectrum}; the ratios $R_\nu$ in Eq.~\eqref{eq:Rnu} therefore anticipate which channels are easiest to suppress. 

Here, $t/t_{\pi}=1$ corresponds approximately to the time at which the two-level model predicts complete population transfer. 
For the $\lvert 2,1\rangle$ curve, the initial population is $P_{\lvert 2,1\rangle}(0) = 1$, which decreases as the pump drives the transition $\lvert 2,1\rangle\rightarrow\lvert 4,1\rangle$. 
Propagation with the complete unexpanded cosine places the first maximum of the target population at $t = 79.10 \, \rm ns$, where $P_{|4,1\rangle}=0.9634$. 
If the system were truly an isolated two-level cell, we would expect $P_{\lvert 4,1\rangle}\approx 1$. The fact that its maximum value is $0.9634$ already indicates dynamics outside the ideal cell.

The dominant coherent correction has a simple ladder interpretation.  
Once the $|4,1\rangle$ rung is populated, the same pump can continue the pair ladder toward $|6,1\rangle$, that is, $|2, 1 \rangle \rightarrow |4, 1 \rangle \rightarrow |6, 1 \rangle$. 
At the first target maximum, as shown by the dashed green line in Fig.~\ref{fig:gate}(a), this population is $0.0317$. The probability, in Fig.~\ref{fig:gate}(b), outside the $n_b=1$ controller sector is much smaller, $0.00243$, and separately initialized neighboring pair transitions remain weak under the same pulse. 
The selected controller sector is therefore well resolved; the main square-pulse error is further pair creation within that sector. 
Therefore, the dominant error is not a change in the controller occupation $n_b=1$, but rather the continued excitation along the same pair ladder.

This ladder structure also gives a direct speed--selectivity tradeoff.  
Reducing the first pump harmonic from its nominal amplitude to $45\%$ lowers the $|6,1\rangle$ population at the first target maximum from $0.0317$ to $0.00641$ and the largest tested neighboring transfer from $0.0120$ to $0.00106$.  
The target population increases from $0.9634$ to $0.9929$, while the first maximum moves from $79.10$ to $168.71$ ns, as shown in Fig.~\ref{fig:gate}(b). 
A weaker pump therefore resolves the nonlinear ladder more sharply, at the cost of a longer operation time.

Pulse shaping provides a second control knob: smooth turn-on and turn-off suppress the higher-rung response without changing the peak pump amplitude.  
Appendix~\ref{app:control} gives this envelope comparison, the nominal leakage anatomy, and a fixed-duration two-parameter scan of amplitude and frequency calibration.

\section{Quantum harmonic controller}
\label{sec:qcontroller}

\begin{figure*}[t]
\centering
\includegraphics[width=0.96\textwidth]{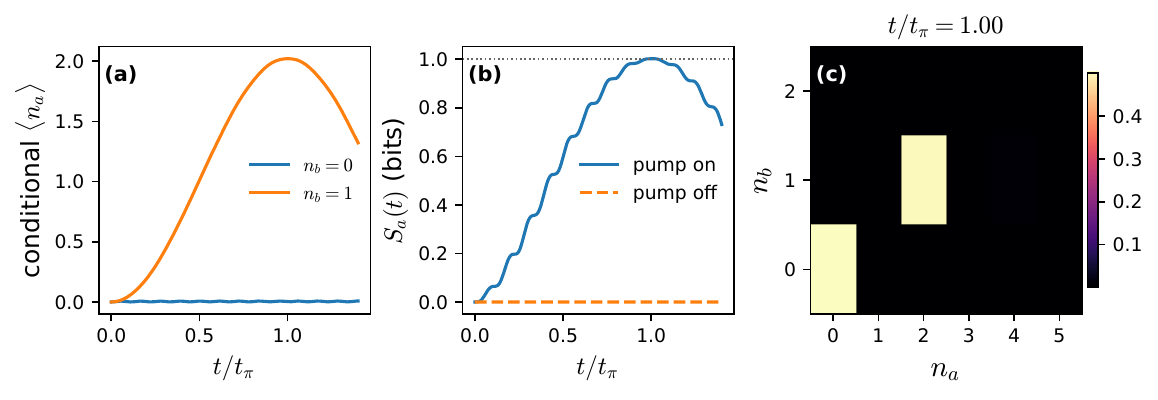}
\caption{\textbf{Quantum harmonic controller.}
(a) Conditional signal trajectories $\langle n_{a} \rangle$ associated with the $n_b = 0$ and $n_b = 1$ components of the dressed superposition in Eq.~\eqref{eq:controllerstate}.  The common pump is resonant with the $n_b=1$ vacuum-pair transition, so this branch develops a substantial pair excitation while the $n_b=0$ branch remains predominantly vacuum-like.  (b) Signal--controller entanglement entropy under the driven evolution and for the pump-off reference.  As the two conditional signal states become distinguishable, the entropy approaches the one-bit value expected for two approximately orthogonal equal-weight branches.  (c) Joint Fock distribution near $t/t_\pi = 1$, showing the correlation between controller occupation and signal excitation directly.}
\label{fig:qcontroller}
\end{figure*}

The previous sections treated the controller occupation $n_b$ as a label for separately prepared Fock sectors.  
The same occupation dependence becomes a genuinely quantum resource when different controller occupations are placed in coherent superposition.  
The physical mechanism is simple: because the pair resonance depends on $n_b$, the different components of the controller superposition generate different signal trajectories under the same pump.  
If the controller coherence is preserved, these conditional trajectories are branches of a single joint quantum state and their separation produces signal--controller entanglement.

We illustrate this mechanism with the initial state
\begin{equation}
    |\Psi(0)\rangle = \frac{|\widetilde{0,0}\rangle + |\widetilde{0,1}\rangle}{\sqrt2},
    \label{eq:controllerstate}
\end{equation}
where $|\widetilde{n_a,n_b}\rangle$ denotes the eigenstate of the static coupled circuit having the largest overlap with the corresponding bare Fock state $|n_a,n_b\rangle$.  
At the operating point, these overlaps are $0.999995$ and $0.999949$ for the two components of Eq.~\eqref{eq:controllerstate}, respectively, so their Fock-state interpretation remains essentially unchanged. Figure~\ref{fig:qcontroller}(a) shows the conditional signal trajectories associated with the $n_b = 0$ and $n_b = 1$ components of the dressed superposition. 
Static dressed states are used because they provide a stationary pump-off reference: any substantial dynamics or entanglement generated after the drive is applied can then be attributed to the driven conditional interaction rather than to the trivial evolution of a bare state that is not an eigenstate of the coupled circuit.

The pump is tuned to the $n_b=1$ vacuum-pair transition
\begin{equation}
    |0,1\rangle \longleftrightarrow |2,1\rangle .
\end{equation}
At the same pump frequency, the corresponding $n_b=0$ transition is detuned by the occupation-dependent nonlinear spectrum.  
Consequently, the two controller components experience different signal dynamics.  
To a good approximation, the evolving state can be viewed schematically as
\begin{equation}
    |\Psi(t)\rangle \simeq \frac{|\psi_0(t)\rangle_a|0\rangle_b + e^{\ii\phi(t)}| \psi_1(t)\rangle_a|1\rangle_b}{\sqrt2},
    \label{eq:conditional-branches}
\end{equation}
where $|\psi_0(t)\rangle_a$ remains predominantly vacuum-like while $|\psi_1(t)\rangle_a$ develops a substantial two-photon component.  
Equation~\eqref{eq:conditional-branches} is not an additional effective model; it is a useful representation of the dominant branch structure of the full driven two-mode state.

The connection between conditional dynamics and entanglement is evident from the overlap of the two signal branches.  
For the idealized equal-weight two-branch state of Eq.~\eqref{eq:conditional-branches}, tracing out the controller gives a signal reduced state whose two nonzero eigenvalues are
\begin{equation}
    \lambda_\pm(t) = \frac{1\pm |\langle\psi_0(t)|\psi_1(t)\rangle|}{2}.
    \label{eq:branch-schmidt}
\end{equation}
For this bipartite pure state, the signal--controller entanglement is quantified by the von Neumann entropy of either reduced state $\rho_a = \mathrm{Tr}_b\rho$~\cite{Bennett1996,Horodecki2009}, which reads as
\begin{eqnarray}
    S_a(t) & = & -\mathrm{Tr}\!\left[ \rho_a(t)\log_2\rho_a(t) \right], \notag \\
           & = & -\sum_{\nu=\pm} \lambda_\nu(t)\log_2\lambda_\nu(t).
    \label{eq:controller-entropy}
\end{eqnarray}
When the two conditional signal states are nearly identical, their overlap is close to unity and $S_a\simeq0$.  
Moreover, when the resonant and off-resonant branches become nearly distinguishable,
$|\langle\psi_0|\psi_1\rangle|\rightarrow0$, the two Schmidt weights approach $1/2$ and the entropy approaches one bit. 
Thus, in this equal-weight pure-state setting, the growth of entanglement has a direct physical interpretation: it reflects the increasing distinguishability of the two signal trajectories conditioned on the harmonic-controller occupation~\cite{BarnettCroke2009,Englert1996}. 
The underlying physical mechanism is closely related to dispersive circuit-QED protocols in which a superconducting qubit becomes correlated with distinguishable coherent-state branches of a bosonic mode~\cite{Vlastakis2013}.

For the full unexpanded-cosine dynamics, Figure~\ref{fig:qcontroller}(b) shows that the entanglement entropy indeed rises to approximately one bit near the first pair-transfer maximum. Under pump-off evolution, by contrast, the same dressed initial superposition remains essentially separable, with entropy below $3\times10^{-4}$ bits over the plotted interval.  
This control calculation shows that the order-unity entanglement is generated by the driven occupation-conditioned dynamics rather than by the weak static hybridization of the two modes.

The entropy can rise slightly above one bit in the full calculation because the driven evolution is not confined exactly to the two branches retained in Eq.~\eqref{eq:conditional-branches}.  
Weak population of additional dressed sectors increases the available Schmidt rank.  
The dominant physics nevertheless remains the correlation between a vacuum-like signal state associated with the off-resonant $n_b=0$ component and a pair-excited signal state associated with the resonant $n_b=1$ component.

In this particular protocol, the principal discrimination between the two controller branches comes from the occupation-dependent resonance frequency.  
The occupation dependence of the pair matrix element also changes the detailed branch evolution, but entanglement alone does not separate the diagonal and off-diagonal contributions.  
That distinction is established independently by the occupation-resolved avoided-crossing gaps and the comparator analysis of Sec.~\ref{sec:comparator}.

The same reasoning extends to larger controller superpositions, but a single monochromatic pump does not generally yield the same transformation across all controller sectors. 
For example,
\begin{equation}
    \frac{|0\rangle+|1\rangle+|2\rangle}{\sqrt3}
\end{equation}
contains three coherent controller components, yet at the present pump frequency only the $n_b=1$ pair transition is strongly resonant.  
The resulting state therefore contains one strongly driven signal branch and two predominantly off-resonant branches rather than three equal conditional rotations.  
More generally, the two-dimensional pair spectrum determines which components of a harmonic-controller superposition participate in a given operation: the pump frequency selects the relevant controller sectors, while its amplitude and bandwidth determine how sharply those sectors are resolved.  
This provides the basis for extending the same mechanism to multitone or shaped conditional operations.  

\section{Controller-conditioned non-Gaussian dynamics}
\label{sec:nonGaussian}

\begin{figure*}[t]
\centering
\includegraphics[width=0.98\textwidth]{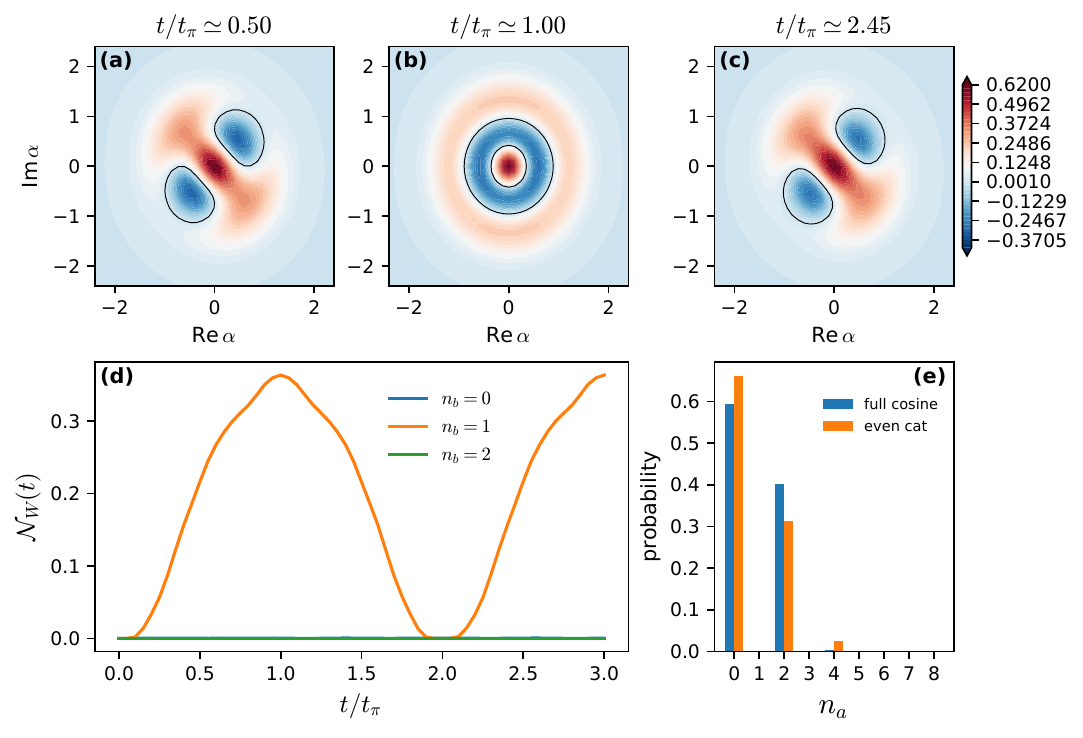}
\caption{\textbf{Controller-conditioned non-Gaussian dynamics.}
(a)--(c) Reduced signal Wigner functions for the resonant $n_b=1$ controller sector at representative times.  Pair driving first populates the even signal ladder and subsequently develops interference structure and negative regions as the state samples the nonlinear Josephson spectrum.  All panels use the same Wigner scale and show the $W=0$ contours.  (d) Integrated Wigner negative volume for separately initialized $n_b=0,1,2$ controller sectors under the same pump.  Strong negativity develops only in the resonant $n_b=1$ sector, whereas the two off-resonant sectors remain nearly Wigner-positive.  (e) Signal Fock distribution at a later point on the resonant trajectory together with the optimized even-cat comparison with $\mathcal F_{\rm cat}=0.967, |\alpha|=0.986$.}
\label{fig:resource}
\end{figure*}

The quantum-controller calculation above used a coherent superposition of controller occupations and asked whether conditional signal evolution produces entanglement.  
We now ask a different question: for a controller prepared in a definite Fock sector, can the same occupation select whether the signal undergoes strongly non-Gaussian dynamics?  
To isolate this effect, the controller sectors are initialized separately rather than in the coherent superposition of Sec.~\ref{sec:qcontroller}.

For each $n_b=0,1,2$, we start from the static dressed state associated with $|0,n_b\rangle$, apply the same pump, and trace out mode $b$.  
The essential point is that pair creation alone is not sufficient to explain the resulting non-Gaussianity.  
A purely quadratic degenerate parametric interaction,
\begin{equation}
\frac{H_{\rm DPA}}{\hbar} = g \,a^{\dagger2} + g^*a^2
\label{eq:quadratic-pair-reference}
\end{equation}
maps Gaussian states to Gaussian states; in particular, the vacuum evolves into a squeezed Gaussian state whose Wigner function remains nonnegative~\cite{Weedbrook2012}. 
This Gaussian squeezed-vacuum regime has been experimentally realized and characterized in Josephson parametric devices~\cite{CastellanosBeltran2008,Mallet2011}.

The present dynamics differs because the Josephson interaction is not purely quadratic. 
The effective structure derived in Sec.~\ref{sec:pairmechanism},
\begin{equation}
    \frac{H_{\rm eff}}{\hbar} = F(n_a,n_b) + \left\{ a^{\dagger2}G(n_a,n_b) + \mathrm{H.c.} \right\},
    \label{eq:nonGaussian-mechanism}
\end{equation}
contains both an occupation-dependent diagonal spectrum and an occupation-dependent pair amplitude.  
As the resonantly driven state climbs the even signal ladder,
\begin{equation}
    |0,n_b\rangle
    \leftrightarrow
    |2,n_b\rangle
    \leftrightarrow
    |4,n_b\rangle
    \leftrightarrow
    |6,n_b\rangle
    \leftrightarrow\cdots ,
    \label{eq:even-pair-ladder}
\end{equation}
successive links have different detunings and different matrix elements. 
Their amplitudes and phases therefore do not follow the uniform Gaussian evolution of Eq.~\eqref{eq:quadratic-pair-reference}.  
Instead, the Josephson nonlinearity distorts the even-Fock-state superposition and can generate phase-space interference and Wigner negativity.

This mechanism is also controller-selective.  
The fixed pump is calibrated to the $n_b=1$ vacuum-pair resonance.  
The $n_b = 1$ state therefore enters the nonlinear pair ladder strongly, whereas the $n_b = 0$ and $n_b = 2$ sectors are detuned by the occupation-resolved spectrum and undergo much weaker excursions.  
A single pump can consequently produce strongly non-Gaussian signal dynamics in one controller sector while leaving neighboring sectors nearly unaffected.

We quantify this behavior through the reduced signal Wigner function $W(\alpha)$, shown in Figs.~\ref{fig:resource}(a)–(c), and its integrated negative volume~\cite{Kenfack2004,Albarelli2018},
\begin{equation}
    \mathcal N_W = \frac12 \left( \int d^2\alpha \, |W(\alpha)| - 1 \right).
    \label{eq:wignerneg}
\end{equation}
A nonnegative Wigner function gives $\mathcal N_W=0$, whereas $\mathcal N_W>0$ certifies Wigner negativity and, consequently, non-Gaussianity~\cite{Weedbrook2012,Albarelli2018}. 
The converse is not generally true: a non-Gaussian state need not possess a negative Wigner function.  
The quantity $\mathcal N_W$ is therefore used here as a stringent witness of the non-Gaussian phase-space structure produced by the conditional pair dynamics.

The three controller sectors, shown in Fig.~\ref{fig:resource}(d), show a pronounced contrast under the same drive. 
Over
\begin{equation}
    0\leq t/t_\pi \leq 3,
\end{equation}
the resonant $n_b = 1$ branch reaches
\begin{equation}
    \mathcal N_W^{\max} \simeq 0.363,
\end{equation}
whereas the largest values are only
\begin{equation}
    5.3\times10^{-4} \qquad (n_b = 0)
\end{equation}
and
\begin{equation}
    \mathcal N_W^{\max} < 2\times10^{-9} \qquad (n_b = 2).
\end{equation}
The contrast follows directly from the two-dimensional pair spectrum: only the $n_b = 1$ branch is resonantly driven deeply enough into the nonlinear signal ladder to accumulate the unequal amplitudes and phases required for substantial Wigner negativity.  
The neighboring controller sectors remain sufficiently off-resonance that their reduced signal states remain nearly Wigner-positive.

A representative later state occurs at
\begin{equation}
    t/t_\pi=2.4393,
\end{equation}
where the resonant signal has
\begin{equation}
    \mathcal N_W=0.179,
    \quad
    \langle\Pi\rangle=0.99943,
    \quad
    \mathrm{Tr}\rho_a^2=0.99942,
    \label{eq:resource-diagnostics}
\end{equation}
with
\begin{equation}
    \Pi=(-1)^{n_a}
\end{equation}
the signal parity operator.  
The nearly unit parity shows that the state remains overwhelmingly in the even sector expected from pair creation, while the nearly unit purity shows that the reduced signal state is only weakly mixed in this separately initialized controller sector.  
Its mean occupation is
\begin{equation}
    \langle n_a\rangle=0.8194.
\end{equation}

The same state has phase-shift quantum Fisher information
\begin{equation}
    F_Q[\rho_a,n_a] = 4.008,
    \label{eq:resource-qfi}
\end{equation}
where $n_a$ is the generator of the signal-mode phase shift. 
This quantity quantifies the distinguishability of neighboring states under the phase transformation $e^{-\ii\theta n_a}$ and determines the ultimate local phase sensitivity through the quantum Cramér--Rao bound~\cite{BraunsteinCaves1994,Paris2009}. 
Its inclusion here provides a complementary, although not uniquely non-Gaussian, characterization of the phase-space structure generated by the nonlinear pair dynamics.

The latter state also has a large overlap with the even coherent-state superposition~\cite{Dodonov1974,YurkeStoler1986}
\begin{equation}
    |C_+(\alpha)\rangle = \frac{|\alpha\rangle+| - \alpha\rangle}{\sqrt{2(1+e^{-2|\alpha|^2})}}.
    \label{eq:even-cat-main}
\end{equation}
Optimizing the amplitude and phase gives
\begin{equation}
    \mathcal F_{\rm cat}=0.9669,
    \,\,\,
    |\alpha|=0.9863,
    \,\,\,
    \arg\alpha=0.8216~{\rm rad}.
\end{equation}
The near-unit even parity and purity make this comparison natural, but the cat state represents only one recognizable point on a more general nonlinear trajectory; Figure~\ref{fig:resource}(e) presents this result.  
The central result is not the preparation of a particular cat state; it is that the occupation of a harmonic controller selects whether a fixed pair pump drives the signal into a strongly Wigner-negative regime.

Together, Secs.~\ref{sec:qcontroller} and \ref{sec:nonGaussian} show two distinct quantum uses of the same occupation-conditioned interaction.  
A coherent superposition of controller occupations converts the conditional signal response into signal--controller entanglement, whereas separately prepared controller Fock sectors select which signal trajectories develop strong non-Gaussian phase-space structure.  
Both effects follow directly from the occupation-resolved pair spectrum and therefore connect the applications to the microscopic interaction derived in Sec.~\ref{sec:pairmechanism}. 
Appendix~\ref{app:resources} gives the detailed dressed-state assignment and numerical entropy calculation.  

\section{Observable comparator and held-out validation}
\label{sec:comparator}

The avoided-crossing gap is directly sensitive to the off-diagonal pair matrix element.  
It therefore provides an observable-level test of whether the controller dependence seen in Fig.~\ref{fig:spectrum} can be explained entirely by an occupation-dependent diagonal spectrum, or whether the pair amplitude itself must depend on $n_b$.  
To make this distinction without relying on a particular perturbative representation, we construct two compact four-state models on the signal ladder $n_a=0,2,4,6$.  
Both models use the same functional form for the pair-transition frequencies,
\begin{equation}
    f(n_a,n_b) = c_0 + c_1 n_a + c_2 n_b + c_3 n_a n_b + c_4 n_a^2 + c_5 n_b^2,
    \label{eq:comparatorfreq}
\end{equation}
where $f(n_a,n_b)$ represents the transition $|n_a,n_b\rangle\leftrightarrow|n_a+2,n_b\rangle$ in cyclic-frequency units.

For each controller sector, the transition-frequency polynomial in Eq.~\eqref{eq:comparatorfreq} is accumulated into a four-state diagonal ladder according to
\begin{align}
    \varepsilon_0(n_b) & = 0, \nonumber\\
    \varepsilon_2(n_b) & = f(0,n_b), \nonumber\\
    \varepsilon_4(n_b) & = \varepsilon_2(n_b) + f(2,n_b), \nonumber\\
    \varepsilon_6(n_b) & = \varepsilon_4(n_b) + f(4,n_b),
    \label{eq:comparator-diagonal}
\end{align}
before transforming to the pair-rotating frame defined by the pump, so that
\begin{equation}
    \varepsilon_{n_a+2}(n_b) - \varepsilon_{n_a}(n_b) = f(n_a,n_b).
\end{equation}
In the pump-rotating pair basis, the diagonal entries are therefore
\begin{equation}
    \varepsilon_{n_a}(n_b) - \frac{n_a}{2}f_p.
    \label{eq:comparator-rotating-diagonal}
\end{equation}
This construction ensures that both comparator models have the same flexible, controller-dependent diagonal spectrum.

The models differ only in the functional form of their off-diagonal pair amplitude.  
The controller-independent model retains signal-occupation dependence,
\begin{equation}
    g_{\rm OI}(n_a) = q_0 + q_1 n_a + q_2 n_a^2,
    \label{eq:gordinary}
\end{equation}
but contains no explicit $n_b$ dependence.  The conditional model adds a single controller coefficient,
\begin{equation}
    g_{\rm C}(n_a,n_b) = q_0 + q_1 n_a + q_2 n_a^2 + q_3 n_b.
    \label{eq:gconditional}
\end{equation}
Adjacent pair states are coupled by $\sqrt{(n_a+1)(n_a+2)}\,g$, so the avoided crossings are obtained from the complete four-state Hamiltonian rather than by identifying the fitted coefficient $g$ directly with a measured gap.

The three controller sectors $n_b=0,1,2$ provide nine training crossings, corresponding to $n_a=0,2,4$ in each row.  
Resonance positions and avoided-crossing gaps are fitted simultaneously.  
For the least-squares objective, resonance residuals are converted to MHz and concatenated with the gap residuals in MHz, so every spectral datum enters with equal numerical weight.  
Each model is fitted globally only once; no controller sector receives an independent fit.  
The complete $n_b=3$ row is excluded from the objective and is evaluated only after the fitted parameters have been fixed.

The resonance and gap entering the objective function are obtained through bounded minimization within a $\pm 10$ MHz window around each crossing, with a frequency tolerance of $10^{-10}$ GHz. The nonlinear fit uses relative termination tolerances of $10^{-10}$ and a maximum of 200 function evaluations. 
Because the resonance and gap data are optimized simultaneously, changing the off-diagonal parameterization can slightly shift the optimal diagonal coefficients $c_k$, even though both models use the same diagonal functional form.  
This accounts for the small difference between their training resonance errors in Table~\ref{tab:comparator}.

The contrast in the avoided-crossing gaps is substantial.  On the training sectors, introducing the single coefficient $q_3$ reduces the gap root-mean-square error from approximately $384.0$ to $6.2$ kHz.  More importantly, the improvement persists outside the fitted controller occupations.  For the completely held-out $n_b=3$ row, the gap RMSE decreases from $970.3$ to $24.5$ kHz, and the maximum gap error falls from $1.383$ MHz to $39.0$ kHz.  No $n_b = 3$ resonance or gap enters the fit, and no sector-specific recalibration is performed before these predictions are evaluated.

The Jacobian condition numbers are approximately $5.63\times10^{2}$ for the controller-independent model and $5.64\times10^{2}$ for the controller-dependent model. 
For the latter, the fitted controller coefficient is
\begin{equation}
    q_3 = (-6.362 \pm 0.363) \times10^{-5}~{\rm GHz},
    \label{eq:q3-comparator}
\end{equation}
where the uncertainty is the linearized least-squares standard error obtained from the local Jacobian and residual variance.  
It is a conditioning diagnostic for this finite comparator model rather than an experimental statistical uncertainty.

The diagonal extrapolation is less accurate.  
The held-out resonance RMSE is approximately $1.6$ MHz for both models, showing that the chosen low-order polynomial in Eq.~\eqref{eq:comparatorfreq} does not reproduce the next controller row with the same accuracy as the off-diagonal trend.  
We leave this limitation explicit rather than absorbing it into an $n_b=3$ refit.  
The central comparison is therefore especially clear at the level of the gaps: once a flexible occupation-dependent diagonal spectrum has already been allowed, a controller-dependent pair amplitude is still required to reproduce the avoided-crossing splittings.
\begin{table}[tbp]
\caption{\textbf{Observable comparator and held-out prediction.}
Errors are in kHz.  The $n_b = 0,1,2$ sectors are fitted globally using both resonance positions and avoided-crossing gaps, while the complete $n_b = 3$ row is withheld.  
The two models use the same diagonal functional form and differ by the presence or absence of the explicit $q_3n_b$ term in the pair amplitude.}
\label{tab:comparator}
\vspace{3pt}
\footnotesize
\begin{ruledtabular}
\begin{tabular}{lcc}
metric & no $n_b$ term & with $n_b$ term\\
\hline
train resonance RMSE & $91.7$ & $62.6$\\
train gap RMSE & $384.0$ & $6.2$\\
held-out resonance RMSE & $1614.5$ & $1603.4$\\
held-out gap RMSE & $970.3$ & $24.5$\\
held-out max gap error & $1383$ & $39.0$
\end{tabular}
\end{ruledtabular}
\end{table}

\begin{figure}[t]
\centering
\includegraphics[width=\columnwidth]{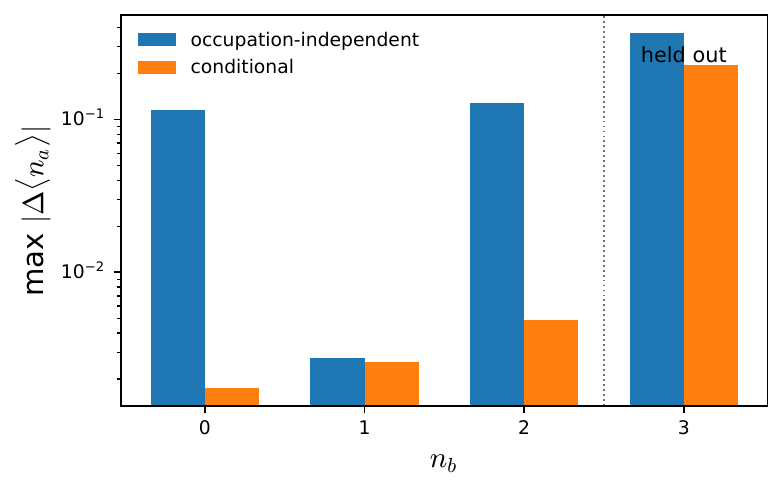}
\caption{\textbf{Dynamical comparator}. Maximum signal-number error through the first operation for the controller-independent and controller-dependent pair-amplitude models, at $t \leq t_{\pi}$. The vertical divider separates the fitted $n_b = 0,1,2$ sectors from the held-out $n_b = 3$ sector.}
\label{fig:comparator-dynamics}
\end{figure}

For the dynamical comparison, each fixed parameter set is evolved at the unexpanded-cosine resonance of the corresponding controller sector. In the fitted sectors $n_b = 0,1,2$, in Fig.~\ref{fig:comparator-dynamics}, the controller-independent model exhibits maximum errors in the signal photon number during the first operation of $0.1153$, $0.00276$, and $0.1279$ photons, respectively. The controller-dependent model keeps the maximum first-operation signal-number error below $4.9\times10^{-3}$ photons, whereas the controller-independent model reaches $0.128$ photons.  In the held-out $n_b = 3$ trajectory, the corresponding maximum errors are $0.226$ and $0.369$ photons.

Unlike the individually calibrated avoided-crossing gaps, this dynamical comparison is sensitive to errors in both the pair amplitude and the extrapolated diagonal spectrum.  
Each reduced model is propagated at the unexpanded-cosine resonance of the corresponding controller sector using its globally fitted parameters, so the remaining held-out discrepancy is partly dominated by the megahertz-scale resonance extrapolation error.  
The spectral gap comparison therefore more cleanly discriminates matrix-element conditioning, while the propagation test shows how the same modeling distinction affects a complete operation.

The resulting distinction separates the present interaction from several familiar forms of conditional nonlinear dynamics.  
Cross-Kerr spectroscopy produces an occupation-dependent resonance without by itself requiring an occupation-dependent resonant pair amplitude.  
Nondegenerate squeezing changes the occupations of both participating bosonic modes, while qubit-conditioned squeezing uses a nonbosonic controller~\cite{DelGrosso2025,Ayyash2024,Ayyash2025,Blumenthal2026,Hope2026}. 
Josephson-photonics and recent high-order mixing experiments demonstrate occupation-dependent nonlinear matrix elements in related processes~\cite{Armour2015,Vanselow2026}. 
The feature isolated here is the simultaneous $n_b$ dependence of the resonance position and avoided-crossing gap of a degenerate pair transition that conserves the controller occupation.

\section{Physical parent architecture and Floquet isolation}
\label{sec:implementation}

The two-mode Hamiltonian of Sec.~\ref{sec:model} is intended as the spectrally selected sector of a multimode Josephson circuit.  
Connecting it to a specific device therefore requires a sequence of experimentally accessible steps: the electromagnetic structure determines the normal-mode frequencies and their zero-point phase participation at the nonlinear element; the pump then generates a set of Floquet-assisted transition channels; and the coupling and detuning of those channels determine which modes can remain perturbative and which must be included explicitly in the driven Hamiltonian.

The individual ingredients of this construction are established in superconducting-circuit platforms.  
Black-box and energy-participation quantization reduce a distributed electromagnetic structure to normal-mode frequencies and zero-point phases at a Josephson coordinate~\cite{Nigg2012,Minev2021}. 
SQUID-terminated resonators demonstrate degenerate and nondegenerate flux-pumped interactions~\cite{Simoen2015,Bengtsson2018}, while multimode parametric cavities show that selected interactions can be addressed in spectra containing many physical modes~\cite{Chang2018,Lee2020,Busnaina2024}.  
Recent energy-participation and finite-element implementations also realize zero-point phase fluctuations comparable with or larger than those used in the benchmark of Table~\ref{tab:benchmark}~\cite{Makihara2024,Vanselow2026}.

\begin{table*}[t]
\caption{\textbf{Physical ingredients supporting the parent architecture.}
Representative theoretical and experimental results establishing the steps from electromagnetic mode quantization to flux-pumped multimode control.}
\label{tab:parents}
\vspace{3pt}
\footnotesize
\begin{ruledtabular}
\begin{tabular*}{0.97\textwidth}{@{\extracolsep{\fill}}llll}
reference & platform/framework & established ingredient & role here\\
\hline
Nigg; Minev~\cite{Nigg2012,Minev2021}
& general Josephson circuits
& \shortstack[l]{normal-mode quantization\\nonlinear participation}
& $\mathrm{EM}\rightarrow\{\omega_\mu,\varphi_\mu\}$\\
Simoen; Bengtsson~\cite{Simoen2015,Bengtsson2018}
& SQUID-terminated resonators
& \shortstack[l]{degenerate/nondegenerate\\flux pumping}
& driven shared coordinate\\
Chang~\cite{Chang2018}
& multimode parametric cavity
& selected microwave-mode interactions
& reduced multimode model\\
Lee; Busnaina~\cite{Lee2020,Busnaina2024}
& highly multimode cavities
& \shortstack[l]{frequency-selective control \\ in dense spectra}
& pump-assisted selection\\
Makihara; Vanselow~\cite{Makihara2024,Vanselow2026}
& multimode nonlinear circuits
& large nonlinear-node ZPF
& accessible phase scale
\end{tabular*}
\end{ruledtabular}
\end{table*}

The benchmark of Table~\ref{tab:benchmark} specifies the two retained modes and the nonlinear scale; a concrete implementation supplies its own measured or simulated spectrum $\{\omega_\mu\}$ and zero-point phases $\{\varphi_\mu\}$.  
Periodic driving adds the requirement that nominally spectator modes remain isolated not only in the static spectrum but also from pump-assisted sum-, difference-, and multipump resonances.

\begin{figure}[b]
\centering
\includegraphics[width=\columnwidth]{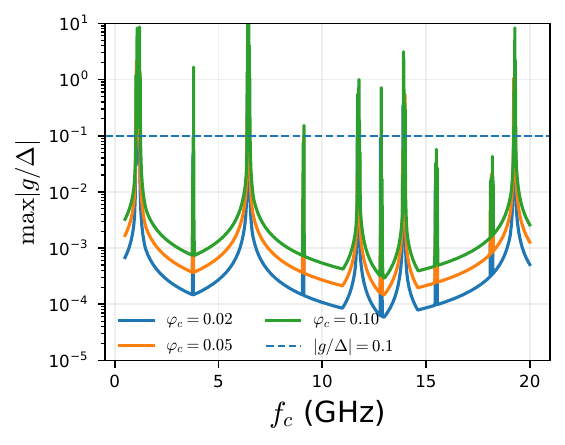}
\caption{\textbf{Floquet spectator-isolation map.}
Maximum direct one-pump coupling-to-detuning ratio over the populated pair-ladder support for a parameterized omitted mode $c$. 
The finite-width resonance windows arise from the occupation dependence of the nonlinear diagonal spectrum.  
For a specific circuit, the measured or simulated spectator frequencies and zero-point phases can be placed directly on this map. 
A mode approaching a pump-assisted exclusion window is no longer treated perturbatively and is instead promoted to the explicit multimode Floquet Hamiltonian.}
\label{fig:spectator}
\end{figure}

Writing the driven Hamiltonian as
\begin{equation}
    H(t) = H_0 + \sum_{m\neq0} V^{(m)}e^{\ii m\Omega t},
\end{equation}
we associate a retained state $|i\rangle$ and an omitted state $|f\rangle$ with the Floquet mismatch
\begin{equation}
    \Delta_{fi}^{(m)} = \frac{E_f-E_i}{\hbar} - m\Omega,
    \qquad
    g_{fi}^{(m)} = \frac{\mel{f}{V^{(m)}}{i}}{\hbar}.
    \label{eq:spectatordef}
\end{equation}
Both $g_{fi}^{(m)}$ and $\Delta_{fi}^{(m)}$ are angular frequencies in this definition.  
If $\mathcal S$ denotes the retained set of states that acquire appreciable population during the intended operation, a controlled closed-system elimination requires
\begin{equation}
    \epsilon_{\rm spec} = \max_{i\in\mathcal S,\,f\notin\mathcal S,\,m} 
    \left| \frac{g_{fi}^{(m)}}{\Delta_{fi}^{(m)}} \right| \ll 1.
    \label{eq:epsspec}
\end{equation}
The criterion is therefore state-support dependent: it must be checked over the Fock-lattice region actually visited by the protocol rather than only from the vacuum spectrum.

For an open implementation, a spectator transition with coherence half-width $\gamma_{fi}$ is characterized by the complex-detuning magnitude
\begin{equation}
    \left| \widetilde{\Delta}_{fi}^{(m)} \right| = \sqrt{\left[\Delta_{fi}^{(m)}\right]^2 + \gamma_{fi}^2},
\end{equation}
and the corresponding isolation parameter becomes
\begin{equation}
    \epsilon_{\rm spec}^{\rm open} = \max_{i\in\mathcal S,\,f\notin\mathcal S,\,m}
    \frac{|g_{fi}^{(m)}|}{\sqrt{[\Delta_{fi}^{(m)}]^2 + \gamma_{fi}^2}} \ll 1.
    \label{eq:epsspec-open}
\end{equation}
Eliminating a weakly coupled lossy spectator also produces dispersive shifts and induced decay with the usual $|g|^2\Delta/(\Delta^2+\gamma^2)$ and $|g|^2\gamma/(\Delta^2+\gamma^2)$ scalings; the explicit expressions and conventions are given in Appendix~\ref{app:spectator}.

The first practical step is to determine which pump harmonics need to be retained. 
At the present bias and modulation amplitude, the exact symmetric-SQUID Fourier coefficients are
\begin{align}
    \frac{|E_J^{(1)}|}{h} & = 80.000~{\rm MHz}, \nonumber\\
    \frac{|E_J^{(2)}|}{h} & = 5.177~{\rm kHz},  \nonumber\\
    \frac{|E_J^{(3)}|}{h} & = 13.6~{\rm Hz}.
\end{align}
The first-harmonic channels therefore dominate overwhelmingly at this operating point.  Importantly, the relevant drive frequency is the calibrated nonlinear resonance of the occupied pair cell,
\begin{equation}
    f_p = 12.845263~{\rm GHz}
\end{equation}
for the target operation, rather than the bare estimate $2f_a$.

To formulate a device-independent exclusion criterion, we introduce a parameterized spectator mode $c$ with frequency $f_c$ and zero-point phase $\varphi_c$ at the same Josephson coordinate.  
The scan uses the three-mode phase
\begin{equation}
    \hat\varphi_J = \varphi_a(a+a^\dagger) + \varphi_b(b+b^\dagger) + \varphi_c(c+c^\dagger)
\end{equation}
and evaluates unexpanded-cosine matrix elements for direct one-pump processes that create one or two spectator quanta, satisfy $|\Delta n_a|\leq2$ and $|\Delta n_b|\leq2$, and have total number-change order no larger than four.  
The initial retained support covers $n_a=0,2,4,6$ and $n_b=0,1,2,3$, with the spectator initially in its vacuum.

The detunings include the occupation-dependent diagonal energy of the same unexpanded static Josephson cosine.  
Consequently, a given spectator process appears as a finite exclusion window rather than as a single frequency: different occupied $(n_a,n_b)$ cells shift the corresponding resonance condition by different amounts.

Figure~\ref{fig:spectator} shows the resulting maximum coupling-to-detuning ratio over the populated support. 
For the representative choice $\varphi_c=0.05$, prominent quadratic-order one-pump windows occur at
\begin{align}
    b^\dagger c^\dagger:
    &\qquad
    f_c = 1.021\text{--} 1.206~{\rm GHz},
    \nonumber\\
    a^\dagger c^\dagger:
    &\qquad
    f_c = 6.382\text{--} 6.486~{\rm GHz},
    \nonumber\\
    c^{\dagger2}:
    &\qquad
    f_c = 6.424\text{--} 6.429~{\rm GHz},
    \nonumber\\
    a c^\dagger:
    &\qquad
    f_c = 19.222\text{--} 19.305~{\rm GHz}.
    \label{eq:spectator-windows}
\end{align}
Quartic mixed processes generate weaker additional structures and are retained in the numerical exclusion envelope.  
Higher pump harmonics are also included in the channel audit, but their strongly suppressed Fourier amplitudes make the first-harmonic envelope dominant for the present modulation.

The implementation prescription is therefore direct.  An electromagnetic model or measured spectrum supplies the frequencies and SQUID participations of additional electromagnetic, plasma, package, and internal modes. 
Their pump-assisted matrix elements and nonlinear detunings are then evaluated over the state support of the intended operation.  
Modes satisfying Eq.~\eqref{eq:epsspec}, or its linewidth-broadened counterpart in Eq.~\eqref{eq:epsspec-open}, can remain perturbative; modes entering an exclusion window are promoted to the explicit Floquet Hamiltonian. 
The electromagnetic and Floquet calculations thus play complementary roles: the first determines which physical modes exist and how strongly they participate in the nonlinear coordinate, while the second determines which of those modes remain dynamically relevant under the applied pump. 
Appendix~\ref{app:spectator} gives the complete channel enumeration, nonlinear diagonal corrections, harmonic hierarchy, exclusion-window data, and open-system extension.

The present analysis focuses on the coherent driven dynamics and on the spectral conditions required for a controlled multimode reduction.
A microscopic treatment of dissipation in the periodically driven circuit, including the transformation of system--bath couplings under the same Floquet or Schrieffer--Wolff dressing used for the Hamiltonian, lies outside the scope of the present work.
Likewise, we will address the use of engineered reservoirs to stabilize or autonomously select the occupation-conditioned states identified here separately.
The linewidth-broadened spectator criteria above are intended only as implementation diagnostics for determining when an omitted mode can remain perturbative.

\section{Conclusion}
\label{sec:conclusion}

A shared Josephson coordinate lets the occupation of one harmonic mode control a nonlinear process in another more deeply than a cross-Kerr shift alone. 
For the flux-pumped interaction studied here, the controller occupation $n_b$ modifies both observables associated with the same degenerate pair transition $|n_a,n_b\rangle \longleftrightarrow |n_a + 2,n_b\rangle$.
The controller therefore shifts the resonance frequency and changes the resonant avoided-crossing gap even though its own occupation is unchanged by the target process.  These two signatures distinguish spectral conditioning from matrix-element conditioning.

The microscopic origin of the off-diagonal effect follows directly from the full Josephson cosine.  
The $b$-number-conserving pair matrix element contains the controller factor $e^{-\varphi_b^2/2}L_{n_b}(\varphi_b^2)$, which makes the pair-transition amplitude depend explicitly on the occupation of a harmonic mode that is not consumed by the transition.  
Expanding the same circuit Hamiltonian identifies how this dependence emerges perturbatively: direct Josephson nonlinearities generate occupation-dependent pair vertices, while virtual processes renormalize both diagonal energies and off-diagonal couplings. 
Time-dependent Schrieffer--Wolff and Magnus constructions organize these corrections from complementary viewpoints—virtual intermediate states and time-ordered Fourier processes, respectively—and yield the same effective physics when compared in a common physical frame.

The unexpanded two-mode cosine confirms that the interaction remains quantitatively significant at the representative mode-level benchmark. 
Its occupation dependence generates a two-dimensional pair spectrum in which both the resonance position and the transition strength vary across the Fock lattice. 
This structure enables selective transfer on an interior pair-ladder link and produces a clear speed--selectivity tradeoff as pump strength and pulse shape change. 
When the harmonic controller is prepared in a coherent superposition, the same conditional dynamics generates nearly one bit of signal--controller entanglement.  
In separately prepared controller sectors, a single fixed pump selects which signal branch develops substantial Wigner negativity, while the off-resonant sectors develop negligible negative volume.  
Finally, a held-out observable comparator shows that occupation-dependent resonance shifts alone are insufficient to reproduce the spectrum: accurate prediction of the $n_b = 3$ avoided-crossing gaps requires an explicit controller-dependent pair amplitude.

The ingredients required for this mechanism--multimode participation in a common pumped SQUID, frequency-selective parametric control, and zero-point phase fluctuations at the relevant nonlinear scale--are already available in superconducting-circuit architectures. 
A concrete implementation additionally requires the measured or simulated multimode spectrum to satisfy the Floquet-isolation conditions developed in Sec.~\ref{sec:implementation}; any spectator approaching a pump-assisted exclusion window must instead be promoted to the explicitly driven model. 
This separates the universal nonlinear mechanism from the device-specific spectral-engineering problem and provides a concrete route toward occupation-controlled bosonic parametric interactions in multimode superconducting circuits.

The present work addresses the coherent interaction and its Floquet isolation. 
Extending the treatment to microscopic open-system dynamics, including dressed dissipation and reservoir-engineered stabilization of the conditional bosonic states, is left for future work.

\section{Acknowledgments}

This work was supported by the Coordenação de Aperfeiçoamento de Pessoal de Nível Superior (CAPES), Finance Code 001, by the National Natural Science Foundation of China (NSFC) underGrant No. 12174346, by Fundação Araucária (Project No. 305), by the São Paulo Research Foundation (FAPESP), Brasil, Processes Number \#2022/00209-6, and by the Brazilian National Council for Scientific and Technological Development (CNPq), Grant No.~405712/2023-5.
In accordance with journal guidelines, the authors declare the use of Artificial Intelligence Generated Content (AIGC) tools during the preparation of this manuscript. Specifically, Gemini/ChatGPT was used for language polishing, text editing, and assistance with in-depth literature research. Additionally, ChatGPT assisted with conceptualizing and drafting the schematic diagrams. After using these tools, the authors rigorously reviewed, revised, and validated all generated text, research insights, and visual content. The authors assume full and sole responsibility for the integrity, accuracy, and originality of the final manuscript and affirm that no AI tool fulfills the role of an author or is listed as one.

\appendix

\section{Circuit reduction, pump harmonics, and benchmark reconstruction}
\label{app:circuit}

The normal-mode reduction used in Sec.~\ref{sec:model} follows directly from the boundary dynamics of a SQUID-terminated resonator. 
Consider a coplanar-waveguide resonator of length $d$, capacitance per unit length $c$, and inductance per unit length $\ell$. 
Its node-flux field is $\phi(x,t)$, with an open boundary at $x=0$ and a symmetric dc SQUID of total boundary capacitance $C_S$ terminating the resonator at $x=d$. 
Neglecting loop inductance, the Lagrangian is
\begin{eqnarray}
\mathcal L & = & \int_0^d dx\left[\frac{c}{2}\dot\phi^2 - \frac{(\partial_x\phi)^2}{2\ell}\right] \nonumber \\
           & + & \frac{C_S}{2}\dot\phi_d^2 + E_J[\Phi_{\rm ext}(t)]\cos\varphi_J,
\label{eq:app-L}
\end{eqnarray}
with $\phi_d=\phi(d,t)$ and $\varphi_J=(2\pi/\Phi_0)\phi_d$.  
Varying Eq.~\eqref{eq:app-L} gives the wave equation in the line, the open boundary $\partial_x\phi(0,t)=0$, and the exact nonlinear SQUID boundary
\begin{equation}
  \frac{1}{\ell}\partial_x\phi(d,t)+C_S\ddot\phi_d + \frac{2\pi}{\Phi_0}E_J[\Phi_{\rm ext}(t)]\sin\varphi_J = 0.
\end{equation}

At zero pump, we define $\theta_{\rm dc} = \pi\Phi_{\rm dc}/\Phi_0$ and $E_0 = 2E_{J0}\cos\theta_{\rm dc}$.  The linearized SQUID inductance is
\begin{equation}
L_J=\frac{(\Phi_0/2\pi)^2}{E_0}.
\end{equation}
For mode functions $u_n(x) = A_n\cos k_nx$, $k_n = \omega_n\sqrt{\ell c}$, the boundary condition gives
\begin{equation}
\frac{k_n}{\ell}\tan(k_nd) = \frac{1}{L_J}-C_S\omega_n^2.
\label{eq:app-modeeq}
\end{equation}
The capacitance normalization is
\begin{equation}
C_n\delta_{nm} = \int_0^d dx\,c\,u_nu_m+C_Su_n(d)u_m(d).
\end{equation}
Quantization with $q_n = \sqrt{\hbar/(2C_n\omega_n)}(a_n+a_n^\dagger)$ gives
\begin{equation}
\varphi_n=\frac{2\pi}{\Phi_0}u_n(d)\sqrt{\frac{\hbar}{2C_n\omega_n}},
\label{eq:app-zpf}
\end{equation}
which is the distributed-circuit expression behind Eq.~\eqref{eq:phase-multimode}.
In a numerical electromagnetic calculation, the same information is obtained by black-box or energy-participation quantization~\cite{Nigg2012,Minev2021}.

For finite modulation, we define $z=\pi\delta\Phi/\Phi_0$. 
The exact Fourier expansion of Eq.~\eqref{eq:EJexact} is
\begin{eqnarray}
E_J(t) & = &   2E_{J0}\cos\theta_{\rm dc}J_0(z)\nonumber\\
       & + & 4E_{J0}\cos\theta_{\rm dc} \sum_{k\ge1}(-1)^kJ_{2k}(z)\cos(2k\Omega t)\nonumber\\
       & - & 4E_{J0}\sin\theta_{\rm dc} \sum_{k\ge0}(-1)^kJ_{2k+1}(z)\nonumber\\
       & \times & \cos[(2k+1)\Omega t].
\label{eq:app-Fourier}
\end{eqnarray}
At the benchmark, the first three nonzero harmonic amplitudes divided by $h$ are $-80.000$ MHz, $-5.177$ kHz, and $13.6$ Hz. 
The numerical Floquet reference uses these exact harmonics rather than the weak-pump approximation of Eq.~\eqref{eq:Epweak}.

A representative two-node lumped realization of the same mode-level benchmark is described by
\begin{align}
C&=\begin{pmatrix}
C_1 + C_S & -\,C_S \\ 
-\,C_S & C_2+C_S
\end{pmatrix},\nonumber\\[1mm]
K&=\begin{pmatrix}
L_1^{-1} + L_J^{-1} & -\,L_J^{-1} \\
-\,L_J^{-1} & L_2^{-1} + L_J^{-1}
\end{pmatrix}.
\end{align}
The normal modes satisfy the generalized eigenproblem $Ku_j = \omega_j^2 Cu_j$ with $u_i^T Cu_j = \delta_{ij}$.  
The component values
\begin{equation}
\begin{gathered}
C_1 = 178.214~\mathrm{fF},\quad C_2 = 65.863~\mathrm{fF},\quad C_S = 4.314~\mathrm{fF},\\
L_1 = 3.6489~\mathrm{nH},\qquad L_2 = 2.7888~\mathrm{nH}
\end{gathered}
\end{equation}
with $L_J = 32.094$ nH reproduce the frequencies and boundary phases in Table~\ref{tab:benchmark}.  The corresponding energy-participation fractions of the shared inductive branch in this particular lumped normalization are approximately $0.0856$ and $0.0814$. 
In an EPR conversion, the corresponding participation is defined with the linearized inductive energy at the DC operating point, which is the energy scale used by this biased normal-mode problem.

Static junction asymmetry can be estimated by writing $E_{J1,2} = E_{J0}(1\pm d_J)$. 
For $d_J = 0.02$ and negligible loop inductance, the two mode frequencies shift by $3.34$ and $5.83$ MHz, while $\varphi_a$ and $\varphi_b$ change by approximately $-0.166\%$ and $+0.019\%$. 
At this level, modest asymmetry therefore acts predominantly by recalibrating the normal-mode frequencies. 
Driven asymmetry and additional electromagnetic modes introduce distinct corrections that must be included in a device-specific implementation.

\section{Exact cosine matrix elements and direct analytical coefficients}
\label{app:exact}

The controller dependence of the resonant pair matrix element can be obtained without expanding the Josephson cosine. 
The required one-mode displacement matrix element is
\begin{equation}
\mel{n'}{e^{\ii\varphi(a+a^\dagger)}}{n} = e^{-\varphi^2/2}(\ii \varphi)^{n'-n} \sqrt{\frac{n!}{n'!}}L_n^{(n'-n)}(\varphi^2),
\end{equation}
for $n'\ge n$, with the $n'<n$ result following by Hermitian conjugation.  
Since $e^{\ii\hat\varphi_J}=e^{\ii\varphi_aX_a}e^{\ii\varphi_bX_b}$, the two-mode matrix element factorizes.  Taking the real part gives the cosine matrix element and directly yields Eq.~\eqref{eq:exactpairmatrix} after the resonant half of $\cos\Omega t$ is retained.

This dependence of the process amplitude on $n_b$ can also be understood from the direct expansion of the exponential operator. Since
\begin{equation}
e^{\ii\varphi_{b}(\hat{b} + \hat{b}^{\dagger})} = 1 + \ii \varphi_{b}(\hat{b} + \hat{b}^{\dagger}) - \frac{\varphi_{b}^{2}}{2} (\hat{b} + \hat{b}^{\dagger})^{2} + \mathcal{O}(\varphi_{b}^{3}),  
\end{equation}
the diagonal matrix element in $\lvert n_b\rangle$ receives a second-order contribution. Using
\begin{equation}
(\hat{b} + \hat{b}^{\dagger})^{2} = \hat{b}^{2} + \hat{b}^{\dagger 2} + 2 \hat{b}^{\dagger} \hat{b} + 1,  
\end{equation}
we obtain
\begin{equation}
\left\langle n_{b} \left| (\hat{b} + \hat{b}^{\dagger})^{2} \right|n_{b} \right\rangle = 2 n_{b} + 1.  
\end{equation}

Consequently,
\begin{align}  
\left\langle n_{b} \left| e^{\ii\varphi_{b}(\hat{b} + \hat{b}^{\dagger})} \right|n_{b} \right\rangle & = 1 - \frac{\varphi_{b}^{2}}{2}(2 n_{b} + 1) + \mathcal{O}(\varphi_{b}^{4}), \notag \\  
& = 1 - \left(n_{b} + \frac{1}{2} \right) \varphi_{b}^{2} + \mathcal{O} (\varphi_{b}^{4}),  
\end{align}  
in exact agreement with the expansion of the exact result.

This last form is particularly useful because it explicitly reveals the origin of the $n_b$ dependence. The term $2\hat{b}^{\dagger}\hat{b} + 1$ contains the number operator $\hat{n}_b = \hat{b}^{\dagger}\hat{b}$, whereas the terms $\hat{b}^{2}$ and $\hat{b}^{\dagger 2}$ do not contribute directly to this diagonal matrix element. 
In the perturbative interpretation, these terms represent connections to intermediate Fock sectors, whereas the exact result expressed in terms of Laguerre polynomials incorporates all these processes simultaneously.

For the parameters considered, the exact factor $C_b(n_b)$ takes the following values for the lowest Fock states:
\begin{equation}
    C_{b}(0) \simeq 0.954,  
\end{equation}
\begin{equation}
    C_{b}(1) \simeq 0.863,  
\end{equation}
\begin{equation}
    C_{b}(2) \simeq 0.777,  
\end{equation}
and
\begin{equation}
    C_{b}(3) \simeq 0.695.  
\end{equation}

Therefore, the amplitude of the pair-creation process decreases progressively as the occupation of mode $b$ increases. 
For example,
\begin{equation}
\frac{M_{n_{a}, 1}^{(1)}} {M_{n_{a}, 0}^{(1)}} \frac{C_{b}(1)}{C_{b}(0)} \simeq 0.905,  
\end{equation}
whereas
\begin{equation}
\frac{M_{n_{a}, 3}^{(1)}}{M_{n_{a}, 0}^{(1)}} \frac{C_{b}(3)}{C_{b}(0)} \simeq 0.729.  
\end{equation}

Thus, increasing the occupation from $n_b = 0$ to $n_b = 3$ reduces the process amplitude by approximately $27\%$, with $n_a$ held fixed. 
Since, to leading order, the transition rates are proportional to the squared modulus of the matrix element,
\begin{equation}
    \Gamma_{n_{a}, n_{b}} \propto |M_{n_{a}, n_{b}}^{(1)}|^{2},  
\end{equation}
the corresponding change in the transition rate can be even more pronounced.

The direct quartic expansion of Eq.~\eqref{eq:fullH} is
\begin{equation}
H^{(4)} = \hbar\omega_an_a + \hbar\omega_bn_b - \frac{E_0}{24}\hat\varphi_J^4 
        + \Ep\cos\Omega t\left(\frac{\hat\varphi_J^2}{2} - \frac{\hat\varphi_J^4}{24}\right).
\label{eq:app-H4}
\end{equation}
The sextic completion adds
\begin{equation}
H^{(6)}-H^{(4)}=\frac{E_0}{720}\hat\varphi_J^6 + \frac{\Ep}{720}\cos\Omega t\, \hat\varphi_J^6.
\label{eq:app-H6}
\end{equation}
The phase and perturbative orders are not interchangeable: a third-order virtual process built from three quadratic pump vertices is also sixth order in the zero-point phases.

Normal ordering Eq.~\eqref{eq:app-H4} gives the leading Kerr coefficients in Eq.~\eqref{eq:kerrleading}.  
In the same convention, the direct pair-amplitude polynomial through the pumped quartic sector is
\begin{align}
g_0 = & \frac{\Ep}{\hbar}\left(\frac{\varphi_a^2}{4} - \frac{\varphi_a^4}{8} - \frac{\varphi_a^2\varphi_b^2}{8}\right),\\
g_a = & -\frac{\Ep}{\hbar}\frac{\varphi_a^4}{12},
\qquad
g_b=-\frac{\Ep}{\hbar}\frac{\varphi_a^2\varphi_b^2}{4}.
\end{align}
The occupation-independent terms proportional to $\varphi_a^4$ and $\varphi_a^2\varphi_b^2$ in $g_0$ arise from vacuum contractions of the pumped quartic operator.  
For comparison with the virtual SW contributions, the direct coefficients are expressed in the same falling-factorial basis.  
At quartic order,
\begin{widetext}
\begin{align}
\zeta_{10}^{[4]} &= -\frac{E_0}{2\hbar}(\varphi_a^4 + \varphi_a^2\varphi_b^2), &
\zeta_{01}^{[4]} &= -\frac{E_0}{2\hbar}(\varphi_b^4 + \varphi_a^2\varphi_b^2),\\
\zeta_{20}^{[4]} &= -\frac{E_0\varphi_a^4}{4\hbar}, &
\zeta_{02}^{[4]} &= -\frac{E_0\varphi_b^4}{4\hbar}, &
\zeta_{11}^{[4]} &= -\frac{E_0\varphi_a^2\varphi_b^2}{\hbar},
\label{eq:app-direct4zeta}
\end{align}
while $\gamma_{00}^{[2+4]} = g_0$, $\gamma_{10}^{[4]} = g_a$, and $\gamma_{01}^{[4]} = g_b$.

The direct sixth-order diagonal coefficients generated by $E_0\hat\varphi_J^6/720$ are
\begin{align}
\zeta_{10}^{[6]} &= \frac{E_0}{\hbar}\left(\frac{\varphi_a^6}{8} + \frac{\varphi_a^4\varphi_b^2}{4}+\frac{\varphi_a^2\varphi_b^4}{8}\right),\\
\zeta_{01}^{[6]} &= \frac{E_0}{\hbar}\left(\frac{\varphi_b^6}{8} + \frac{\varphi_a^2\varphi_b^4}{4}+\frac{\varphi_a^4\varphi_b^2}{8}\right),\\
\zeta_{20}^{[6]} & = \frac{E_0}{8\hbar}(\varphi_a^6 + \varphi_a^4\varphi_b^2),&
\zeta_{02}^{[6]} & = \frac{E_0}{8\hbar}(\varphi_b^6 + \varphi_a^2\varphi_b^4),\\
\zeta_{11}^{[6]} & = \frac{E_0}{2\hbar}(\varphi_a^4\varphi_b^2 + \varphi_a^2\varphi_b^4),\\
\zeta_{30}^{[6]} & = \frac{E_0\varphi_a^6}{36\hbar},&
\zeta_{03}^{[6]} & = \frac{E_0\varphi_b^6}{36\hbar},\\
\zeta_{21}^{[6]} & = \frac{E_0\varphi_a^4\varphi_b^2}{4\hbar},&
\zeta_{12}^{[6]} & = \frac{E_0\varphi_a^2\varphi_b^4}{4\hbar}.
\label{eq:app-direct6zeta}
\end{align}
The resonant $q_a=+2$ component of the pumped sextic term gives
\begin{align}
\gamma_{00}^{[6]} & = \frac{\Ep}{\hbar}\left(\frac{\varphi_a^6}{32} + \frac{\varphi_a^4\varphi_b^2}{16}
                    + \frac{\varphi_a^2\varphi_b^4}{32}\right),\\
\gamma_{10}^{[6]} & = \frac{\Ep}{24\hbar}(\varphi_a^6 + \varphi_a^4\varphi_b^2),&
\gamma_{01}^{[6]} & = \frac{\Ep}{8\hbar}(\varphi_a^4\varphi_b^2 + \varphi_a^2\varphi_b^4),\\
\gamma_{20}^{[6]} & = \frac{\Ep\varphi_a^6}{96\hbar},&
\gamma_{11}^{[6]} & = \frac{\Ep\varphi_a^4\varphi_b^2}{12\hbar},&
\gamma_{02}^{[6]} & = \frac{\Ep\varphi_a^2\varphi_b^4}{16\hbar}.
\label{eq:app-direct6gamma}
\end{align}
\end{widetext}
Equations~\eqref{eq:app-direct4zeta}--\eqref{eq:app-direct6gamma} contain the direct Josephson contributions. 
The SW elimination in Appendix~\ref{app:sw} generates virtual corrections that renormalize the same $\zeta_{ru}$ and $\gamma_{ru}$ coefficients.

\section{Schrieffer--Wolff, Magnus, and physical-frame reconstruction}
\label{app:sw}

The effective interaction in Sec.~\ref{sec:pairmechanism} follows by retaining the resonant pair ladder while eliminating the remaining Fourier-resolved vertices perturbatively. 
The construction below fixes the retained sector, the physical energy mismatches, the SW recursion through third order, the corresponding Magnus expansion, and the transformation back to laboratory states and observables.

\subsection{Fourier vertices and resonant projector}

Write the phase-expanded Hamiltonian as $H(t) = H_0 + V(t)$ with
\begin{equation}
H_0 = \hbar \omega_a n_a + \hbar \omega_b n_b.
\end{equation}
A normal-ordered monomial is labeled by its number changes and pump harmonic,
\begin{equation}
O_\mu(t) = (a^\dagger)^r a^s(b^\dagger)^u b^v e^{\ii m\Omega t},
\quad
\mu=(r,s,u,v,m),
\label{eq:app-monomial}
\end{equation}
with mismatch
\begin{equation}
\Delta_\mu = (r-s) \omega_a + (u-v) \omega_b + m \Omega.
\label{eq:app-mismatch}
\end{equation}
Equivalently, a component with number changes $(q_a,q_b)$ has $\Delta_{m;q_aq_b} = q_a\omega_a + q_b\omega_b + m\Omega$.

The retained slow set is
\begin{equation}
\mathcal R=\{(0;0,0),\,(-1;+2,0),\,(+1;-2,0)\}.
\label{eq:app-retained}
\end{equation}
It contains every static number-conserving operator together with the complete signal-pair creation and annihilation ladders.  
Let $\mathcal R$ also denote the projector onto these monomials and $\mathcal O = 1 - \mathcal R$ its complement, and write $V = V_R + V_O$. 
Since the near-resonant $a^{\dagger2}e^{-\ii\Omega t}$ term belongs to $V_R$, no denominator proportional to $2\omega_a-\Omega$ is introduced by the SW transformation.

At quadratic pump order, the independent off-resonant vertices are
\begin{center}
\footnotesize
\begin{ruledtabular}
\begin{tabular}{lll}
vertex & coefficient in $V$ & mismatch\\
\hline 
$a^{\dagger2}e^{+\ii\Omega t}$ & $\Ep\varphi_a^2/4$ & $2\omega_a+\Omega$\\
$(2n_a+1)e^{+\ii\Omega t}$ & $\Ep\varphi_a^2/4$ & $\Omega$\\
$b^{\dagger2}e^{\pm\ii\Omega t}$ & $\Ep\varphi_b^2/4$ & $2\omega_b\pm\Omega$\\
$(2n_b+1)e^{+\ii\Omega t}$ & $\Ep\varphi_b^2/4$ & $\Omega$\\
$a^\dagger b^\dagger e^{\pm\ii\Omega t}$ & $\Ep\varphi_a\varphi_b/2$ & $\omega_a+\omega_b\pm\Omega$\\
$a^\dagger b e^{\pm\ii\Omega t}$ & $\Ep\varphi_a\varphi_b/2$ & $\omega_a-\omega_b\pm\Omega$
\end{tabular}
\end{ruledtabular}
\end{center}
The static and pumped quartic and sextic terms are decomposed by the same rule, with their direct coefficients given in Appendix~\ref{app:exact}.

\subsection{Time-dependent SW transformation}

The SW transformation changes the Hamiltonian representation via a unitary transformation. 
Introduce the anti-Hermitian periodic generator
\begin{equation}
S(t) = S_1(t) + S_2(t) + S_3(t) + \cdots, \qquad S^\dagger = -S,
\end{equation}
where $S_1$ is first order in the interaction, $S_2$ is second order, and so on, and the normal-form state $|\psi_{\rm NF}\rangle=e^{S(t)}|\psi\rangle$. 
The transformed Hamiltonian is
\begin{equation}
H_{\rm NF} = e^S H e^{-S} - \ii \hbar e^S \partial_t e^{-S}.
\label{eq:app-tdsw}
\end{equation}

We require that, after the transformation, the terms coupling off-resonant sectors be eliminated up to the desired perturbative order. 
Define
\begin{equation}
\mathcal L X = [H_0,X] - \ii \hbar \dot X.
\label{eq:app-Lsw}
\end{equation}
For the monomial in Eq.~\eqref{eq:app-monomial},
\begin{equation}
\mathcal L O_ \mu = \hbar \Delta_ \mu O_\mu.
\end{equation}
The first-order cancellation equation is therefore
\begin{equation}
\mathcal L S_1 = V_O,
\label{eq:first_cancellation}
\end{equation}
This may seem abstract, but it becomes clearer when we consider a single off-resonant process. 
Suppose that we have
\begin{equation}
    V_{\alpha} (t) = V_{\alpha} e^{\ii t \Delta_{\alpha} / \hbar},
\end{equation}
here, $\Delta_{\alpha}$ represents the energy detuning of the process. 
For example, for the two-photon creation transition $\lvert n_a\rangle\rightarrow\lvert n_a+2\rangle$, the detuning may involve a combination such as
\begin{equation}
    \Delta_{\alpha} \sim \hbar (\Omega + 2 \omega_{a}).
\end{equation}

If this quantity is large, the process is far off resonance. 
Solving Eq.~\eqref{eq:first_cancellation}, we obtain schematically
\begin{equation}
    S_{1, \alpha} = \frac{V_{\alpha}}{\hbar \Delta_{\alpha}}, 
\end{equation}
The key physical feature of this expression is the factor $1/\Delta_{\alpha}$ in the denominator. 
The farther the process is from resonance, the smaller its virtual contribution.

Solving
\begin{equation}
    \mathcal{L} S_{1} = V_{\mathrm{off}}, \quad \mathcal{L} X = [H_{0}, X] - \ii \hbar \dot{X},
\end{equation}
we obtain the explicit generator
\begin{equation}
S_1(t) = \sum_{\mu\in O}\frac{v_\mu}{\hbar\Delta_\mu}O_\mu(t).
\label{eq:app-S1}
\end{equation}
Hermiticity of $V$ pairs every term with one of opposite mismatch, so Eq.~\eqref{eq:app-S1} is automatically anti-Hermitian.

For the counter-rotating local pair and the pump-induced number modulation of mode $a$, the first-order generator contains
\begin{align}
S_{1,a^2}^{[2]}(t) & = \frac{\Ep\varphi_a^2}{4\hbar(\Omega+2\omega_a)} \left(a^{\dagger2}e^{+\ii\Omega t}-a^2e^{-\ii\Omega t}\right),
\label{eq:app-S1pair}\\
S_{1,n_a}^{[2]}(t) & = \frac{\Ep\varphi_a^2(2n_a+1)}{4\hbar\Omega} \left(e^{+\ii\Omega t} 
                     - e^{-\ii\Omega t}\right).
\label{eq:app-S1number}
\end{align}
Every remaining off-resonant contribution follows the same ratio of microscopic vertex to physical mismatch.

An important point arises here. 
The pump contains two components:
\begin{equation}
    \cos (\Omega t) = \frac{1}{2} (e^{\ii \Omega t} + e^{-\ii \Omega t}).
\end{equation}

Moreover, the pair-creation operator has its own evolution associated with $2\omega_a$. 
Depending on the temporal component under consideration, denominators such as $\Omega-2\omega_a$ or $\Omega+2\omega_a$ arise. 
The former corresponds to the near-resonant pair-creation sector, $\Omega\simeq2\omega_a$, and must therefore be retained. 
The latter, $\Omega+2\omega_a$, is always far off resonance for positive frequencies and corresponds to the counter-rotating component. 
This is precisely the type of term that the SW transformation is designed to eliminate.

\subsection{Second-order Hamiltonian and virtual Stark shifts}

We now seek to determine the second-order effects induced by the virtual processes. 
Expanding Eq.~\eqref{eq:app-tdsw} to second order gives
\begin{align}
A_2 & = [S_1,V_R]+\frac12[S_1,V_O], \nonumber\\
H_{\rm SW}^{(2)} & = \mathcal R A_2,
\qquad
S_2 = \mathcal L^{-1} \mathcal O A_2,
\label{eq:app-A2}
\end{align}
here, $\mathcal{R}$ denotes the operation of retaining the terms that belong to the effective sector, whereas $\mathcal{O}$ denotes the off-resonant terms. 
Conceptually, $A_2$ therefore contains both new effective terms and additional off-resonant terms. 
The former enter the effective Hamiltonian, whereas the latter are eliminated using $S_2$. 
Writing $V_R=\sum_{\rho\in R}v_\rho O_\rho$, $V_O=\sum_{\mu\in O}v_\mu O_\mu$, this simply means that
\begin{align}
V_{\mathrm{res}} & = v_{1} O_{1} + v_{2} O_{2} + \ldots,\\
V_{\mathrm{off}} & = v_{3} O_{3} + v_{4} O_{4} + \ldots.
\end{align}
Each $O_{\mu}$ is a specific operator, such as $\hat{a}^{\dagger 2}$, $\hat{a}\hat{b}^{\dagger}$, or $\hat{a}^{\dagger}\hat{b}$, together with its associated time dependence. 
Since
\begin{equation}
    S_{1} = \sum_{\mu \in O} s_{\mu}^{(1)} O_{\mu},
\end{equation}
with $s_\mu^{(1)}=v_\mu/(\hbar\Delta_\mu)$, we can evaluate the commutators term by term. 
For example,
\begin{equation}
    [S_{1}, V_{\mathrm{res}}] = \sum_{\mu, \rho} s_{\mu}^{(1)} v_{\rho} [O_{\mu}, O_{\rho}].
\end{equation}

If
\begin{equation}
    [O_{\mu}, O_{\rho}] = \sum_{\gamma} \Gamma_{\mu \rho}^{\gamma} O_{\gamma},
\end{equation}
the coefficient of the operator $O_{\gamma}$ is therefore $\sum_{\mu,\rho}s_{\mu}^{(1)}v_{\rho}\Gamma_{\mu\rho}^{\gamma}$. 
Applying the same procedure to $\frac{1}{2}[S_1,V_{\mathrm{off}}]$, we obtain the coefficient of any monomial $O_\gamma$ in $A_2$ is
\begin{equation}
 a_\gamma^{(2)} =  \sum_{\mu\in O}\sum_{\rho\in R}s_\mu^{(1)}v_\rho\Gamma_{\mu\rho}^{\gamma}
                +  \frac12\sum_{\mu\in O}\sum_{\nu\in O}s_\mu^{(1)}v_\nu\Gamma_{\mu\nu}^{\gamma}.
\label{eq:app-a2coeff}    
\end{equation}
Therefore, $a_{\gamma}^{(2)}$ is simply the coefficient that remains for the operator $O_{\gamma}$ after all second-order commutators have been evaluated. 
For a retained monomial, this is its second-order energy coefficient. 
In the falling-factorial basis of Eq.~\eqref{eq:app-ledger},
\begin{equation}
\zeta_{ru}^{(2)}  = \frac{a_{(r,r,u,u,0)}^{(2)}}{\hbar},
\qquad
\gamma_{ru}^{(2)} = \frac{a_{(r+2,r,u,u,-1)}^{(2)}}{\hbar}.
\label{eq:app-zeta-gamma2}
\end{equation}
Normal ordering reduces all products to the same monomial basis. 
For one mode,
\begin{equation}
(a^\dagger)^r a^s(a^\dagger)^R a^S = \sum_{k=0}^{\min(s,R)}k!\binom{s}{k}\binom{R}{k} (a^\dagger)^{r+R-k}a^{s+S-k},
\label{eq:app-normalorder}
\end{equation}
and contractions in modes $a$ and $b$ factorize. 
The index $k$ denotes the number of contractions between the operators $a$ and $a^{\dagger}$.

For two modes, define $M_{rsuv} = (a^\dagger)^ra^s(b^\dagger)^ub^v$.  
Applying Eq.~\eqref{eq:app-normalorder} independently to $a$ and $b$ gives
\begin{widetext}
\begin{equation}
M_{rsuv}M_{RSUV} = \sum_{k=0}^{\min(s,R)}\sum_{l=0}^{\min(v,U)} k!l!\binom{s}{k}\binom{R}{k}\binom{v}{l}\binom{U}{l} M_{r+R-k,\,s+S-k,\,u+U-l,\,v+V-l},
\label{eq:app-twomodeproduct}
\end{equation}
\end{widetext}
here, $k$ counts the contractions involving mode $a$, whereas $l$ counts those involving mode $b$. Including the pump harmonics, write
\begin{equation}
[O_\mu,O_\nu]=\sum_\gamma\Gamma_{\mu\nu}^{\gamma}O_\gamma.
\label{eq:app-GammaDef}
\end{equation}
The structure constants are obtained by subtracting Eq.~\eqref{eq:app-twomodeproduct} with the two monomials exchanged. Explicitly,
\begin{widetext}
\begin{equation}
\Gamma_{\mu\nu}^{\gamma} = \sum_{k,l} k!l!\binom{s}{k}\binom{R}{k}\binom{v}{l}\binom{U}{l}\,\delta_{\gamma,\gamma_{\mu\nu}^{kl}} 
- \sum_{k,l} k!l!\binom{S}{k}\binom{r}{k}\binom{V}{l}\binom{u}{l}\,\delta_{\gamma,\gamma_{\nu\mu}^{kl}},
\label{eq:app-GammaExplicit}   
\end{equation}
where
\begin{eqnarray}
\gamma_{\mu\nu}^{kl} &=& (r+R-k,s+S-k,u+U-l,v+V-l,m+M), \\
\gamma_{\nu\mu}^{kl} &=& (R+r-k,S+s-k,U+u-l,V+v-l,M+m),
\end{eqnarray}
\end{widetext}
and the sums use the contraction ranges of Eq.~\eqref{eq:app-twomodeproduct}.  
The SW and Magnus commutators are therefore reduced to finite sums of normal-ordered monomials.

The elementary commutators read
\begin{align}
[a^2,a^{\dagger2}] & = 4n_a + 2, \qquad [ab,a^\dagger b^\dagger]=n_a + n_b + 1,\\
[ab^\dagger,a^\dagger b] & = n_b - n_a.
\end{align}

We can now understand the “virtual loops.” For this purpose, consider the process $\hat{a}^{\dagger}\hat{b}$ which induces the transition
\begin{equation}
    |n_{a}, n_{b} \rangle \rightarrow |n_{a} + 1, n_{b} - 1 \rangle.
\end{equation}

If this process is off-resonance, it is virtual. 
The reverse path is $\hat{a}\hat{b}^{\dagger}$. 
Therefore,
\begin{equation}
    |n_{a}, n_{b} \rangle \rightarrow |n_{a} + 1, n_{b} - 1 \rangle \rightarrow |n_{a}, n_{b} \rangle.
\end{equation}

Note that the resulting effective contribution is proportional to
\begin{equation}
[a^{\dagger} b, a b^{\dagger}] = n_{a} - n_{b},
\end{equation}
and, therefore, this loop does not permanently create or annihilate photons. 
It only shifts the energy of the state with
\begin{equation}
    \delta E \propto (n_{a} - n_{b}),
\end{equation}
this show why second-order virtual loops are predominantly diagonal. 
Using only two quadratic pump vertices gives
\begin{equation}
\frac{H_{\rm SW}^{(2,\varphi^4)}}{\hbar} = \delta\omega_a^{(2,4)}n_a + \delta\omega_b^{(2,4)}n_b + \mathrm{const.},
\label{eq:app-H2phi4}
\end{equation}
where, with $D_{\sigma\tau}=\Omega + \sigma\omega_a + \tau\omega_b$,
\begin{widetext}
\begin{align}
\delta\omega_a^{(2,4)} = & \frac{\Ep^2}{4\hbar^2}\Bigg[-\frac{\varphi_a^4}{\Omega + 2\omega_a}
+ \varphi_a^2\varphi_b^2\left(\frac1{D_{--}} + \frac1{D_{+-}} - \frac1{D_{-+}} - \frac1{D_{++}}\right)\Bigg],
\label{eq:app-dwa} \\
\delta\omega_b^{(2,4)}= & \frac{\Ep^2}{4\hbar^2}\Bigg[\varphi_b^4\left(\frac1{\Omega - 2\omega_b} 
- \frac1{\Omega + 2\omega_b}\right) + \varphi_a^2\varphi_b^2\left(\frac1{D_{--}} + \frac1{D_{-+}} - \frac1{D_{+-}} 
- \frac1{D_{++}}\right)\Bigg].
\label{eq:app-dwb}
\end{align}
\end{widetext}
For example, the $a^\dagger b$ path with amplitude $v_{ab} = \Ep \varphi_a\varphi_b/2$ contributes as
\begin{equation}
   \frac{v_{ab}^2}{\hbar D_{+-}}[a^\dagger b,ab^\dagger]  = \frac{\Ep^2\varphi_a^2\varphi_b^2}{4\hbar D_{+-}} (n_a - n_b), 
\end{equation}
which identifies the corresponding contribution in Eqs.~\eqref{eq:app-dwa} and \eqref{eq:app-dwb}.  
The resonant denominator $2\omega_a - \Omega$ is absent because that channel belongs to the retained pair sector and is never eliminated.

At sixth-order phase, the retained operator basis can contain diagonal falling-factorial polynomials up to cubic order and pair operators multiplied by quadratic occupation polynomials,
\begin{eqnarray}
\!\!\!\!\!\!\! &&   \frac{H_{\rm eff}^{(\le6)}}{\hbar}  =  \sum_{r+u\le3}\zeta_{ru}\Nfa{r}\Nfb{u} \nonumber \\
\!\!\!\!\!\!\! && + \left\{a^{\dagger2}\sum_{r+u\le2}\gamma_{ru}\Nfa{r}\Nfb{u}+\mathrm{H.c.}\right\} + \mathrm{const.}
\label{eq:app-ledger}  
\end{eqnarray}
where $\mathcal N_j^{(0)}=1$ and $\mathcal N_j^{(1)}=n_j$.  
The direct sixth-order coefficients are given in Appendix~\ref{app:exact}, while the virtual coefficients are finite sums over the Fourier vertices and normal-order contractions above.

The first term in Eq.~\eqref{eq:app-ledger} contains all diagonal shifts up to sixth order. 
For example, $\zeta_{10} n_a$ gives the frequency contribution of mode $a$, $\zeta_{01} n_b$ gives the frequency contribution of mode $b$, $\zeta_{20} n_a(n_a - 1)$ is a Kerr contribution, $\zeta_{11}n_a n_b$ is a cross-Kerr contribution, and $\zeta_{30} n_a (n_a - 1)(n_a - 2)$ is a higher-order correction to the Kerr nonlinearity.

The second term explicitly contains $\hat{a}^{\dagger 2}$ and therefore belongs to the pair-creation sector. 
For example, $\gamma_{00}\hat{a}^{\dagger 2}$ is the usual pairing term, whereas $\gamma_{10}\hat{a}^{\dagger 2}n_a$ indicates that the pairing amplitude depends on the occupation of mode $a$ itself. 
Similarly, $\gamma_{01}\hat{a}^{\dagger 2} n_b$ indicates that mode $b$ controls the pair-creation amplitude in mode $a$. 
The latter term is directly related to $g_b$.

Moreover, Eq.~\eqref{eq:app-ledger} shows how the expression can be systematically generalized to
\begin{equation}
    G(n_{a}, n_{b}) \simeq g_{0} + g_{a} n_{a} + g_{b} n_{b},
\end{equation}
where we arrive at
\begin{eqnarray}
G(n_{a}, n_{b}) & = & \gamma_{00} + \gamma_{10} n_{a} + \gamma_{01} n_{b}  + \gamma_{20} n_{a}(n_{a} - 1) \notag \\ 
                & + &  \gamma_{11} n_{a} n_{b} + \gamma_{02} n_{b}(n_{b} - 1) + \ldots,
\end{eqnarray}
we can then obtain
\begin{align}
g_{0} & = \gamma_{00},\\
g_{a} & = \gamma_{10},\\
g_{b} & = \gamma_{01},
\end{align}
the terms $\gamma_{20}$, $\gamma_{11}$, and $\gamma_{02}$ are precisely the higher-order corrections that begin to appear when the expansion is carried out to sixth order.

\subsection{Third order and the pump-parity rule}

The off-resonant part of Eq.~\eqref{eq:app-A2} determines $S_2$. 
The third-order expression is
\begin{eqnarray}
A_3 & = & [S_2,V_R] + \frac12[S_2,V_O] - \frac12[S_1,\mathcal L S_2] \nonumber \\
    & + & \frac12[S_1,[S_1,V_R]] + \frac13[S_1,[S_1,V_O]],
\label{eq:app-A3}
\end{eqnarray}
with
\begin{equation}
H_{\rm SW}^{(3)} = \mathcal R A_3,\qquad S_3 = \mathcal L^{-1}\mathcal O A_3.
\label{eq:app-H3}
\end{equation}
Equations~\eqref{eq:app-A3} and \eqref{eq:app-H3} describe paths with two virtual intermediate steps that return to the retained pair sector.
Define the off-resonant second-order generator coefficient $s_\eta^{(2)} = a_\eta^{(2)}/(\hbar\Delta_\eta)$ for $\eta\in O$.  
Using the same structure constants as Eq.~\eqref{eq:app-a2coeff}, the coefficient of $O_\gamma$ in $A_3$ is
\begin{widetext}
\begin{eqnarray}
a_\gamma^{(3)} & = &  \sum_{\mu\in O,\rho\in R}s_\mu^{(2)}v_\rho\Gamma_{\mu\rho}^{\gamma}
  + \frac12\sum_{\mu,\nu\in O}s_\mu^{(2)}v_\nu\Gamma_{\mu\nu}^{\gamma}
  - \frac12\sum_{\mu,\eta\in O}s_\mu^{(1)}a_\eta^{(2)}\Gamma_{\mu\eta}^{\gamma} \nonumber \\
& + & \frac12\sum_{\mu,\nu\in O}\sum_{\rho\in R}\sum_\eta s_\mu^{(1)}s_\nu^{(1)}v_\rho\Gamma_{\nu\rho}^{\eta}\Gamma_{\mu\eta}^{\gamma}
  + \frac13\sum_{\mu,\nu,\rho\in O}\sum_\eta s_\mu^{(1)}s_\nu^{(1)}v_\rho\Gamma_{\nu\rho}^{\eta}\Gamma_{\mu\eta}^{\gamma}.
\label{eq:app-a3coeff}
\end{eqnarray}
\end{widetext}
For retained monomials,
\begin{equation}
\zeta_{ru}^{(3)} = \frac{a_{(r,r,u,u,0)}^{(3)}}{\hbar},
\quad
\gamma_{ru}^{(3)} = \frac{a_{(r+2,r,u,u,-1)}^{(3)}}{\hbar}.
\label{eq:app-zeta-gamma3}
\end{equation}
Together, Eqs.~\eqref{eq:app-GammaExplicit}, \eqref{eq:app-a2coeff}, and \eqref{eq:app-a3coeff} determine every retained second- and third-order coefficient from the circuit vertices.

The pump harmonic index imposes a selection rule before evaluating the energy denominators. 
Every first-harmonic pump vertex carries $m = \pm 1$ and every static vertex carries $m=0$. A product of $j$ pump vertices therefore satisfies
\begin{equation}
\Delta m\equiv j\pmod 2.
\label{eq:app-parity}
\end{equation}
A purely pump-generated diagonal term has $\Delta m=0$ and consequently uses an even number of pump vertices.  
A one-pump pair term has $\Delta m=\pm1$ and uses an odd number. 
This is why two quadratic pump vertices give the Stark shifts above, while three quadratic pump vertices can renormalize the resonant pair coefficient.

The single-mode quadratic-pump sector gives a compact example. 
With $v_a=\Ep\varphi_a^2/4$, the first generator is
\begin{eqnarray}
S_1^{(a)}(t) & = & \frac{v_a(2n_a+1)}{\hbar\Omega}(e^{+\ii \Omega t} - e^{-\ii \Omega t})\nonumber\\
             & + & \frac{v_a}{\hbar(\Omega + 2\omega_a)}(a^{\dagger2}e^{+\ii \Omega t} - a^2e^{-\ii \Omega t}).    
\end{eqnarray}
Carrying Eq.~\eqref{eq:app-A3} through the $a^{\dagger2}e^{-\ii \Omega t}$ sector gives
\begin{equation}
\gamma_{00,a}^{\rm SW3(p2^3)} = -\frac{\Ep^3\varphi_a^6(\Omega+\omega_a)(5\Omega + 4\omega_a)}{96\hbar^3 \Omega^2 (\Omega - \omega_a)(\Omega+2\omega_a)}.
\label{eq:app-gamma3example}
\end{equation}
Because all three vertices in this example are quadratic, the result remains a Gaussian pair coupling and renormalizes only $\gamma_{00}$. 
Kerr and number-dependent pair operators enter when at least one nonlinear Josephson vertex participates, as in the static-quartic--quadratic-pump paths already present at second order.

\subsection{Magnus expansion in the same process basis}

In the interaction picture,
\begin{equation}
V_I(t)=\sum_\mu v_\mu\widetilde O_\mu e^{\ii \Delta_\mu t},
\label{eq:app-VI}
\end{equation}
so the Magnus expansion begins with the same list of circuit vertices and the same physical mismatches as SW.
Writing
\begin{equation}
U_I(t)=\exp[\Omega_1(t) + \Omega_2(t) + \Omega_3(t) + \cdots],
\end{equation}
the first three terms are
\begin{align}
\Omega_1(t) & = - \frac{\ii}{\hbar}\int_0^t V_I(t_1)\,dt_1,\\
\Omega_2(t) & = -\frac{1}{2\hbar^2}\int_0^t dt_1 \int_0^{t_1}dt_2\,[V_I(t_1),V_I(t_2)],
\label{eq:app-M2}\\
\Omega_3(t) & = \frac{\ii}{6\hbar^3} \int_{0<t_3<t_2<t_1<t}dt_1dt_2dt_3 \nonumber \\
& \times\big([V_1,[V_2,V_3]] + [V_3,[V_2,V_1]]\big).
\label{eq:app-M3}
\end{align}
For two frequencies,
\begin{equation}
I_2(x,y;t) = \frac{1}{\ii y} \left[\frac{e^{\ii (x+y)t} - 1}{\ii (x+y)}-\frac{e^{\ii xt} - 1}{\ii x}\right],
\label{eq:app-I2}
\end{equation}
and the ordered three-vertex integral is
\begin{equation}
I_3(x,y,z;t) = \frac{I_2(x,y+z;t) - I_2(x,y;t)}{\ii z}.
\label{eq:app-I3}
\end{equation}
The denominators in Eqs.~\eqref{eq:app-I2} and \eqref{eq:app-I3} are therefore the same virtual energy mismatches that appear in Eq.~\eqref{eq:app-S1}.

A laboratory-frame high-frequency truncation based only on $[H_{-1},H_{+1}]/\Omega$ is insufficient for the present resonance problem. 
Because the pump is intentionally resonant with a circuit transition, the relevant virtual denominators are $q_a\omega_a+q_b\omega_b+m\Omega$, not simply $m\Omega$. 
The interaction-picture construction retains the resonant pair channel and averages only genuinely off-resonant paths.

\begin{figure}[tbp]
\centering
\includegraphics[width=0.96\columnwidth]{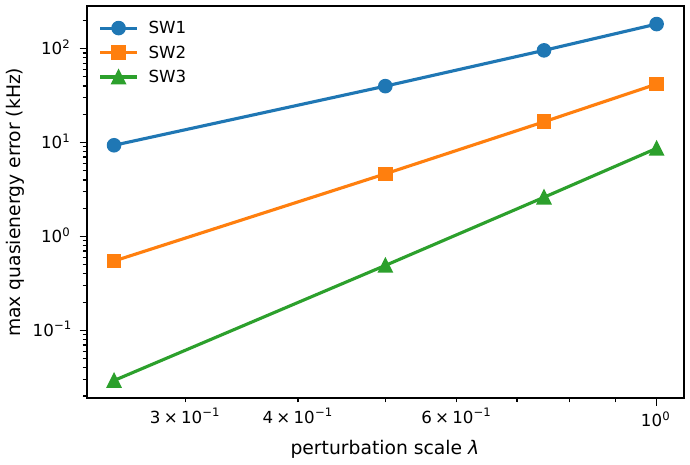}\par\vspace{1mm}
\includegraphics[width=0.96\columnwidth]{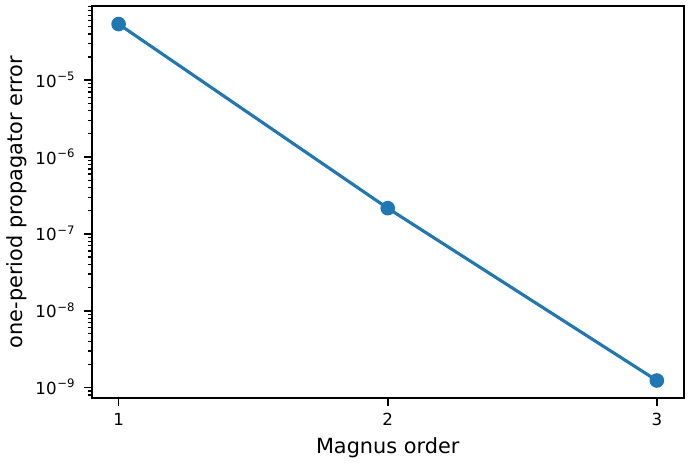}
\caption{Perturbative convergence of the SW and Magnus constructions. 
The upper panel shows the decrease of the quasienergy error as the SW normal form is extended from first to third order over the perturbation-scale sweep. 
The lower panel shows the corresponding Magnus-order convergence of the one-period propagator.  Both calculations use the same Fourier-resolved circuit vertices and physical mismatches.}
\label{fig:app-sw-magnus-scaling}
\end{figure}

The two perturbative organizations encode the same physical paths differently: SW labels virtual intermediate operators and their mismatches, whereas Magnus orders the corresponding vertices in time. 
The sequence $a^{\dagger2}e^{+\ii\Omega t_1}$ followed by $a^2e^{-\ii\Omega t_2}$ produces $[a^{\dagger2},a^2] = -(4 n_a + 2)$ and hence the local Stark shift. 
At third order, the harmonic sequence $(+1) + (+1) + (-1) = +1$ is the time-domain counterpart of Eq.~\eqref{eq:app-parity} and can return to the pair sector. 
Figure~\ref{fig:app-sw-magnus-scaling} shows the corresponding convergence with perturbative order.

\subsection{Common physical frame}

The same unitary transformation used to simplify the Hamiltonian must also act on states and observables.  
The SW normal-form and laboratory states satisfy
\begin{equation}
|\psi_{\rm NF}(t)\rangle = e^{S(t)}|\psi_{\rm lab}(t)\rangle.
\end{equation}
After the pair-rotating transformation $R(t)$,
\begin{equation}
|\psi_{\rm lab}(t)\rangle = e^{-S(t)}R(t)|\psi_{\rm eff}(t)\rangle,
\label{eq:app-state-reconstruction}
\end{equation}
and a laboratory observable $O$ is represented in the effective frame by
\begin{equation}
O_{\rm eff}(t) = R^\dagger(t) e^{S(t)} O e^{-S(t)}R(t).
\label{eq:app-observable-reconstruction}
\end{equation}
The dressing begins as
\begin{equation}
e^S O e^{-S} = O + [S_1,O] + [S_2,O] + \frac12[S_1,[S_1,O]] + \cdots.
\end{equation}
Hence a measured quadrature or number operator can acquire sideband and nonlinear pieces even when the normal-form Hamiltonian is static in the rotating frame.

The periodic SW dressing plays the role of micromotion in the corresponding Floquet representation.  
Effective coefficients can therefore differ between SW and a stroboscopic Magnus gauge, while quasienergies, states, and observables reconstructed in the same physical frame agree to the retained perturbative order. 
The same retained and eliminated sectors can also be formulated within quasi-degenerate perturbation theory, as implemented in Pymablock~\cite{ArayaDay2025Pymablock}.

\section{Floquet extraction, convergence, and state-support diagnostics}
\label{app:floquet}

The quantitative reference is the unexpanded two-mode cosine with the exact pump Fourier components $E_J^{(m)}$.  
Its avoided crossings are extracted from the Floquet Hamiltonian in Sambe space,
\begin{equation}
\mathcal H_F = \sum_m H^{(m)} \otimes T_m + \hbar\Omega\,\mathbb I \otimes N_F,
\label{eq:app-sambe}
\end{equation}
where $T_m$ shifts the Fourier index by $m$ and $N_F$ is the diagonal Fourier-number operator. 
For each pair cell, the resonance is located by minimizing the splitting of the two Floquet eigenstates with largest joint weight in the selected bare pair subspace. 
The resulting minimum splitting is reported as the avoided-crossing gap. 
The assignment overlap quoted in Sec.~\ref{sec:spectrum} is the mean selected-subspace weight of those two states.

Fock and Fourier cutoffs are increased until the extracted resonances, gaps, and dynamical observables are stable. 
The spectrum in Fig.~\ref{fig:spectrum} uses a signal basis containing the displayed $n_a = 6$ support and a controller basis extending beyond the $n_b = 3$ row; both are enlarged in the convergence calculations. 
The strong suppression of higher pump harmonics keeps the required Sambe range modest at the benchmark, but convergence is established numerically rather than inferred from the modulation amplitude alone.

\begin{figure}[tbp]
\centering
\includegraphics[width=0.96\columnwidth]{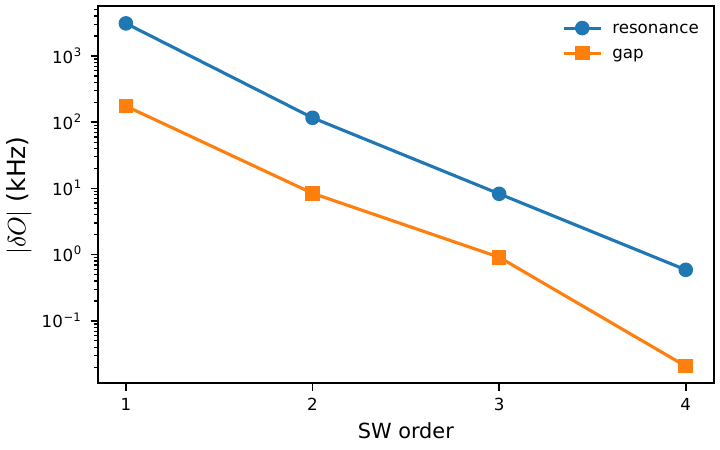}
\caption{Phase-resummed SW-order convergence for the benchmark pair transition. Once the Josephson phase is kept unexpanded, increasing SW order systematically improves both the resonance and the avoided-crossing gap.}
\label{fig:swconvergence-main}
\end{figure}

The compact $H_{\rm eff}^{(6,3)}$ and the phase-resummed SW hierarchy probe different approximation axes. 
The latter retains the full matrix function of the Josephson phase while increasing the SW order and the retained pair-ladder size, allowing finite phase order, finite SW order, and finite Hilbert-space truncation to be separated. 
Extended coherent and squeezed states probe several pair cells simultaneously, and their high-Fock tails are monitored to distinguish genuine dynamical disagreement from basis truncation.  
Figure~\ref{fig:swconvergence-main} shows the phase-resummed convergence with SW order.

\section{Control protocol and coherent error definitions}
\label{app:control}

The control analysis distinguishes the local two-state transfer scale from the optimum of the complete nonlinear ladder. 
For a selected pair crossing, Eq.~\eqref{eq:tpi} defines $t_\pi$ from the \emph{cyclic-frequency} avoided-crossing gap, while the pulse duration used in the full dynamics is taken from the first maximum of the unexpanded-cosine target population. 
Their small difference reflects weak coupling to additional pair-ladder states.

For the target transition $|2,1\rangle\leftrightarrow|4,1\rangle$, same-ladder leakage is measured by the dressed-state population of $|\widetilde{6,1}\rangle$. 
Controller-sector leakage is defined from the total bare projector onto states with $n_b\ne1$ after transforming the state to the common physical basis. 
Off-target response is measured in separately initialized neighboring pair transitions under the same physical pump. 
These quantities separate continued climbing within the selected pair ladder, leakage between controller sectors, and response of neighboring transitions.

\begin{figure}[tbp]
\centering
\includegraphics[width=0.98\columnwidth]{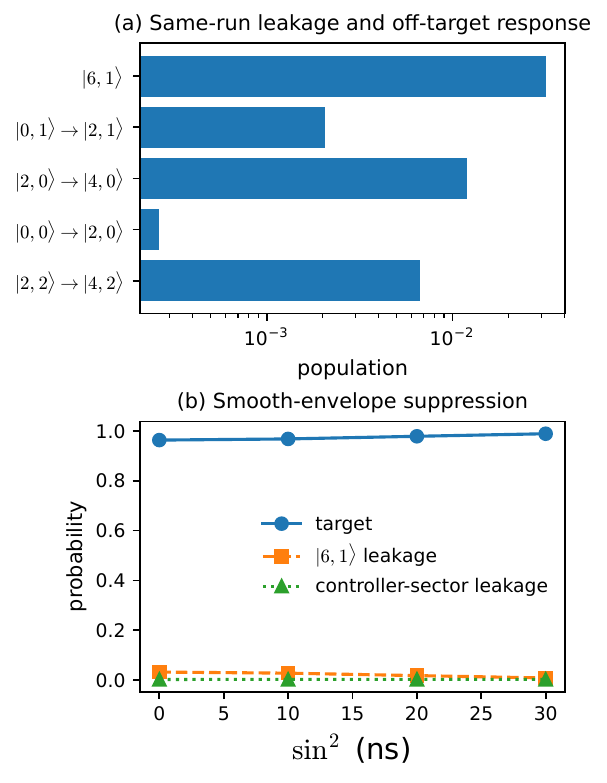}
\caption{Leakage channels and pulse shaping. (a) Population in $|6,1\rangle$ at the nominal target maximum together with transfer probabilities for separately initialized neighboring transitions under the same pump. (b) Target transfer, same-ladder leakage, and controller-sector leakage as the $\sin^2$ ramp duration is increased at fixed peak amplitude.}
\label{fig:control-aux}
\end{figure}

The amplitude sweep rescales the physical first pump harmonic and recalibrates the first target maximum at each amplitude. 
Figure~\ref{fig:control-aux} resolves the associated leakage channels and the effect of pulse shaping. 
The smooth envelope uses $\sin^2$ turn-on and turn-off ramps at fixed peak amplitude, with the total duration calibrated numerically. 
Increasing the ramp duration suppresses continued pair-ladder climbing while preserving near-unit target transfer.

\begin{figure}[tbp]
\centering
\includegraphics[width=0.98\columnwidth]{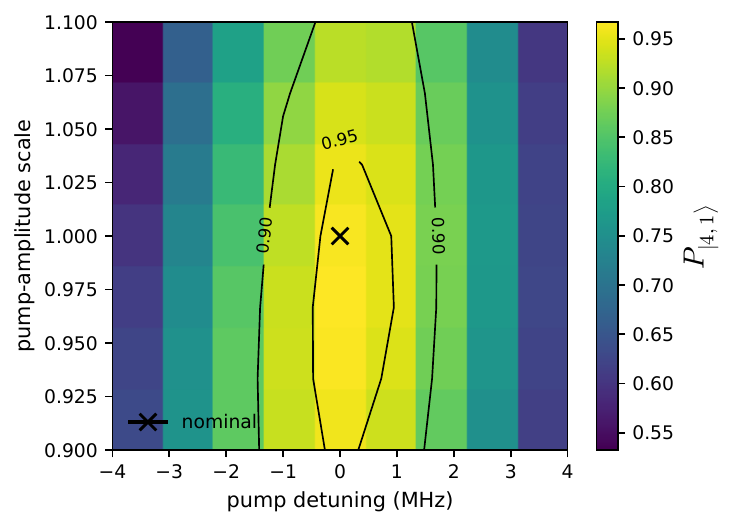}
\caption{Deterministic calibration map for the selected pair operation. The target population is evaluated at the nominal pulse duration while pump detuning and amplitude are varied. The cross marks the calibrated point; contours indicate target populations $0.90$ and $0.95$.}
\label{fig:calibration-main}
\end{figure}

At fixed pulse duration, varying the pump-frequency offset and pump-amplitude scale gives the calibration landscape in Fig.~\ref{fig:calibration-main}.

\section{Quantum-controller and non-Gaussian diagnostics}
\label{app:resources}

The static dressed states used in Secs.~\ref{sec:qcontroller} and \ref{sec:nonGaussian} are obtained by diagonalizing the pump-off two-mode Hamiltonian and assigning each eigenstate the bare Fock label $|n_a,n_b\rangle$ for which its overlap is maximal. 
The controller superposition in Eq.~\eqref{eq:controllerstate} is constructed from these dressed eigenstates before the periodic drive is applied. 
The same preparation is used for the pump-off reference, so that the comparison isolates dynamics induced by the periodic modulation rather than by a mismatch between bare and static dressed states.

For the driven state $\rho(t)$, the reduced signal and controller density operators are
\begin{equation}
\rho_a(t) = \mathrm{Tr}_b\rho(t),
\qquad
\rho_b(t) = \mathrm{Tr}_a\rho(t).
\end{equation}
The signal--controller entanglement entropy reported in Sec.~\ref{sec:qcontroller} is evaluated from $\rho_a(t)$ using Eq.~\eqref{eq:controller-entropy}. 
For the closed two-mode evolution, the joint state remains pure, so the reduced entropies of the two modes are equal.

For the fixed-controller calculations of Sec.~\ref{sec:nonGaussian}, the initial states are the static dressed eigenstates associated with $|0,n_b\rangle$ for $n_b = 0,1,2$. 
Each sector is propagated independently under the same physical pump, after which mode $b$ is traced out before evaluating the signal diagnostics.

The Wigner function is calculated from $\rho_a$ on a finite phase-space grid and normalized according to
\begin{equation}
\int d^2\alpha\,W(\alpha) = 1.
\end{equation}
The negative volume is then obtained from Eq.~\eqref{eq:wignerneg}. 
Both the phase-space extent and the grid resolution are increased until the integrated negative volume is stable. 
The signal Hilbert-space cutoff is enlarged independently to verify that the negative regions are not artifacts of Fock-space truncation.

The signal parity and purity are evaluated as
\begin{equation}
\langle \Pi \rangle = \mathrm{Tr}\!\left[\rho_a(-1)^{n_a}\right],
\qquad
\mathcal P = \mathrm{Tr}\rho_a^2,
\end{equation}
whereas the mean signal occupation reads
\begin{equation}
\langle n_a\rangle = \mathrm{Tr}(\rho_a n_a).
\end{equation}

The even-cat comparison uses the state defined in Eq.~\eqref{eq:even-cat-main}.  
Its fidelity with the reduced signal state is
\begin{equation}
\mathcal F_{\rm cat}(\alpha) = \langle C_{+}(\alpha)|\rho_a|C_{+}(\alpha)\rangle,
\end{equation}
and the reported value is
\begin{equation}
\mathcal F_{\rm cat} = \max_{|\alpha|,\,\arg\alpha} \mathcal F_{\rm cat}(\alpha).
\end{equation}
The optimization therefore determines both the magnitude and phase of the coherent amplitude rather than restricting the comparison to a fixed quadrature axis.

For phase shifts generated by the signal number operator,
\begin{equation}
\rho_a(\theta) = e^{-\ii\theta n_a} \rho_a e^{+\ii\theta n_a},
\end{equation}
the mixed-state quantum Fisher information quoted in Eq.~\eqref{eq:resource-qfi} is evaluated from the eigendecomposition
\begin{equation}
\rho_a = \sum_i\lambda_i|i\rangle\langle i|
\end{equation}
as
\begin{equation}
F_Q[\rho_a,n_a] = 2\sum_{i,j;\,\lambda_i + \lambda_j>0}
\frac{(\lambda_i-\lambda_j)^2}{\lambda_i + \lambda_j}\left|\langle i|n_a|j\rangle\right|^2.
\label{eq:app-qfi}
\end{equation}

The latter state highlighted in Sec.~\ref{sec:nonGaussian} is evaluated with the same reduced-state construction used throughout the trajectory. 
The Wigner negative volume, parity, purity, mean occupation, cat-state fidelity, and quantum Fisher information are recomputed after enlarging the signal Fock cutoff.  
We also check the Wigner quantities when enlarging the phase-space window and grid resolution.
The reported diagnostics are retained only after these numerical truncations cease to produce appreciable changes.

\section{Floquet spectator closure and implementation domain}
\label{app:spectator}

The two-mode description is controlled only when additional electromagnetic and internal circuit modes remain sufficiently far from all pump-assisted transitions explored during the intended operation. 
Because periodic driving can activate resonances that are absent from the static spectrum, spectator isolation must be assessed using both the Fourier spectrum of the flux-modulated SQUID and the occupations reached by the retained modes. 
The relevant criterion is therefore dynamical rather than purely spectral.

\begin{figure}[tbp]
\centering
\includegraphics[width=0.98\columnwidth]{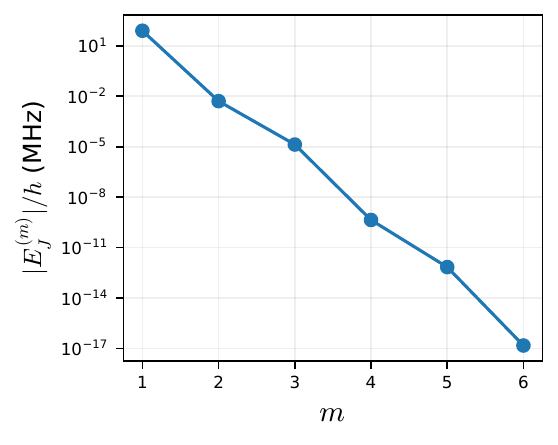}
\caption{\textbf{Pump-harmonic hierarchy.}
Absolute Fourier amplitudes of the symmetric-SQUID Josephson energy at the benchmark dc bias and modulation amplitude. 
The first harmonic dominates strongly, so the spectator structure at this operating point is governed primarily by direct one-pump processes.}
\label{fig:app-pumpharmonics}
\end{figure}

For the symmetric SQUID, the exact flux dependence is evaluated before the spectator channels are enumerated. 
Figure~\ref{fig:app-pumpharmonics} shows the resulting Fourier hierarchy at the benchmark bias and modulation amplitude. 
The first harmonic is dominant by several orders of magnitude, while the higher harmonics provide progressively weaker multipump channels.

For a retained state $|i\rangle$, an omitted state $|f\rangle$, and pump harmonic $m$, Sec.~\ref{sec:implementation} defines the angular-frequency mismatch and coupling as
\begin{equation}
\Delta_{fi}^{(m)} = \frac{E_f-E_i}{\hbar} - m\Omega,
\qquad
g_{fi}^{(m)} = \frac{\mel{f}{V^{(m)}}{i}}{\hbar}.
\end{equation}
A mode can remain perturbative only when the corresponding ratios $|g_{fi}^{(m)}/\Delta_{fi}^{(m)}|$ are small over the complete retained-state support relevant to the operation.

Finite linewidths modify this isolation condition. 
If a spectator transition has coherence half-width $\gamma_{fi}$, written in the same angular-frequency convention as $\Delta_{fi}^{(m)}$, a useful perturbative scale is obtained from
\begin{equation}
\left| \widetilde{\Delta}_{fi}^{(m)} \right| = \sqrt{\left[\Delta_{fi}^{(m)}\right]^2 + \gamma_{fi}^2}.
\end{equation}
The corresponding dimensionless ratio is
\begin{equation}
\frac{|g_{fi}^{(m)}|}{\left|\widetilde{\Delta}_{fi}^{(m)}\right|}.
\end{equation}
For a weakly coupled damped spectator, perturbative elimination also gives the characteristic scales
\begin{align}
|\delta\omega_{fi}^{\rm ind}|
&\sim
\frac{|g_{fi}|^2|\Delta_{fi}|}{\Delta_{fi}^2 + \gamma_{fi}^2},
\nonumber\\
\Gamma_{fi}^{\rm ind}
& \sim\frac{|g_{fi}|^2\gamma_{fi}}{\Delta_{fi}^2 + \gamma_{fi}^2}.
\end{align}
The first quantity estimates the spectator-induced coherent frequency shift, while the second estimates the associated dissipative rate.  
In a concrete circuit, these scales should be compared with the spectral resolution required to distinguish neighboring transitions and with the inverse duration of the intended operation.

These linewidth-dependent expressions are used here only as perturbative implementation criteria for spectator isolation. 
The entanglement, state-transfer, and non-Gaussian dynamics analyzed in the main text are coherent closed-system calculations. 
A microscopic treatment of dissipation in the periodically driven circuit would also require transforming the system--bath couplings and dissipative channels consistently with the Floquet or Schrieffer--Wolff dressing of the Hamiltonian.  
Such an open-system treatment, together with reservoir-engineered stabilization or autonomous preparation of the occupation-conditioned states, is left for future work.

To determine the coherent spectator-exclusion domain, we introduce an additional mode $c$ with frequency $f_c$ and zero-point phase $\varphi_c$ at the same Josephson coordinate.  
The three-mode phase entering the full cosine is
\begin{equation}
\hat\varphi_J  = \varphi_a X_a + \varphi_b X_b + \varphi_c X_c,
\qquad
X_j\equiv j+j^\dagger.
\end{equation}
The exclusion map in Fig.~\ref{fig:spectator} is constructed from exact cosine matrix elements of direct one-pump processes that populate the spectator mode. 
The channel set contains processes that create one or two spectator quanta and satisfy
\begin{equation}
|\Delta n_a|\leq2,
\qquad
|\Delta n_b|\leq2,
\end{equation}
and have total number-change order at most four. 
The initial retained support is
\begin{equation}
n_a=0,2,4,6,
\qquad
n_b=0,1,2,3,
\qquad
n_c=0,
\end{equation}
which covers the pair-ladder region used in the occupation-resolved and selective-control calculations.

The spectator detunings are not evaluated from the bare harmonic frequencies alone. 
They include the direct nonlinear diagonal energy of the same unexpanded static Josephson cosine used in the microscopic model. 
For a three-mode Fock state $|n_a,n_b,n_c\rangle$,
\begin{align}
\frac{E_{\rm diag}}{h} ={}& f_a n_a + f_b n_b + f_c n_c \nonumber\\
&+\frac{E_0}{h}\left[1 - \prod_{j=a,b,c} e^{-\varphi_j^2/2} L_{n_j}(\varphi_j^2) \right. \nonumber\\
&\left.\hspace{18mm} - \frac12 \sum_{j=a,b,c} \varphi_j^2(2n_j+1)\right].
\label{eq:spectator-diag-energy}
\end{align}
The quadratic contribution has already been absorbed into the linear normal-mode problem, so the second line retains only the residual nonlinear Josephson energy.

For each allowed channel and each valid initial state, the transition mismatch is evaluated in cyclic-frequency units as
\begin{equation}
\delta f_{fi}^{(1)} = \frac{ E_{\rm diag}(f) - E_{\rm diag}(i)}{h} - f_p ,
\end{equation}
with $f_p=\Omega/(2\pi)$. 
The direct coupling is converted to the same cyclic-frequency convention before forming the dimensionless ratio. 
Consequently,
\begin{equation}
\left| \frac{g_{fi}^{(1)}}{\Delta_{fi}^{(1)}} \right| = \left| \frac{g_{fi}^{(1)}/2\pi} {\delta f_{fi}^{(1)}} \right|,
\end{equation}
provided the numerator and denominator are expressed consistently.

Because Eq.~\eqref{eq:spectator-diag-energy} depends on $n_a$ and $n_b$, a single operator channel does not correspond to one sharp spectator frequency. 
Instead, it produces a family of nearby resonance conditions across the occupied Fock support.  These occupation-dependent shifts are the origin of the finite-width exclusion windows in Fig.~\ref{fig:spectator}. 
The plotted envelope takes the maximum exact coupling-to-detuning ratio over all retained initial states and all included direct one-pump channels.

For the representative value $\varphi_c=0.05$, the strongest quadratic-order one-pump windows are
\begin{center}
\begin{ruledtabular}
\begin{tabular}{lcc}
process & $f_c$ window (GHz) & $g_{\rm max}/2\pi$ (MHz)\\
\hline
$b^\dagger c^\dagger$
& $1.021$--$1.206$
& $0.984$\\
$a^\dagger c^\dagger$
& $6.382$--$6.486$
& $0.966$\\
$c^{\dagger2}$
& $6.424$--$6.429$
& $0.066$\\
$a c^\dagger$
& $19.222$--$19.305$
& $0.921$\\
\end{tabular}
\end{ruledtabular}
\end{center}
Here $g_{\rm max}$ is the largest exact direct one-pump matrix element over the retained pair-ladder support.  
The quoted interval is not a linewidth: it is the spread of nonlinear resonance centers produced by the different retained occupations. 
The widths therefore reflect the state dependence of the Josephson spectrum rather than dissipation.

The first three windows have a simple interpretation. 
The $b^\dagger c^\dagger$ channel corresponds to nondegenerate creation of one controller and one spectator excitation; the $a^\dagger c^\dagger$ channel creates one signal and one spectator excitation; and $c^{\dagger2}$ corresponds to degenerate spectator-pair creation. 
The higher-frequency $a c^\dagger$ window is a difference-conversion process in which a signal excitation is converted into a spectator excitation with assistance from the pump. 
These examples show why examining only modes close to the pump frequency is insufficient: sum and difference conditions can bring spectators at substantially different frequencies into resonance.

Quartic mixed processes generate additional, weaker structures because the full Josephson cosine contains vertices involving more than two mode operators. 
They are retained in the numerical envelope of Fig.~\ref{fig:spectator}. 
Higher pump harmonics likewise generate additional multipump resonance conditions. 
At the benchmark modulation amplitude, however, their Fourier amplitudes are sufficiently small that the first-harmonic channels dominate the exclusion landscape.

The exclusion map should therefore be interpreted as an implementation-domain test rather than as the spectrum of a particular fabricated device. 
For a concrete circuit, electromagnetic simulation or spectroscopy provides the frequencies of additional resonator, package, plasma, and internal modes together with their phase participation at the SQUID. 
Each physical mode can then be placed at its corresponding $(f_c,\varphi_c)$ point and tested against the pump-assisted resonance structure. 
Modes lying far from all relevant windows and satisfying the perturbative criterion of Sec.~\ref{sec:implementation} can remain outside the explicit dynamical model. 
A mode approaching a window must instead be promoted to the retained multimode Floquet Hamiltonian.

The electromagnetic and driven calculations thus address complementary questions. 
The electromagnetic model determines which modes exist and how strongly they participate in the shared Josephson coordinate; the Floquet analysis determines which of those modes become dynamically relevant under the applied pump. 
This separation provides the appropriate device-level closure test for the coherent two-mode theory developed here, while the quantitative effects of dissipation and the use of engineered reservoirs constitute a separate open-system problem to be addressed in future work.

\section{Numerical methods and convergence}
\label{app:numerics}

The numerical calculations combine direct diagonalization, time-dependent propagation, Floquet analysis, and independent perturbative block-diagonalization. 
Superconducting-circuit spectra and matrix elements are constructed and cross-checked with \texttt{scqubits}
\cite{Groszkowski2021Scqubits,Chitta2022Scqubits}, while the finite Hilbert-space Hamiltonians and time-dependent quantum dynamics are implemented with QuTiP~5 \cite{Lambert2026QuTiP5}. 
The analytical Schrieffer--Wolff construction is additionally compared with
quasi-degenerate perturbation theory generated by \texttt{Pymablock}~\cite{ArayaDay2025Pymablock}.  These calculations provide complementary checks rather than independent physical models: all use the same circuit parameters, pump waveform, mode frequencies, and zero-point phases.

The unexpanded-cosine calculations are performed in truncated Fock spaces for the signal and controller modes. 
Convergence is established by increasing the individual mode cutoffs until the resonances, avoided-crossing gaps, occupation dynamics, and state diagnostics used in the main text no longer change appreciably. 
For states extending over several pair-ladder cells, the probability accumulated near the upper Fock boundary is monitored independently to distinguish genuine physical deviations from truncation effects.

Floquet spectra are obtained from the periodically driven Hamiltonian using a finite Sambe basis. 
Both the Fock-space cutoffs and the number of retained Fourier sectors are increased independently. For each pair transition, the resonance frequency is determined from the minimum splitting of the two Floquet eigenstates with the largest weight in the corresponding bare pair subspace.
The avoided-crossing gap is the minimum quasienergy splitting obtained from the same branch-tracking procedure. 
Convergence therefore requires stability of both the quasienergies and the pair-subspace assignment as the Hilbert and Sambe cutoffs are enlarged.

The perturbative calculations are tested separately with respect to Josephson-phase order and Schrieffer--Wolff order.  
In the compact effective Hamiltonian, the cosine is expanded to finite order in the zero-point phase, and the normal form is truncated at finite perturbative order. 
In the phase-resummed calculation, the full matrix function of the Josephson cosine is retained while the Schrieffer--Wolff order is increased. 
Comparing these calculations with the unexpanded Floquet reference separates errors associated with phase truncation, virtual process order, and finite Hilbert-space support.

As an independent check of the effective-Hamiltonian construction, the relevant quasi-degenerate block-diagonalization is also evaluated with \texttt{Pymablock}. 
Agreement is assessed at the level of retained spectral quantities and effective matrix elements after accounting for the unitary convention used by each perturbative representation. 
The comparison is therefore made between physical quasienergies and transition amplitudes rather than between gauge-dependent intermediate coefficients.

Time-domain convergence is verified by repeating the propagation with enlarged Fock spaces and tighter numerical integration tolerances. 
The reported transfer probabilities, signal occupations, controller leakage, entanglement entropy, parity, purity, and quantum Fisher information are retained only after these quantities are stable under the corresponding numerical enlargements. 
Wigner-function calculations are additionally repeated with increased phase-space extent and grid resolution so that the integrated negative volume is insensitive to both the Fock cutoff and the phase-space discretization.

The three numerical frameworks therefore serve distinct roles:
\texttt{scqubits} connects the circuit parameters to superconducting mode spectra and matrix elements, QuTiP provides the finite-dimensional quantum dynamics and Floquet calculations, and \texttt{Pymablock} provides an independent perturbative block-diagonalization check. 
The agreement of these complementary calculations, together with systematic Hilbert-space, Sambe-space, perturbative-order, and phase-space convergence tests, defines the numerical accuracy of the results reported in the main text.

\bibliography{ref}

@article{Ma2021,
  author  = {Ma, Wen-Long and Puri, Shruti and Schoelkopf, Robert J.
             and Devoret, Michel H. and Girvin, S. M. and Jiang, Liang},
  title   = {Quantum control of bosonic modes with superconducting circuits},
  journal = {Science Bulletin},
  volume  = {66},
  number  = {17},
  pages   = {1789--1805},
  year    = {2021},
  doi     = {10.1016/j.scib.2021.05.024}
}

@article{Joshi2021,
  author  = {Joshi, Atharv and Noh, Kyungjoo and Gao, Yvonne Y.},
  title   = {Quantum information processing with bosonic qubits in circuit QED},
  journal = {Quantum Science and Technology},
  volume  = {6},
  number  = {3},
  pages   = {033001},
  year    = {2021},
  doi     = {10.1088/2058-9565/abe989}
}

@article{Reagor2016,
  author  = {Reagor, Matthew and Pfaff, Wolfgang and Axline, Christopher
             and Heeres, Reinier W. and Ofek, Nissim and Sliwa, Katrina
             and Holland, Eric and Wang, Chen and Blumoff, Jacob
             and Chou, Kevin and Hatridge, Michael J. and Frunzio, Luigi
             and Devoret, Michel H. and Jiang, Liang
             and Schoelkopf, Robert J.},
  title   = {Quantum memory with millisecond coherence in circuit QED},
  journal = {Physical Review B},
  volume  = {94},
  pages   = {014506},
  year    = {2016},
  doi     = {10.1103/PhysRevB.94.014506}
}

@article{Milul2023,
  author  = {Milul, Ofir and Guttel, Barkay and Goldblatt, Uri
             and Hazanov, Sergey and Joshi, Lalit M.
             and Chausovsky, Daniel and Kahn, Nitzan
             and {\c{C}}ifty{\"u}rek, Engin and Lafont, Fabien
             and Rosenblum, Serge},
  title   = {Superconducting Cavity Qubit with Tens of Milliseconds
             Single-Photon Coherence Time},
  journal = {PRX Quantum},
  volume  = {4},
  number  = {3},
  pages   = {030336},
  year    = {2023},
  doi     = {10.1103/PRXQuantum.4.030336}
}

@article{Damas2026,
  author = {Damas, Gabriella G. and Diniz, Ciro Micheletti and de Almeida, Norton G. and Villas-B\^oas, Celso J. and de Moraes Neto, G. D.},
  title = {Engineered {Kerr} nonlinearities for precise quantum control of {Fock} states},
  journal = {Phys. Rev. Applied},
  volume = {25},
  pages = {034097},
  year = {2026},
  doi = {10.1103/2q95-sfjs}
}

@article{Johansson2009,
  author = {Johansson, J. R. and Johansson, G. and Wilson, C. M. and Nori, F.},
  title = {Dynamical {Casimir} effect in a superconducting coplanar waveguide},
  journal = {Phys. Rev. Lett.},
  volume = {103},
  pages = {147003},
  year = {2009},
  doi = {10.1103/PhysRevLett.103.147003}
}

@article{Wilson2011,
  author = {Wilson, C. M. and Johansson, G. and Pourkabirian, A. and Simoen, M. and Johansson, J. R. and Duty, T. and Nori, F. and Delsing, P.},
  title = {Observation of the dynamical {Casimir} effect in a superconducting circuit},
  journal = {Nature},
  volume = {479},
  pages = {376--379},
  year = {2011},
  doi = {10.1038/nature10561}
}

@article{Boutin2017,
  author = {Boutin, Samuel and Toyli, David M. and Venkatramani, Aditya V. and Eddins, Andrew W. and Siddiqi, Irfan and Blais, Alexandre},
  title = {Effect of higher-order nonlinearities on amplification and squeezing in Josephson parametric amplifiers},
  journal = {Phys. Rev. Applied},
  volume = {8},
  pages = {054030},
  year = {2017},
  doi = {10.1103/PhysRevApplied.8.054030}
}

@article{Baskov2025,
  author = {Baskov, Roman and Weiss, Daniel K. and Girvin, Steven M.},
  title = {Exact amplitudes of parametric processes in driven Josephson circuits},
  journal = {Phys. Rev. Applied},
  volume = {24},
  pages = {054038},
  year = {2025},
  doi = {10.1103/zpl5-lztx}
}

@article{Bourassa2012,
  author = {Bourassa, J. and Beaudoin, F. and Gambetta, J. M. and Blais, A.},
  title = {Josephson-junction-embedded transmission-line resonators: From {Kerr} medium to in-line transmon},
  journal = {Phys. Rev. A},
  volume = {86},
  pages = {013814},
  year = {2012},
  doi = {10.1103/PhysRevA.86.013814}
}

@article{Nigg2012,
  author = {Nigg, Simon E. and Paik, Hanhee and Vlastakis, Brian and Kirchmair, Gerhard and Shankar, S. and Frunzio, L. and Devoret, M. H. and Schoelkopf, R. J. and Girvin, S. M.},
  title = {Black-box superconducting circuit quantization},
  journal = {Phys. Rev. Lett.},
  volume = {108},
  pages = {240502},
  year = {2012},
  doi = {10.1103/PhysRevLett.108.240502}
}

@article{Minev2021,
  author = {Minev, Zlatko K. and Leghtas, Zaki and Mundhada, Shantanu O. and Christakis, Ioannis M. and Pop, Ioan M. and Devoret, Michel H.},
  title = {Energy-participation quantization of Josephson circuits},
  journal = {npj Quantum Inf.},
  volume = {7},
  pages = {131},
  year = {2021},
  doi = {10.1038/s41534-021-00461-8}
}

@article{DelGrosso2025,
  author = {Del Grosso, Nicol\'as F. and Corti\~nas, Rodrigo G. and Villar, Paula I. and Lombardo, Fernando C. and Paz, Juan Pablo},
  title = {Controlled-squeeze gate in superconducting quantum circuits},
  journal = {Phys. Rev. A},
  volume = {111},
  pages = {042606},
  year = {2025},
  doi = {10.1103/PhysRevA.111.042606}
}

@article{Ayyash2024,
  author = {Ayyash, Mohammad and Xu, Xicheng and Ashhab, Sahel and Mariantoni, Matteo},
  title = {Driven multiphoton qubit-resonator interactions},
  journal = {Phys. Rev. A},
  volume = {110},
  pages = {053711},
  year = {2024},
  doi = {10.1103/PhysRevA.110.053711}
}

@article{Ayyash2025,
  author = {Ayyash, Mohammad and Ashhab, Sahel},
  title = {Dispersive regime of multiphoton qubit-oscillator interactions},
  journal = {Phys. Rev. A},
  volume = {112},
  pages = {023713},
  year = {2025},
  doi = {10.1103/hrc2-7nqg}
}

@article{Blumenthal2026,
  author = {Blumenthal, E. and Gutman, N. and Kaminer, I. and Hacohen-Gourgy, S.},
  title = {Single- and two-mode squeezing by modulated coupling to a {Rabi}-driven qubit},
  journal = {Adv. Quantum Technol.},
  volume = {9},
  pages = {e00490},
  year = {2026},
  doi = {10.1002/qute.202500490}
}

@article{Hope2026,
  author = {Hope, Marius K. and Lidal, Jonas and Massel, Francesco},
  title = {Preparation of conditionally squeezed states in qubit-oscillator systems},
  journal = {Phys. Rev. Research},
  volume = {8},
  pages = {L012046},
  year = {2026},
  doi = {10.1103/jrrb-hymw}
}

@article{Wang2021PND,
  author = {Wang, Chiao-Hsuan and Noh, Kyungjoo and Lebreuilly, Jos\'e and Girvin, S. M. and Jiang, Liang},
  title = {Photon-number-dependent Hamiltonian engineering for cavities},
  journal = {Phys. Rev. Applied},
  volume = {15},
  pages = {044026},
  year = {2021},
  doi = {10.1103/PhysRevApplied.15.044026}
}

@article{Heeres2015,
  author = {Heeres, Reinier W. and Vlastakis, Brian and Holland, Eric and Krastanov, Stefan and Albert, Victor V. and Frunzio, Luigi and Jiang, Liang and Schoelkopf, Robert J.},
  title = {Cavity state manipulation using photon-number selective phase gates},
  journal = {Phys. Rev. Lett.},
  volume = {115},
  pages = {137002},
  year = {2015},
  doi = {10.1103/PhysRevLett.115.137002}
}

@article{Vanselow2026,
  author = {Vanselow, A. and Beauseigneur, B. and Lattier, L. and Villiers, M. and Denis, A. and Morfin, P. and Leghtas, Z. and Campagne-Ibarcq, P.},
  title = {Dissipating quartets of excitations in a superconducting circuit},
  journal = {Phys. Rev. X},
  volume = {16},
  pages = {011032},
  year = {2026},
  doi = {10.1103/bjpc-8xcf}
}

@article{Blais2021,
  author = {Blais, Alexandre and Grimsmo, Arne L. and Girvin, S. M. and Wallraff, Andreas},
  title = {Circuit quantum electrodynamics},
  journal = {Rev. Mod. Phys.},
  volume = {93},
  pages = {025005},
  year = {2021},
  doi = {10.1103/RevModPhys.93.025005}
}

@article{Schneider2020,
  author = {Schneider, B. H. and Bengtsson, A. and Svensson, I. M. and Aref, T. and Johansson, G. and Bylander, J. and Delsing, P.},
  title = {Observation of Broadband Entanglement in Microwave Radiation from a Single Time-Varying Boundary Condition},
  journal = {Phys. Rev. Lett.},
  volume = {124},
  pages = {140503},
  year = {2020},
  doi = {10.1103/PhysRevLett.124.140503}
}

@article{Wilson2010,
  author  = {Wilson, C. M. and Duty, T. and Sandberg, M. and
             Persson, F. and Shumeiko, V. and Delsing, P.},
  title   = {Photon Generation in an Electromagnetic Cavity with a
             Time-Dependent Boundary},
  journal = {Phys. Rev. Lett.},
  volume  = {105},
  pages   = {233907},
  year    = {2010},
  doi     = {10.1103/PhysRevLett.105.233907}
}

@article{Puri2017,
  author  = {Puri, Shruti and Boutin, Samuel and Blais, Alexandre},
  title   = {Engineering the Quantum States of Light in a
             {Kerr}-Nonlinear Resonator by Two-Photon Driving},
  journal = {npj Quantum Inf.},
  volume  = {3},
  pages   = {18},
  year    = {2017},
  doi     = {10.1038/s41534-017-0019-1}
}

@article{Simoen2015,
  author = {Simoen, Micha\"el and Chang, C. W. Sandbo and Krantz, Philip and Bylander, Jonas and Wustmann, Waltraut and Shumeiko, Vitaly and Delsing, Per and Wilson, C. M.},
  title = {Characterization of a multimode coplanar waveguide parametric amplifier},
  journal = {J. Appl. Phys.},
  volume = {118},
  pages = {154501},
  year = {2015},
  doi = {10.1063/1.4933265}
}

@article{Bengtsson2018,
  author = {Bengtsson, Andreas and Krantz, Philip and Simoen, Micha\"el and Svensson, Ida-Maria and Schneider, Ben and Shumeiko, Vitaly and Delsing, Per and Bylander, Jonas},
  title = {Nondegenerate parametric oscillations in a tunable superconducting resonator},
  journal = {Phys. Rev. B},
  volume = {97},
  pages = {144502},
  year = {2018},
  doi = {10.1103/PhysRevB.97.144502}
}

@article{Chang2018,
  author = {Chang, C. W. Sandbo and Simoen, M. and Aumentado, Jos\'e and Sab\'in, Carlos and Forn-D\'iaz, P. and Vadiraj, A. M. and Quijandr\'ia, Fernando and Johansson, G. and Fuentes, I. and Wilson, C. M.},
  title = {Generating multimode entangled microwaves with a superconducting parametric cavity},
  journal = {Phys. Rev. Applied},
  volume = {10},
  pages = {044019},
  year = {2018},
  doi = {10.1103/PhysRevApplied.10.044019}
}

@article{Lee2020,
  author = {Lee, Nathan R. A. and Pechal, Marek and Wollack, E. Alex and Arrangoiz-Arriola, Patricio and Wang, Zhaoyou and Safavi-Naeni, Amir H.},
  title = {Propagation of microwave photons along a synthetic dimension},
  journal = {Phys. Rev. A},
  volume = {101},
  pages = {053807},
  year = {2020},
  doi = {10.1103/PhysRevA.101.053807}
}

@article{Makihara2024,
  author = {Makihara, Takuma and Lee, Nathan and Guo, Yudan and Guan, Wenyan and Safavi-Naeini, Amir H.},
  title = {A parametrically programmable delay line for microwave photons},
  journal = {Nat. Commun.},
  volume = {15},
  pages = {4640},
  year = {2024},
  doi = {10.1038/s41467-024-48975-x}
}

@article{Busnaina2024,
  author = {Busnaina, Jamal H. and Shi, Zheng and McDonald, Alexander and Dubyna, Dmytro and Nsanzineza, Ibrahim and Hung, Jimmy S. C. and Chang, C. W. Sandbo and Clerk, Aashish A. and Wilson, Christopher M.},
  title = {Quantum simulation of the bosonic Kitaev chain},
  journal = {Nat. Commun.},
  volume = {15},
  pages = {3065},
  year = {2024},
  doi = {10.1038/s41467-024-47186-8}
}

@article{Armour2015,
  author = {Armour, A. D. and Kubala, B. and Ankerhold, J.},
  title = {Josephson photonics with a two-mode superconducting circuit},
  journal = {Phys. Rev. B},
  volume = {91},
  pages = {184508},
  year = {2015},
  doi = {10.1103/PhysRevB.91.184508}
}

@article{Chang2020ThreePhoton,
  author = {Chang, C. W. Sandbo and Sab\'in, Carlos and Forn-D\'iaz, P. and Quijandr\'ia, Fernando and Vadiraj, A. M. and Nsanzineza, I. and Johansson, G. and Wilson, C. M.},
  title = {Observation of three-photon spontaneous parametric down-conversion in a superconducting parametric cavity},
  journal = {Phys. Rev. X},
  volume = {10},
  pages = {011011},
  year = {2020},
  doi = {10.1103/PhysRevX.10.011011}
}

@article{Wustmann2017,
  author = {Wustmann, Waltraut and Shumeiko, Vitaly},
  title = {Nondegenerate parametric resonance in a tunable superconducting cavity},
  journal = {Phys. Rev. Applied},
  volume = {8},
  pages = {024018},
  year = {2017},
  doi = {10.1103/PhysRevApplied.8.024018}
}

@article{SchriefferWolff1966,
  author = {Schrieffer, J. R. and Wolff, P. A.},
  title = {Relation between the Anderson and Kondo Hamiltonians},
  journal = {Phys. Rev.},
  volume = {149},
  pages = {491--492},
  year = {1966},
  doi = {10.1103/PhysRev.149.491}
}

@article{Bravyi2011,
  author = {Bravyi, Sergey and DiVincenzo, David P. and Loss, Daniel},
  title = {Schrieffer--Wolff transformation for quantum many-body systems},
  journal = {Ann. Phys.},
  volume = {326},
  pages = {2793--2826},
  year = {2011},
  doi = {10.1016/j.aop.2011.06.004}
}

@article{Malekakhlagh2022,
  author = {Malekakhlagh, Moein and Magesan, Easwar and Govia, Luke C. G.},
  title = {Time-dependent Schrieffer--Wolff--Lindblad perturbation theory: Measurement-induced dephasing and second-order Stark shift in dispersive readout},
  journal = {Phys. Rev. A},
  volume = {106},
  pages = {052601},
  year = {2022},
  doi = {10.1103/PhysRevA.106.052601}
}

@article{Magnus1954,
  author = {Magnus, Wilhelm},
  title = {On the exponential solution of differential equations for a linear operator},
  journal = {Commun. Pure Appl. Math.},
  volume = {7},
  pages = {649--673},
  year = {1954},
  doi = {10.1002/cpa.3160070404}
}

@article{Blanes2009,
  author = {Blanes, Sergio and Casas, Fernando and Oteo, Jos\'e A. and Ros, Jos\'e},
  title = {The {Magnus} expansion and some of its applications},
  journal = {Phys. Rep.},
  volume = {470},
  pages = {151--238},
  year = {2009},
  doi = {10.1016/j.physrep.2008.11.001}
}

@article{Ma2021BosonicControl,
  author = {Ma, Wen-Long and Puri, Shruti and Schoelkopf, Robert J. and Devoret, Michel H. and Girvin, S. M. and Jiang, Liang},
  title = {Quantum control of bosonic modes with superconducting circuits},
  journal = {Science Bulletin},
  volume = {66},
  number = {17},
  pages = {1789--1805},
  year = {2021},
  doi = {10.1016/j.scib.2021.05.024}
}

@article{CampagneIbarcq2020,
  author = {Campagne-Ibarcq, P. and Eickbusch, A. and Touzard, S. and Zalys-Geller, E. and Frattini, N. E. and Sivak, V. V. and Reinhold, P. and Puri, S. and Shankar, S. and Schoelkopf, R. J. and Frunzio, L. and Mirrahimi, M. and Devoret, M. H.},
  title = {Quantum error correction of a qubit encoded in grid states of an oscillator},
  journal = {Nature},
  volume = {584},
  pages = {368--372},
  year = {2020},
  doi = {10.1038/s41586-020-2603-3}
}

@article{Sivak2023,
  author = {Sivak, V. V. and Eickbusch, A. and Royer, B. and Singh, S. and Tsioutsios, I. and Ganjam, S. and Miano, A. and Brock, B. L. and Ding, A. Z. and Frunzio, L. and Girvin, S. M. and Schoelkopf, R. J. and Devoret, M. H.},
  title = {Real-time quantum error correction beyond break-even},
  journal = {Nature},
  volume = {616},
  pages = {50--55},
  year = {2023},
  doi = {10.1038/s41586-023-05782-6}
}

@article{Franca2001,
  author = {Fran{\c c}a Santos, M. and Solano, E. and de Matos Filho, R. L.},
  title = {Conditional large {Fock} state preparation and field state reconstruction in cavity {QED}},
  journal = {Phys. Rev. Lett.},
  volume = {87},
  pages = {093601},
  year = {2001},
  doi = {10.1103/PhysRevLett.87.093601}
}

@article{Solano2005,
  author = {Solano, E.},
  title = {Selective interactions in trapped ions: State reconstruction and quantum logic},
  journal = {Phys. Rev. A},
  volume = {71},
  pages = {013813},
  year = {2005},
  doi = {10.1103/PhysRevA.71.013813}
}

@article{Vrajitoarea2020,
  author = {Vrajitoarea, Andrei and Huang, Ziwen and Groszkowski, Peter and Koch, Jens and Houck, Andrew A.},
  title = {Quantum control of an oscillator using a stimulated {Josephson} nonlinearity},
  journal = {Nat. Phys.},
  volume = {16},
  pages = {211--217},
  year = {2020},
  doi = {10.1038/s41567-019-0703-5}
}

@article{You2024Crosstalk,
  author = {You, Xinyuan and Lu, Yunwei and Kim, Taeyoon and K{\"u}rk{\c{c}}{\"u}o{\u g}lu, Do{\u g}a Murat and Zhu, Shaojiang and van Zanten, David and Roy, Tanay and Lu, Yao and Chakram, Srivatsan and Grassellino, Anna and Romanenko, Alexander and Koch, Jens and Zorzetti, Silvia},
  title = {Crosstalk-robust quantum control in multimode bosonic systems},
  journal = {Phys. Rev. Applied},
  volume = {22},
  pages = {044072},
  year = {2024},
  doi = {10.1103/PhysRevApplied.22.044072}
}

@article{Gao2019,
  author = {Gao, Yvonne Y. and Lester, Brian J. and Chou, Kevin S. and Frunzio, Luigi and Devoret, Michel H. and Jiang, Liang and Girvin, S. M. and Schoelkopf, Robert J.},
  title = {Entanglement of bosonic modes through an engineered exchange interaction},
  journal = {Nature},
  volume = {566},
  pages = {509--512},
  year = {2019},
  doi = {10.1038/s41586-019-0970-4}
}

@article{Lambert2026QuTiP5,
  author  = {Neill Lambert and Eric Gigu{\`e}re and Paul Menczel and
             Boxi Li and Patrick Hopf and Gerardo Su{\'a}rez and
             Marc Gali and Jake Lishman and Rushiraj Gadhvi and
             Rochisha Agarwal and Asier Galicia and Nathan Shammah and
             Paul Nation and J. R. Johansson and Shahnawaz Ahmed and
             Simon Cross and Alexander Pitchford and Franco Nori},
  title   = {QuTiP 5: The Quantum Toolbox in Python},
  journal = {Physics Reports},
  volume  = {1153},
  pages   = {1--62},
  year    = {2026},
  doi     = {10.1016/j.physrep.2025.10.001}
}

@article{Groszkowski2021Scqubits,
  author  = {Peter Groszkowski and Jens Koch},
  title   = {scqubits: a Python package for superconducting qubits},
  journal = {Quantum},
  volume  = {5},
  pages   = {583},
  year    = {2021},
  doi     = {10.22331/q-2021-11-17-583}
}

@article{Chitta2022Scqubits,
  author  = {Sai Pavan Chitta and Tianpu Zhao and Ziwen Huang and
             Ian Mondragon-Shem and Jens Koch},
  title   = {Computer-aided quantization and numerical analysis of
             superconducting circuits},
  journal = {New Journal of Physics},
  volume  = {24},
  pages   = {103020},
  year    = {2022},
  doi     = {10.1088/1367-2630/ac94f2}
}

@article{ArayaDay2025Pymablock,
  author  = {Araya Day, Isidora and Miles, Sebastian and
             Kerstens, Hugo K. and Varjas, Daniel and
             Akhmerov, Anton R.},
  title   = {{Pymablock}: An Algorithm and a Package for
             Quasi-Degenerate Perturbation Theory},
  journal = {SciPost Phys. Codebases},
  volume  = {50},
  year    = {2025},
  doi     = {10.21468/SciPostPhysCodeb.50}
}

@article{Felicetti2014,
  author  = {Felicetti, S. and Sanz, M. and Lamata, L. and
             Romero, G. and Johansson, G. and Delsing, P. and
             Solano, E.},
  title   = {Dynamical {Casimir} Effect Entangles Artificial Atoms},
  journal = {Phys. Rev. Lett.},
  volume  = {113},
  pages   = {093602},
  year    = {2014},
  doi     = {10.1103/PhysRevLett.113.093602}
}

@article{Grimm2020,
  author  = {Grimm, A. and Frattini, N. E. and Puri, S. and
             Mundhada, S. O. and Touzard, S. and Mirrahimi, M. and
             Girvin, S. M. and Shankar, S. and Devoret, M. H.},
  title   = {Stabilization and Operation of a {Kerr}-Cat Qubit},
  journal = {Nature},
  volume  = {584},
  number  = {7820},
  pages   = {205--209},
  year    = {2020},
  doi     = {10.1038/s41586-020-2587-z}
}

@article{Eriksson2024,
  author  = {Eriksson, Axel M. and S{\'e}pulcre, Th{\'e}o and
             Kervinen, Mikael and Hillmann, Timo and Kudra, Marina and
             Dupouy, Simon and Lu, Yong and Khanahmadi, Maryam and
             Yang, Jiaying and Castillo-Moreno, Claudia and
             Delsing, Per and Gasparinetti, Simone},
  title   = {Universal Control of a Bosonic Mode via Drive-Activated
             Native Cubic Interactions},
  journal = {Nat. Commun.},
  volume  = {15},
  pages   = {2512},
  year    = {2024},
  doi     = {10.1038/s41467-024-46507-1}
}

@article{Diringer2024,
  author  = {Diringer, Asaf A. and Blumenthal, Eliya and
             Grinberg, Avishay and Jiang, Liang and
             Hacohen-Gourgy, Shay},
  title   = {Conditional-{NOT} Displacement: Fast Multioscillator
             Control with a Single Qubit},
  journal = {Phys. Rev. X},
  volume  = {14},
  pages   = {011055},
  year    = {2024},
  doi     = {10.1103/PhysRevX.14.011055}
}

@article{Mundhada2019,
  author  = {Mundhada, S. O. and Grimm, A. and Venkatraman, J. and
             Minev, Z. K. and Touzard, S. and Frattini, N. E. and
             Sivak, V. V. and Sliwa, K. and Reinhold, P. and
             Shankar, S. and Mirrahimi, M. and Devoret, M. H.},
  title   = {Experimental Implementation of a Raman-Assisted
             Eight-Wave Mixing Process},
  journal = {Phys. Rev. Applied},
  volume  = {12},
  pages   = {054051},
  year    = {2019},
  doi     = {10.1103/PhysRevApplied.12.054051}
}

@article{Schuster2007,
  author  = {Schuster, D. I. and Houck, A. A. and Schreier, J. A. and
             Wallraff, A. and Gambetta, J. M. and Blais, A. and
             Frunzio, L. and Majer, J. and Johnson, B. and
             Devoret, M. H. and Girvin, S. M. and Schoelkopf, R. J.},
  title   = {Resolving Photon Number States in a Superconducting Circuit},
  journal = {Nature},
  volume  = {445},
  pages   = {515--518},
  year    = {2007},
  doi     = {10.1038/nature05461}
}

@article{Sandberg2008,
  author  = {Sandberg, M. and Wilson, C. M. and Persson, F. and
             Bauch, T. and Johansson, G. and Shumeiko, V. and
             Duty, T. and Delsing, P.},
  title   = {Tuning the Field in a Microwave Resonator Faster than
             the Photon Lifetime},
  journal = {Applied Physics Letters},
  volume  = {92},
  pages   = {203501},
  year    = {2008},
  doi     = {10.1063/1.2929367}
}

@article{Wustmann2013,
  author  = {Wustmann, Waltraut and Shumeiko, Vitaly},
  title   = {Parametric Resonance in Tunable Superconducting Cavities},
  journal = {Physical Review B},
  volume  = {87},
  pages   = {184501},
  year    = {2013},
  doi     = {10.1103/PhysRevB.87.184501}
}

@article{Kochetov2015,
  author  = {Kochetov, Bogdan A. and Fedorov, Arkady},
  title   = {Higher-Order Nonlinear Effects in a Josephson Parametric
             Amplifier},
  journal = {Physical Review B},
  volume  = {92},
  pages   = {224304},
  year    = {2015},
  doi     = {10.1103/PhysRevB.92.224304}
}

@article{Bennett1996,
  author  = {Bennett, Charles H. and Bernstein, Herbert J. and
             Popescu, Sandu and Schumacher, Benjamin},
  title   = {Concentrating Partial Entanglement by Local Operations},
  journal = {Phys. Rev. A},
  volume  = {53},
  pages   = {2046--2052},
  year    = {1996},
  doi     = {10.1103/PhysRevA.53.2046}
}

@article{Horodecki2009,
  author  = {Horodecki, Ryszard and Horodecki, Pawe{\l} and
             Horodecki, Micha{\l} and Horodecki, Karol},
  title   = {Quantum Entanglement},
  journal = {Rev. Mod. Phys.},
  volume  = {81},
  pages   = {865--942},
  year    = {2009},
  doi     = {10.1103/RevModPhys.81.865}
}

@article{BarnettCroke2009,
  author  = {Barnett, Stephen M. and Croke, Sarah},
  title   = {Quantum State Discrimination},
  journal = {Adv. Opt. Photon.},
  volume  = {1},
  pages   = {238--278},
  year    = {2009},
  doi     = {10.1364/AOP.1.000238}
}

@article{Englert1996,
  author  = {Englert, Berthold-Georg},
  title   = {Fringe Visibility and Which-Way Information:
             An Inequality},
  journal = {Phys. Rev. Lett.},
  volume  = {77},
  pages   = {2154--2157},
  year    = {1996},
  doi     = {10.1103/PhysRevLett.77.2154}
}

@article{Vlastakis2013,
  author  = {Vlastakis, Brian and Kirchmair, Gerhard and Leghtas, Zaki
             and Nigg, Simon E. and Frunzio, Luigi and Girvin, S. M.
             and Mirrahimi, Mazyar and Devoret, M. H.
             and Schoelkopf, R. J.},
  title   = {Deterministically Encoding Quantum Information Using
             100-Photon Schr{\"o}dinger Cat States},
  journal = {Science},
  volume  = {342},
  pages   = {607--610},
  year    = {2013},
  doi     = {10.1126/science.1243289}
}

@article{Weedbrook2012,
  author  = {Weedbrook, Christian and Pirandola, Stefano and
             Garc{\'i}a-Patr{\'o}n, Ra{\'u}l and Cerf, Nicolas J.
             and Ralph, Timothy C. and Shapiro, Jeffrey H.
             and Lloyd, Seth},
  title   = {Gaussian Quantum Information},
  journal = {Rev. Mod. Phys.},
  volume  = {84},
  pages   = {621--669},
  year    = {2012},
  doi     = {10.1103/RevModPhys.84.621}
}

@article{CastellanosBeltran2008,
  author  = {Castellanos-Beltran, M. A. and Irwin, K. D. and
             Hilton, G. C. and Vale, L. R. and Lehnert, K. W.},
  title   = {Amplification and Squeezing of Quantum Noise with a
             Tunable Josephson Metamaterial},
  journal = {Nat. Phys.},
  volume  = {4},
  pages   = {929--931},
  year    = {2008},
  doi     = {10.1038/nphys1090}
}

@article{Mallet2011,
  author  = {Mallet, F. and Castellanos-Beltran, M. A. and Ku, H. S.
             and Glancy, S. and Knill, E. and Irwin, K. D.
             and Hilton, G. C. and Vale, L. R. and Lehnert, K. W.},
  title   = {Quantum State Tomography of an Itinerant Squeezed
             Microwave Field},
  journal = {Phys. Rev. Lett.},
  volume  = {106},
  pages   = {220502},
  year    = {2011},
  doi     = {10.1103/PhysRevLett.106.220502}
}

@article{Kenfack2004,
  author  = {Kenfack, Anatole and {\.Z}yczkowski, Karol},
  title   = {Negativity of the {Wigner} Function as an Indicator
             of Non-Classicality},
  journal = {J. Opt. B: Quantum Semiclass. Opt.},
  volume  = {6},
  pages   = {396--404},
  year    = {2004},
  doi     = {10.1088/1464-4266/6/10/003}
}

@article{Albarelli2018,
  author  = {Albarelli, Francesco and Genoni, Marco G. A.
             and Paris, Matteo G. A. and Ferraro, Alessandro},
  title   = {Resource Theory of Quantum Non-Gaussianity and
             {Wigner} Negativity},
  journal = {Phys. Rev. A},
  volume  = {98},
  pages   = {052350},
  year    = {2018},
  doi     = {10.1103/PhysRevA.98.052350}
}

@article{BraunsteinCaves1994,
  author  = {Braunstein, Samuel L. and Caves, Carlton M.},
  title   = {Statistical Distance and the Geometry of Quantum States},
  journal = {Phys. Rev. Lett.},
  volume  = {72},
  pages   = {3439--3443},
  year    = {1994},
  doi     = {10.1103/PhysRevLett.72.3439}
}

@article{Paris2009,
  author  = {Paris, Matteo G. A.},
  title   = {Quantum Estimation for Quantum Technology},
  journal = {Int. J. Quantum Inf.},
  volume  = {7},
  number  = {supp01},
  pages   = {125--137},
  year    = {2009},
  doi     = {10.1142/S0219749909004839}
}

@article{Dodonov1974,
  author  = {Dodonov, V. V. and Malkin, I. A. and Man'ko, V. I.},
  title   = {Even and Odd Coherent States and Excitations of a
             Singular Oscillator},
  journal = {Physica},
  volume  = {72},
  number  = {3},
  pages   = {597--615},
  year    = {1974},
  doi     = {10.1016/0031-8914(74)90215-8}
}

@article{YurkeStoler1986,
  author  = {Yurke, B. and Stoler, D.},
  title   = {Generating Quantum Mechanical Superpositions of
             Macroscopically Distinguishable States via
             Amplitude Dispersion},
  journal = {Phys. Rev. Lett.},
  volume  = {57},
  pages   = {13--16},
  year    = {1986},
  doi     = {10.1103/PhysRevLett.57.13}
}
\end{document}